\documentclass[apj,twocolumn,floatfix,twocolappendix,numberedappendix,appendixfloats,colorlinks]{openjournal}
\usepackage{hyperref}
\usepackage{nameref} 
\hypersetup{linkcolor=red,citecolor=Blue,filecolor=cyan,urlcolor=magenta}
\usepackage{booktabs,caption}
\usepackage[flushleft]{threeparttable}
\usepackage{tabularx}
\usepackage[dvipsnames]{xcolor}
\usepackage{soul}
\usepackage{amsmath}
\usepackage{textcomp,gensymb}
\usepackage{orcidlink}
\usepackage[T1]{fontenc}
\usepackage{rotating}
\usepackage{sidecap} 
\sidecaptionvpos{figure}{m}
\usepackage{subcaption}

\renewcommand{\baselinestretch}{1.05} 
\newcommand{\avgsnr}{$\langle$SNR$\rangle$}{}
\newcommand{\avgsnrm}{\langle\rm{SNR}\rangle}

\newcommand{\reseff}{$\mathcal{R}_{\rm{eff}}$}{}
\newcommand{\reseffm}{\mathcal{R}_{\rm{eff}}}
\newcommand{\res}{$\mathcal{R}$}{}
\newcommand{\ts}{\textsuperscript}

\newcommand{\new}{\textcolor{black}}

\shorttitle{Morphological metrics and their biases}
\shortauthors{Sazonova et al.}
\graphicspath{{./}{}}

\begin{document}

\title{\texttt{statmorph-lsst}: quantifying and correcting morphological biases in galaxy surveys}

\author{\vspace{-1.3cm}Elizaveta Sazonova\,\orcidlink{0000-0001-6245-5121}$^{1,2}$}
\author{Cameron R. Morgan\,\orcidlink{0009-0009-2522-3685}$^{1,2}$}
\author{Michael Balogh\,\orcidlink{0000-0003-4849-9536}$^{1,2}$}
\author{Matías Blaña\,\orcidlink{0000-0003-2139-0944}$^{3}$}
\author{Carlos G. Bornancini\,\orcidlink{0000-0001-6800-3329}$^{4,5}$}
\author{Aidan P. Cotter\,\orcidlink{0009-0004-1910-4990}$^{6}$}
\author{Darko Donevski\,\orcidlink{0000-0001-5341-2162}$^{6,7}$}
\author{Alister W. Graham\,\orcidlink{0000-0002-6496-9414
}$^{8}$}
\author{Hector M. Hernandez Toledo\,\orcidlink{0000-0001-9601-7779}$^{9}$}
\author{Benne W. Holwerda\,\orcidlink{0000-0002-4884-6756}$^{10}$}
\author{Jeyhan S. Kartaltepe\,\orcidlink{0000-0001-9187-3605}$^{11}$}
\author{Garreth Martin\,\orcidlink{0000-0003-2939-8668}$^{12}$}
\author{William J. Pearson\,\orcidlink{0000-0002-7300-2213}$^{6}$}
\author{Rossella Ragusa\,\orcidlink{0009-0000-6680-523X}$^{13}$}
\author{Vicente Rodriguez-Gomez\,\orcidlink{0000-0002-9495-0079}$^{14}$}
\author{Michael J. Rutkowski\,\orcidlink{0000-0001-7016-5220}$^{15}$}
\author{Jose Antonio Vázquez-Mata\,\orcidlink{0000-0001-8694-1204}$^{9}$}
\author{Rogier A. Windhorst,\orcidlink{0000-0001-8156-6281}$^{16}$}
\author{Jacob Yuzovitskiy,\orcidlink{0009-0007-4440-2000}$^{17}$}

\affiliation{$^{1}$Waterloo Centre for Astrophysics, University of Waterloo, Waterloo, ON, N2L 3G1 Canada}\email{liza.sazonova@uwaterloo.ca}
\affiliation{$^{2}$Department of Physics and Astronomy, University of Waterloo, Waterloo, ON N2L 3G1, Canada}

\affiliation{$^{3}$Vicerrectoría de Investigación y Postgrado, Universidad de La Serena, La Serena 1700000, Chile}

\affiliation{$^{4}$Instituto de Astronomía Teórica y Experimental, (IATE, CONICET-UNC), Córdoba, Argentina}
\affiliation{$^{5}$Universidad Nacional de Córdoba, Observatorio Astronómico de Córdoba, Laprida 854, X5000BGR, Córdoba, Argentina}

\affiliation{$^{6}$National Center for Nuclear Research, Pasteura 7, 02-093 Warsaw, Poland}

\affiliation{$^{7}$SISSA, Via Bonomea 265, 34136 Trieste, Italy}

\affiliation{$^{8}$Centre for Astrophysics and Supercomputing, Swinburne University of Technology, Hawthorn, VIC 3122, Australia}

\affiliation{$^{9}$Universidad Nacional Autónoma de México, Instituto de Astronomía, A.P. 70-264, 04510 Ciudad de México, México}

\affiliation{$^{10}$University of Louisville, Department of Physics and Astronomy, 102 Natural Science Building, 40292 KY Louisville, USA}

\affiliation{$^{11}$Laboratory for Multiwavelength Astrophysics, School of Physics and
Astronomy, Rochester Institute of Technology, 84 Lomb Memorial Drive, Rochester, NY 14623, USA}

\affiliation{$^{12}$School of Physics and Astronomy, University of Nottingham, University Park, Nottingham NG7 2RD, UK}

\affiliation{$^{13}$INAF – Osservatorio Astronomico di Capodimonte, Salita Moiariello 16, 80131 Napoli, Italy}

\affiliation{$^{14}$Instituto de Radioastronomía y Astrofísica, Universidad Nacional Autónoma de México, A.P. 72-3, 58089 Morelia, México}

\affiliation{$^{15}$Minnesota State University, Mankato, Department of Physics and Astronomy, 141 Trafton Science Center N, Mankato, MN 56001, USA}

\affiliation{$^{16}$School of Earth and Space Exploration, Arizona State University,
Tempe, AZ 85287-6004, USA}

\affiliation{$^{17}$CUNY College of Staten Island, Department of Physics \& Astronomy, 2800 Victory Blvd., Staten Island, NY 10314, USA}

\begin{abstract}

Quantitative morphology provides a key probe of galaxy evolution across cosmic time and environments. However, these metrics can be biased by changes in imaging quality --  resolution and depth -- either across the survey area or the sample. To prepare for the upcoming Rubin LSST data, we investigate this bias for \textit{all} metrics measured by \texttt{statmorph} and single-component S\'ersic fitting with \textsc{Galfit}. We find that geometrical measurements (\new{centroid,} ellipticity, axis ratio, Petrosian radius, and effective radius) are \new{robust within 10\%} at most depths and resolutions. Light concentration measurements ($C$, Gini, $M_{20}$) systematically decrease with resolution, leading low-mass or high-redshift bulge-dominated sources to appear indistinguishable from disks. S\'ersic index $n$, while unbiased, suffers from a 20-40\% uncertainty due to degeneracies in the S\'ersic fit. Disturbance measurements ($A$, $A_S$, $D$) depend on the signal-to-noise \new{ratio} and are thus affected by noise and surface brightness dimming. We quantify this dependence for each parameter, offer empirical correction functions, and show that the evolution in $C$ observed in JWST galaxies can be explained purely by observational biases. We propose two new measurements -- isophotal asymmetry $A_X$ and substructure $St$ -- that aim to resolve some of these biases. Finally, we provide a Python package \texttt{statmorph-lsst}\footnote{\href{https://github.com/astro-nova/statmorph-lsst}{\texttt{statmorph-lsst} GitHub respository}} implementing these changes and a full dataset that enables tests of custom functions\footnote{\href{https://zenodo.org/records/17585609}{Dataset DOI: 10.5281/zenodo.17585608}}. 

\end{abstract}


\maketitle

\section{Introduction} \label{sec:intro}

The evolution of galaxies is driven by two main channels. The first is the evolution of the galaxies’ stellar population and gas content, driven by star formation in clouds of molecular gas. The second is the evolution of galactic structure, shaped both by star formation altering the potential well and gravitational interactions that can redistribute the stars after they formed. In the local Universe, the two are closely linked: star-forming galaxies have primarily disk morphologies while most quiescent galaxies are either elliptical or lenticular with a prominent central bulge \citep[e.g.,][]{Strateva2001,Schawinski2014}. However, the reason the two are linked is still debated: what physical processes tie star formation and galactic structure together so closely, and at what point in the evolution of the Universe do they become correlated?

One way to answer these questions is to determine when and where the color-morphology bimodality is first established: by looking at the changes in structure in different environments \citep[e.g.,][]{Dressler1980, Holden2007} and across redshifts \citep[e.g.,][]{Mortlock2013, Kartaltepe2023}. To carry out these studies, we must robustly quantify the galaxy light distribution. 

This is traditionally done using either a parametric approach, such as fitting a S\'ersic profile to the light distribution \citep{Sersic1963,Graham2005}, or a suite of non-parametric measurements: concentration, asymmetry, smoothness \citep[CAS;][]{Conselice2003}, Gini-$M_{20}$ \citep{Lotz2004}, and others. These metrics are capable of quantifying the bulge strength\footnote{We use the term ``bulge strength'' to refer to any measure of how centrally concentrated the light is, rather than specifically the prominence of a physical bulge component or a S\'ersic index. Similarly we use the term ``early-type'' to mean a galaxy with a centrally concentrated light profile regardless of its colour or the presence of a true separate bulge component, and ``late-type'' a galaxy with relatively more flux away from the center. We do not attempt to classify galaxies kinematically using these projected structural quantities.} \citep[e.g.,][]{Bruce2014}, the degree of disturbance \citep[e.g.,][]{Lotz2004}, and more broadly the impact of a diverse range of internal and external processes, such as ram pressure stripping \citep[e.g.,][]{Roberts2020,Bellhouse2022}, star formation knots \citep{Ferreira2022a}, dust lanes \citep{Rutkowski2013,Sazonova2021}, gas outflows \citep{Deg2023}, and even the cumulative baryonic feedback \citep{Watkins2025,Martin2025}. Since each parameter can be used for a wide variety of studies, these quantitative measurements are an essential tool in studying the evolution of galactic structure.

In the last few decades, the data from the Hubble Space Telescope (HST) and, more recently, the James Webb Space Telescope (JWST), allowed us to finally probe these topics across cosmic time -- leading to more heated debates. Several recent studies found that high-redshift galaxies, in general, are more disk-like than bulge-like \citep[e.g.,][]{Ferreira2022b,Ferreira2023,Kartaltepe2023,Morishita2024}. On the other hand, a new population of massive compact galaxies emerged first at $z$$>$1 and then a different population at $z$$\sim$5, both non-existent in the local Universe \citep[e.g.,][]{Poggianti2013,Matthee2024}. However, quantitative morphological measurements make it difficult to distinguish between disks and spheroids in the early Universe, considering some high-redshift sources are essentially unresolved \citep[e.g.,][]{Whalen2025}. Finally, some morphological analyses point at the relative dearth of mergers at the highest redshifts -- first at $z$$\sim$3 with HST \citep{Mortlock2013}, then at $z$$\sim$5 and even up to $z$$\sim$$10$ with JWST \citep{Ferreira2023, Ono2025}. By nature of research, we are pushing our available data to the most distant, most compact, lowest mass galaxies, probing the extremes of our ability to image and resolve galaxies themselves or their faint outskirts. While these studies demonstrate the potential of morphology metrics have in helping us to understand the structure of galaxies in the early Universe, they also have to be considered with caution.


Most quantitative morphology metrics have been shown to be dependent in some way on the properties of the image, such as resolution and depth \citep[e.g.,][]{Graham2005,Holwerda2011,Bottrell2019}. For example, \cite{Thorp2021} showed that the asymmetry parameter is sensitive to depth, and so it is difficult to detect mergers in galaxies with a low signal-to-noise \new{ratio}. These are especially important considerations in the era of large surveys, such as Vera Rubin Legacy Survey of Space and Time (LSST), and deep observations, such as JWST. Today we are probing the most extreme objects -- the faintest, smallest, and the highest-redshift ones -- pushing our metrics to their limit as galaxies span smaller angular sizes and suffer from cosmological surface brightness dimming, effectively making us unable to detect and measure faint features.

Moreover, quantitative morphology metrics are sensitive not only to the image itself, but also the implementation of the metric -- the definition of the centre, the segmentation map, and the algorithm itself. For example, galaxies observed as part of SDSS survey have different asymmetry values in \cite{Pawlik2016} and \cite{Sazonova2021} catalogs, which use \texttt{PawlikMorph} and \texttt{statmorph} \citep{statmorph}, respectively, so it is important to test the performance of individual codes as well as metrics. Since \texttt{statmorph} is the most commonly used code to measure non-parametric morphology that is free and open source, we decided to test the performance of all morphological measurements available in \texttt{statmorph} against changes in resolution and depth.

It is therefore crucial to come up with a common set of morphological metrics for all users, which 1) have a consistent implementation, 2) have a well-understood relationship with image quality, and 3) are publicly available and easy to implement. With such a set of metrics, catalogs can be compared across different instruments, surveys, or epochs of a single survey; and we can robustly study the evolution of galaxy structure across cosmic time without simply observing how redshift biases the metrics we chose. 

This work is the first step in this task. In this paper, we systematically quantify how morphology metrics measured
with \texttt{statmorph} and \textsc{Galfit} depend on image resolution and depth. We explain in detail how \texttt{statmorph} metrics are calculated and the important caveats and pitfalls in their algorithm, and provide empirical corrections that can mitigate the image biases. 

To do this, we used a dataset of 190 nearby galaxies observed by HST in the optical \textit{I}-band filter, which provides an exquisite resolution of under 25 parsec/px at their distances. We chose objects with sufficiently deep imaging, reaching the \textit{I}-band surface brightness limit \new{at least} $\mu_0 = 24$ mag/arcsec$^2$. We then degraded these images to varying resolutions and surface brightness limits to produce 60,000 augmented images, and evaluated the performance of the morphology metrics provided in \texttt{statmorph}.

The paper is organized as follows. Our dataset and the data processing steps we took prior to evaluating morphology are described in Section \ref{sec:data}. Section \ref{sec:morphology} makes up the bulk of the paper: we define each morphology metric provided by statmorph, discuss its implementation details, show its dependence on resolution and depth, and for some provide an empirical correction term. In Section \ref{sec:discussion}, we discuss our results and potential implications for studies of galaxy morphology across cosmic time, showing how some results can be seen as simply caused by a measurement bias. We summarize our results in Section \ref{sec:summary}. Throughout this work, we use Planck 2018 $\Lambda$CDM cosmology \citep{Planck2018} and AB magnitudes. 

For convenience, Table \ref{tab:toc} provides a look-up table listing each parameter studied in this work and the bias of this parameter across our sample, as well as in low-resolution and low-depth regimes. We encourage the readers to use this table to quickly jump to the parameters they are most interested in. A visual summary of our results is in Fig. \ref{fig:summary} at the end of this paper.


{\renewcommand{\arraystretch}{1.5}
\begin{table*}
\centering
\footnotesize                     
\setlength{\tabcolsep}{2pt}       
\begin{threeparttable}
\caption{Table of Section \ref{sec:morphology} contents}
    \begin{tabular}{lllllrrr}
        \toprule
            $p$ & Parameter Nxame & Sec. & $x$ & $y$ & Error [all data] & Error [low-res] & Error [low-SNR]\\
        \midrule
            \multicolumn{8}{l}{Geometric measurements (Sec. \ref{sec:moments})}\\
        \midrule
            \footnotemark[1]$\mathbf{x}_0^A$ & Centroid &
                    \ref{sec:moments} & \reseff{}& \avgsnr{} &
                    $4.40^{+12.8}_{-3.53}$& $16.6^{+11.1}_{-4.47}$& $4.34^{+12.3}_{-3.32}$\\

            $e$ & Ellipticity &
                \ref{sec:moments} & \reseff{}& \avgsnr{} &
                $-0.02^{+0.09}_{-0.12}$& $-0.13^{+0.18}_{-0.15}$& $-0.02^{+0.14}_{-0.15}$\\
                
            \footnotemark[1]$\theta$ & Orientation &
                \ref{sec:moments} & \reseff{}& \avgsnr{} &
                $0.30^{+16.2}_{-14.6}$& $1.20^{+29.6}_{-22.3}$& $0.65^{+30.1}_{-25.2}$\\

            \footnotemark[1]$\theta_{e>0.3}$ & Orientation ($e>0.3$) &
                \ref{sec:moments} & \reseff{}& \avgsnr{} &
                $0.08^{+8.44}_{-6.17}$& $0.63^{+27.7}_{-20.5}$& $0.53^{+21.3}_{-13.8}$\\

            \footnotemark[1]$\theta^{\rm{Sersic}}$ & S\'ersic orientation &
                \ref{sec:sersic} & \reseff{}& \avgsnr{} &
                $-0.02^{+4.70}_{-4.82}$& $-0.21^{+25.0}_{-18}$& $-0.02^{+8.71}_{-7.09}$\\

            \footnotemark[1]$\theta^{\rm{Sersic}}_{e>0.3}$ & S\'ersic orientation ($e>0.3$) &
                \ref{sec:sersic} & \reseff{}& \avgsnr{} &
                $-0.04^{+1.98}_{-3.07}$& $-0.50^{+10.8}_{-11.5}$& $-0.07^{+4.00}_{-5.74}$\\
            
            $e_{\rm{Sersic}}$ & S\'ersic ellipticity &
                \ref{sec:sersic} & \reseff{}& \avgsnr{} &
                $0.00^{+0.06}_{-0.04}$& $0.05^{+0.15}_{-0.10}$& $0.00^{+0.07}_{-0.05}$\\
        
        \midrule
            \multicolumn{8}{l}{Radius measurements (Sec. \ref{sec:radii})}\\
        \midrule
            $R_{p, \circ}$\footnotemark[2] & Petrosian radius (circular) &
                \ref{sec:rpet} & \res{}& \avgsnr{} &
                $0.08^{+1.07}_{-0.86}$ & $0.70^{+1.40}_{-1.29}$ & $-0.12^{+1.08}_{-1.68}$\\
            $R_{p, e}$ & Petrosian radius (elliptical) &
                \ref{sec:rpet} & \res{}& \avgsnr{} &
                $0.01^{+1.23}_{-1.36}$ & $0.47^{+1.81}_{-1.77}$ & $0.01^{+1.23}_{-1.36}$\\
            $R_{20}$ & 20\% light radius &
                \ref{sec:r20} & \res{}& \avgsnr{} &
                $0.10^{+0.49}_{-0.14}$& $0.51^{+0.43}_{-0.28}$& $0.05^{+0.48}_{-0.19}$\\
            $R_{50}$ & Half-light radius &
                \ref{sec:r20} & \res{}& \avgsnr{} &
                $0.09^{+0.70}_{-0.24}$& $0.59^{+0.75}_{-0.48}$& $0.03^{+0.68}_{-0.48}$\\
            $R_{80}$ & 80\% light radius &
                \ref{sec:r20} & \reseff{}& \avgsnr{} &
                $0.06^{+0.82}_{-0.48}$& $1.07^{+1.42}_{-0.60}$& $-0.02^{+0.82}_{-1.00}$\\
            $R_{\rm{max}}$ & Maximum radius &
                \ref{sec:r20} & \res{}& \avgsnr{} &
                $-0.75^{+3.81}_{-5.12}$& $-0.47^{+5.19}_{-5.71}$& $-4.09^{+2.60}_{-8.44}$\\
            $R_{0.5}^{\rm{Sersic}}$ & S\'ersic half-light radius &
                \ref{sec:sersic} & \res{}& \avgsnr{} &
                $-0.04^{+0.71}_{-1.13}$& $-0.18^{+0.95}_{-1.32}$& $0.04^{+1.48}_{-1.06}$\\
        \midrule
            \multicolumn{8}{l}{Bulge strength measurements (Sec. \ref{sec:bulge})}\\
        \midrule

            $n$ & S\'ersic index &
                \ref{sec:sersic} & \reseff{}& \avgsnr{} &
                $-0.07^{+0.54}_{-1.09}$& $-0.81^{+1.30}_{-2.13}$& $0.01^{+0.87}_{-0.86}$\\
                
            $C$ & Concentration &
                \ref{sec:concentration} & \reseff{}& \avgsnr{} &
                $-0.22^{+0.29}_{-0.79}$& $-1.14^{+0.62}_{-0.56}$& $-0.22^{+0.32}_{-0.72}$\\

            $G$ & Gini coefficient &
                \ref{sec:gini} & \reseff{}& \avgsnr{} &
                $-0.03^{+0.04}_{-0.07}$& $-0.08^{+0.08}_{-0.05}$& $-0.05^{+0.06}_{-0.07}$\\

            $M_{20}$ & Moment of the brightest 20\%&
                \ref{sec:m20} & \reseff{}& \avgsnr{} &
                $0.15^{+0.49}_{-0.22}$& $0.63^{+0.37}_{-0.42}$& $0.20^{+0.52}_{-0.27}$\\

            $B(G,M_{20})$ & $G-M_{20}$ bulge strength &
                \ref{sec:bgm20} & \reseff{}& \avgsnr{} &
                $-0.25^{+0.30}_{-0.55}$& $-0.82^{+0.50}_{-0.37}$& $-0.39^{+0.42}_{-0.60}$\\

        \midrule
            \multicolumn{8}{l}{Disturbance measurements (Sec. \ref{sec:disturbance})}\\
        \midrule
            \footnotemark[3]$S(G,M_{20})$ & $G-M_{20}$ disturbance &
                \ref{sec:sgm20} & \reseff{}& $\mu_0$ &
                $0.01^{+0.47}_{-0.52}$& $-0.01^{+1.02}_{-0.59}$& $-0.01^{+0.46}_{-0.64}$\\
            $A_{\rm{RMS}}$ & RMS asymmetry &
                \ref{sec:aiso} & \reseff{}\footnotemark[1]& \avgsnr{} &
                $-0.01^{+0.83}_{-1.46}$& $0.15^{+1.05}_{-1.67}$& $-0.11^{+2.24}_{-2.19}$\\
            $A_{\rm{CAS}}$ & CAS asymmetry &
                \ref{sec:asymmetry} & \reseff{}& \avgsnr{} &
                $-0.06^{+0.10}_{-0.21}$ & $-0.03^{+0.08}_{-0.20}$ & $-0.21^{+0.12}_{-0.18}$\\

            \footnotemark[3]$A_{S}$ & Shape asymmetry &
                \ref{sec:ashape} & \reseff{}& $\mu_0$ &
                $0.03^{+0.20}_{-0.14}$& $0.05^{+0.23}_{-0.17}$& $0.12^{+0.20}_{-0.18}$\\
                
            $A_{X}$ & Isophotal asymmetry &
                \ref{sec:aiso} & \reseff{}& $\mu_0$ &
                -- & -- & --\\
            $A_{o}$ & Outer asymmetry &
                \ref{sec:asymmetry} & \reseff{}& \avgsnr{} &
                $-0.05^{+0.13}_{-0.20}$& $-0.02^{+0.12}_{-0.18}$& $-0.18^{+0.11}_{-0.23}$\\

        \midrule
            \multicolumn{8}{l}{Other measurements (Sec. \ref{sec:other})}\\
        \midrule
            \footnotemark[3]$M$ & Multimode index &
                \ref{sec:multimode} & \reseff{}& $\mu_0$ &
               $-0.01^{+0.44}_{-0.30}$& $-0.05^{+0.26}_{-0.16}$& $0.12^{+1.27}_{-0.38}$\\

            $I$ & Intensity index &
                \ref{sec:intensity}& \reseff{}& $\mu_0$ &
                $-0.00^{+0.08}_{-0.14}$& $0.06^{+0.23}_{-0.19}$& $-0.00^{+0.10}_{-0.17}$\\

            \footnotemark[3]$D$ & Deviation index &
                \ref{sec:deviation}& \reseff{}& \avgsnr{} &
                $0.17^{+0.81}_{-0.41}$& $0.70^{+0.92}_{-0.58}$& $0.30^{+1.06}_{-0.77}$\\
                
            $S$ & Smoothness index &
                \ref{sec:smoothness} & \reseff{}& \avgsnr{} &
                $-0.00^{+0.03}_{-0.13}$& $-0.02^{+0.17}_{-0.35}$& $-0.06^{+0.05}_{-0.45}$\\

            $St$ & Substructure index &
                \ref{sec:substructure} & \reseff{}& $\mu_0$ &
                $-0.01^{+0.04}_{-0.11}$& $-0.01^{+0.03}_{-0.07}$& $-0.05^{+0.05}_{-0.13}$\\
            
        \bottomrule
    \end{tabular}
    \begin{tablenotes}
        \item \textit{Notes} -- for each parameter, $x$ and $y$ are the most predictive resolution and signal-to-noise \new{ratio} metrics found using mutual information regression (see Sec. \ref{sec:morphology}). The error is the average bias from the baseline measurement obtained approximately at 100 pc/px depth and 23.5 mag/arcsec$^2$ surface brightness limit (see Sec. \ref{sec:morphology} for details). The final two columns give the bias in low-resolution (fewer than 5 elements) and low-SNR (fewer than 2 per pixel) subsamples.
        \item \footnotemark[1] Here we quantify the offset form the baseline in a random direction, so a correction is not possible.
        \item \footnotemark[2] We define the Petrosian radius in Sec. \ref{sec:rpet} setting $\eta=0.2$.
        \item \footnotemark[3] The error quantiles are multiplied by 10.
        
    \end{tablenotes}
    \label{tab:toc}
\end{threeparttable}
\end{table*}
}

\section{Data processing} \label{sec:data}

\subsection{Sample selection}

The goal of this work is to test how the measured structure of galaxies depends on the image properties. One common approach is to use simulated galaxies \citep[e.g.,][]{Thorp2021,Sazonova2024}, which allows creating noiseless images to robustly test the impact of the sky background. However, simulated galaxies with large enough samples are produced either in cosmological simulations \citep[e.g., Illustris TNG;][]{Nelson2019} or by modelling simple analytic profiles \citep[e.g., GalSim;][]{galsim}. The first approach has a limited resolution defined by the softening length, while the second approach models only the large-scale light distribution, and so these simulations do not adequately resolve small-scale features present in a galaxy (e.g., star clusters, dust, foreground stars, and many more), which may affect the measurements. Instead, we opted to use real observations, similarly to \cite{Yu2023}. Space-based imaging provides exquisite resolution of local galaxies, so we turned to archival HST observations of nearby systems. The depth of HST imaging is heterogeneous\new{, depending on the exposure time}, so we built a sample of local galaxies with \new{sufficiently} deep HST imaging, thereby testing both a wide range of resolutions and depths.

Our analysis applies to changes in resolution and depth both across different surveys or instruments, or equivalently across different redshifts due to cosmological dimming and angular scale change. Traditionally, morphological studies focus on optical morphology, so we chose to restrict this study to the \textit{HST} F814W (\textit{I}) band. We do not apply, nor discuss, morphological K-corrections \citep[e.g.,][]{Windhorst2002} here -- but we stress the importance of choosing filters that map to the same rest-frame band when comparing morphologies of galaxies at different redshifts. While there is great potential in extending morphological studies to other wavelengths, it is beyond the scope of the current work.

The Revised New General Catalog catalog \citep[RNGC/IC;][]{Dreyer1888,Dreyer1910,Sinnott1988,Steinicke2010} provides a list of large nearby galaxies and nebulae found within 200 Mpc \new{to ensure a better than 50 pc/px resolution at the HST pixel scale}. We cross-matched the RNGC/IC catalog with the Hubble Source Catalog V3 \citep[HSCv3;][]{Whitmore2016} to see which of these objects have been observed by the HST\footnote{We note that HSCv3 is not perfect and some of the galaxies from RNGC/IC did not have an HSCv3 cross-match, even though they have been observed by HST. We have included a couple of objects we knew about that had HST observations, but there may be more.}.  We then queried the Mikulski Archive for Space Telescopes\footnote{\href{https://archive.stsci.edu/}{MAST archive: https://archive.stsci.edu/}} (MAST) for these observations, and further restricted our sample to galaxies with imaging deeper than the \textit{I}-band surface brightness limit $\mu_0 = 24$ mag/arcsec$^2$\footnote{We define the depth or surface brightness limit, $\mu_0$, as a surface brightness of the 1$\sigma$ level of the background. While it is common to define the depth as a n$\sigma$ detection of a point source for surveys, since morphological analyses aim to characterize extended emission, the surface brightness depth is more appropriate.}, which resulted in 169 sources. We downloaded the \new{F814W} HST observations from the Hubble Legacy Archive\footnote{\href{https://hla.stsci.edu/}{Hubble Legacy Archive: https://hla.stsci.edu/}} (HLA). 
%

We then visually inspected the images. We removed any sources that were clearly on the edge of the footprint and where the majority of the galaxy was not imaged, galaxies where the chip gap of the ACS detector passed through the nucleus of the source, as it makes any morphological measurements less reliable, and galaxies with severe imaging artifacts \new{or} invalid header information (25 galaxies). This left us with 146 galaxies from RNGC/IC.  We then looked at each HST footprint to see if there are any other local galaxies in the image that are not included in the RNGC/IC, and cross-matched the positions of the candidate sources with the NED database\footnote{\href{https://ned.ipac.caltech.edu/}{NED Database: https://ned.ipac.caltech.edu/}}. We added all sources with known spectroscopic redshifts that are within 200 Mpc to our catalog (41 source\new{s}, primarily from the Coma and Perseus clusters). These additional objects are almost all dwarf galaxies and satellites that were too faint to be included in the original RNGC/IC catalog, and allow us to study a wider range of possible morphologies, including satellite galaxies potentially undergoing gas stripping. Finally, this resulted in a catalog of 187 galaxies at distances smaller than 200 Mpc, imaged with HST with a depth of 24 mag/arcsec$^2$ or better \new{(Sec. \ref{sec:reduction})}. The sample spans a wide range of stellar masses, morphological types, and environments. We queried NED to get any missing \textit{r}-band Petrosian radii $R_p$ (see Sec. \ref{sec:rpet} for a definition), and spectroscopic distances.  Our sample is summarized in Table \ref{tab:sample}. Appendix \ref{app:sample} shows the distributions of magnitudes and morphologies of the galaxies in our sample, \new{including Hubble classifications,} confirming that we have a diverse set of galaxies spanning \new{a range of structures between early-type, late-type, and merging galaxies}.

{\renewcommand{\arraystretch}{1.5}
\setlength{\tabcolsep}{3pt} 
\begin{table}
\begin{threeparttable}
\caption{An excerpt of the local galaxy sample}
    \begin{tabular}{@{}llllccccc@{}}
        \toprule
        \toprule
            Name &
            R.A. &
            Dec. &
            Dist. &
            \new{$M_I$} &
            \new{Type} &
            \res{}\footnotemark[1] &
            $R_p$\footnotemark[2]  \\
             &
            deg &
            deg &
            Mpc &
            \new{mag} &
            &
            pc/px &
            \arcsec
             \\
        \midrule
        NGC 17 & 2.7775 & -12.1078 & 83 & -21.8 & S/P & 20 & 12 \\
        NGC 59 & 3.8554 & -21.4447 & 5 & -16.7 & S0 & 1 & 49 \\
        NGC 201 & 9.8954 & 0.8603 & 62 & -21.7 & SBc & 15 & 50 \\
        NGC 406 & 16.8517 & -69.8761 & 21 & -19.7 & Sc & 5 & 73 \\
        NGC 384 & 16.8542 & 32.2928 & 60 & -20.4 & E3 & 14 & 15 \\
        IC 1623 A & 16.9450 & -17.5067 & 85 & -22.2 & Im & 21 & 20 \\
        NGC 404 & 17.3621 & 35.7183 & 4 & -14.5 & S0 & 1 & 110 \\
        NGC 449 & 19.0312 & 33.0889 & 67 & -19.6 & Sb & 16 & 10 \\
        NGC 520 & 21.1446 & 3.7942 & 32 & -21.7 & Im & 8 & 75 \\
        NGC 695 & 27.8092 & 22.5828 & 137 & -22.9 & S0 & 32 & 13 \\
        IC 1776 & 31.3129 & 6.1056 & 48 & -19.4 & SBc & 12 & 34 \\
        NGC 1084 & 41.4992 & -7.5778 & 20 & -21.2 & Sc & 5 & 76 \\
        NGC 1222 & 47.2371 & -2.9556 & 34 & -19.7 & S0 & 7 & 22 \\
        NGC 1260 & 49.3633 & 41.4053 & 81 & -21.3 & Sa & 15 & 23 \\
        NGC 1289 & 49.7075 & -1.9731 & 40 & -20.3 & SB0 & 10 & 27 \\
        NGC 1271 & 49.7967 & 41.3536 & 81 & -20.6 & SB0 & 15 & 9 \\
        NGC 1272 & 49.8387 & 41.4906 & 54 & -22.4 & E1 & 13 & 54 \\
        NGC 1281 & 50.0262 & 41.6297 & 61 & -20.2 & E5 & 11 & 20 \\
        NGC 1311 & 50.0279 & -52.1872 & 8 & -17.2 & SBm & 2 & 68 \\
        NGC 1344 & 52.0812 & -31.0681 & 16 & -21.2 & E5 & 4 & 68 \\
        NGC 1345 & 52.3813 & -17.7792 & 22 & -18.7 & SBc & 5 & 30 \\
        IC 1959 & 53.3004 & -50.4122 & 9 & -17.3 & SB & 2 & 64 \\
        IC 391 & 74.3421 & 78.1897 & 22 & -19.7 & Sc & 6 & 34 \\
        LEDA 16571 & 75.3973 & -4.2571 & 446 & -21.4 & SBb & 90 & 4 \\
        NGC 1741 B & 75.3975 & -4.2633 & 58 & -19.1 & Scm & 15 & 12 \\
        NGC 1741 & 75.4091 & -4.2596 & 61 & -20.5 & Im & 14 & 21 \\
        IC 399 & 75.4333 & -4.2883 & 56 & -19.5 & Scm & 14 & 10 \\
        NGC 2082 & 85.4621 & -64.3014 & 17 & -19.7 & SBb & 4 & 49 \\
         &  & & ... &  &  &  &  \\
    \bottomrule
    \end{tabular}
    \begin{tablenotes}
        \item \textit{Notes} -- the full table of 187 galaxies included in this analysis is available in the online supplementary materials. 
        \item \footnotemark[1] Angular resolution of the image in units of parsecs per pixel. The instrumental resolution is 0.05$\arcsec$/px for ACS and 0.04$\arcsec$/px for WFC3.
        \item \footnotemark[2] Petrosian radius \citep[Sec. \ref{sec:rpet};][]{Petrosian1976}
    \end{tablenotes}
    \label{tab:sample}
\end{threeparttable}
\end{table}
}

\subsection{Image processing}\label{sec:reduction}

For each galaxy, we created cutouts and ran the analysis semi-interactively. By default, we made cutouts spanning 3$R_p$, increasing this size for some objects where the diffuse light extended close to the edges of the image. Since this is a local sample imaged by HST, in many cases the field of view was smaller than 3$R_p$, so we used the entire footprint instead. We used the context information provided by the HLA pipeline to create masks for bad pixels and detector gaps, and the exposure time map alongside gain to get electron counts and hence the 1$\sigma$ Poisson uncertainty map.

We estimated the background level of the images using \textsc{photutils} \citep{photutils} \new{as follows. Many} of our sources have a large angular extent comparable to the footprint of the detector, and so the image is dominated in some cases by the galaxy's light, making a simple sigma-clipping procedure unreliable. Instead, we estimate the background iteratively. First, we estimate the background $\sigma$ with sigma-clipping. Then, we run a 1$\sigma$ source detection on the image and mask out any detected sources and re-run the sigma-clipping to get a new estimate for $\sigma$. We repeat this until the standard deviation estimates converge to within 5\% (typically 4-5 times). Appendix \ref{app:bg} shows an example of this procedure on one galaxy. Finally, we use the \texttt{Background2D} with a box size spanning 1\% of the image and a 5 pixel filter size to see if there are any large-scale gradients in the background. Typically, subtracting a single median background value is desirable in low-surface-brightness science to retain as much of faint diffuse light as possible. However, some HST images have variations in gain across the mosaic and hence effective depth across the detector. We visually inspect the \texttt{Background2D} fit and if we find background discontinuities across mosaicked chips we subtract the fit; otherwise we simply subtract a flat background value. We define the surface brightnes of the image $\mu$ using $1\sigma$ of the background obtained with this procedure.

HLA does not provide variance maps for most images, so we constructed our own. The Poisson error of a raw image is a square root of the electron counts. To convert an image to counts we use the exposure time map, $t_{\textrm{exp}}$, gain $g$, and our background estimate via $f_{\textrm{sky}} \sim (gt_{\textrm{exp}} \times \sigma)^2$. We convert the image into electrons to calculate the Poisson error, and re-normalize the error array by the exposure time. We also save the exposure time map and the sky map to use in later steps.

\begin{figure*}
    \centering
    \includegraphics[width=\linewidth]{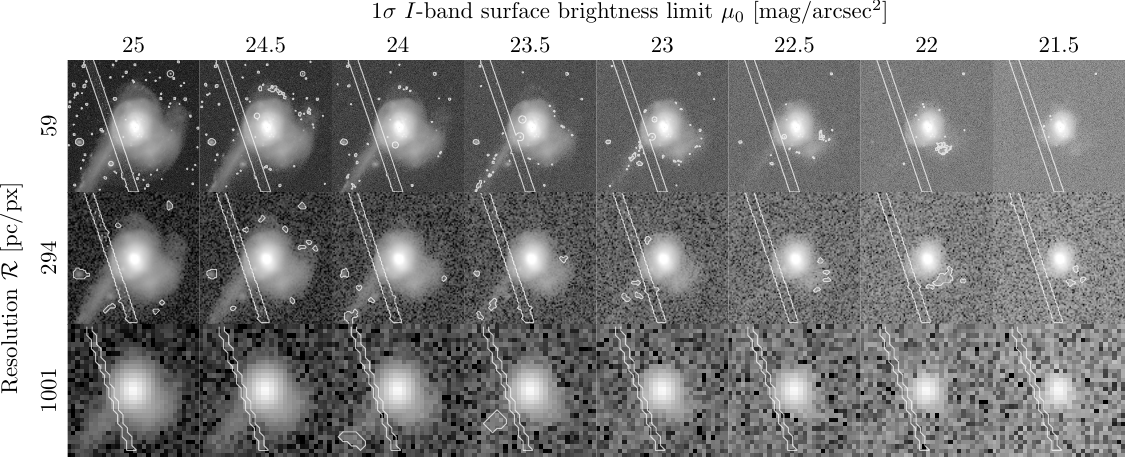}
    \caption{A subset of augmentations performed for an example galaxy, NGC 17, observed in F814W (\textit{I} band). The $1\sigma$ surface brightness limit $\mu_0$ decreasing left to right, and the resolution $\mathcal{R}$ is degrading top to bottom. Foreground sources detected on each image are shown in white contours. As depth degrades, the large tidal tail to the southwest is lost in noise, while the internal disturbances are invisible in low-resolution imaging.}
    \label{fig:augmentations}
\end{figure*}

We then created segmentation maps \new{using a \textsc{Photutils}-based custom pipeline} to mask out other foreground/background sources. Since our sources are large, there are more likely to be other sources overlapping with their light, and so an accurate segmentation map is crucial. We used a hot-cold approach described in \cite{Sazonova2021}, similar to \cite{Galametz2013}. \cite{Sazonova2021} and \cite{Morgan2024} described the algorithm in more detail, but we summarize it in Appendix \ref{app:segmentation}. 

Creating good-quality segmentation maps is particularly challenging for high-resolution observations of local galaxies, since a bright clump may be a foreground star or a star cluster within the galaxy itself. We aimed to mask any obvious stars and smaller background galaxies. To ensure a reliable source mask, we ran the segmentation procedure interactively, visually inspecting the resulting masks and changing the segmentation parameters when necessary until a good map was produced, decided by our visual examination and looking that 1) obvious stars are masked, 2) the source is not over-deblended, and 3) foreground objects away from the source are masked.

The goal of this work is to study how morphological parameters change based on image quality, and the segmentation procedure can dramatically change the quantitative morphology, e.g., if the main source is over-deblended, or if some contaminant is not masked. As the image quality changes, so does our ability to clearly distinguish different sources, and therefore the estimated segmentation maps. If only low-resolution imaging is available, it would be unrealistic to use a segmentation map obtained with higher-resolution data. Therefore, to mimic a real use case most closely, we create new segmentation maps each time the image quality is degraded. In most cases, we keep the same segmentation parameters from the original run, although in a few cases we change them to help deblending at the lowest resolutions. In cases of merging galaxies, the two sources are deblended in shallow imaging but not in deeper observations where a tidal feature connects them. This is an intrinsic complication of trying to separate associated sources, and we did not adjust the algorithm to produce the same deblending results (see Sec. \ref{sec:multimode} for an example). We note that this adds an additional source of uncertainty to morphological measurements, since at different image depths what is defined as the ``source`` may change.

\subsection{Image quality \new{degradation}}

Finally, for each galaxy, we create a series of images with a degraded image quality, changing both the resolution and depth. \new{Following machine learning nomenclature, we call these augmentations.} An example is shown in Fig. \ref{fig:augmentations} for a subset of augmentations for NGC 17. 

Resolution is defined both by two effects: the effective size of the point-spread function (PSF) and the physical pixel scale, and both of these can affect the measurements independently in different ways. However, most instruments set the pixel scale to sample the PSF by approximately two pixels for the PSF full width at half maximum (FWHM) to ensure Nyquist sampling. The exact ratio between the pixel scale and the PSF size varies from 1 (e.g., JWST) to 3$\sim$4 in surveys such as Sloan Digital Sky Survey (SDSS), LSST, or the Ultraviolet Near Infrared Optical Northern Survey (UNIONS). We chose to fix the PSF size to be twice the pixel scale (which is similar to the HST resolution). Therefore, when we say \textit{resolution}, we interchangeably mean both the number of pixels, and the number of resolution elements set by the PSF. Investigating the relationship between the pixel size and the PSF size \textit{independently} would be valuable, but is beyond the scope of this paper. We are planning to release an additional set of augmentations using an empirical LSST PSF in the GitHub repository.

To change the image resolution, we first choose the new pixel size in parsecs per pixel ranging from 25 pc/px to 2,000 pc/px. Then, we convolve the original image with a Moffat PSF with a $\beta=2$ and a FWHM that would correspond to two pixels at the new resolution. The convolution step smooths over the noise, which is unrealistic, as in real instruments the seeing is applied by the atmosphere or the telescope before the photons hit the detector with some Poissonian probability. Therefore, after the convolution, we resample each pixel from a Poisson distribution about that pixel's value to mimic this effect. This degrades the raw signal-to-noise \new{ratio} of the image, but the images are still deep enough for the rest of our analysis. Finally, we use \textsc{Scikit-Image} \citep{skimage} to resample the image from its original resolution to the new one. We also resample the mask, and calculate the new error array from the counts in the resampled image. Finally, we convolve the instrumental \textit{HST} PSF simulated with \textsc{TinyTim} \citep{Krist2011} with the Moffat PSF to get the effective PSF of the new image, and store it for later analysis.

After the image is resampled to the new resolution, we calculate the new 1$\sigma$ surface brightness limit ($\mu_0$), since downscaling the image effectively increases the image depth, again using the iterative masking and sigma-clipping to calculate $\sigma$. \new{We then selected a series of surface-brightness limits on an evenly spaced grid from 20 to 26 mag arcsec$^{-2}$. For images that did not originally reach a depth of 26 mag arcsec$^{-2}$, we stopped at the deepest limit achievable. We note, however, that after image resizing the effective central surface brightness, $\mu_0$, increased beyond the 24 mag arcsec$^{-2}$ threshold used in the initial sample selection.} To decrease the signal-to-noise \new{ratio}, we add a Gaussian white noise field with 0 mean and a standard deviation $\sigma'$ to match the desired depth. This effectively emulates a brighter sky with a higher Poisson error, rather than a lower exposure time. In the optical regime we are interested in, the impact of these two approaches is essentially the same, since the Poisson noise from the source is not a significant fraction of the total error. However, in UV or X-ray observations, there would be an important difference between larger exposure time or a brighter background, and so our results should be interpreted with caution in those regimes. Given the original background standard deviation $\sigma^{\rm{og}}$, the original depth $\mu_\sigma^{\rm{og}}$, and the desired depth $\mu_\sigma^{\rm{new}}$, the amount of additional noise needed is given by:

\begin{equation}
    (\sigma')^2 = \sigma^2 \left( 10^{2\left(\mu_\sigma^{\rm{new}} - \mu_\sigma^{\rm{og}}\right)/2.5} - 1 \right).
\end{equation}

We add this Gaussian noise to the rescaled image and the error array, re-run the segmentation routine.

For each galaxy, we attempt to create images for 18 physical resolutions from 25 to 2500 pc/px, and for 25 surface brightness limits from 20 to 26 mag/arcsec$^2$. If the image is too shallow, we limit this to the deepest surface brightness limit possible. If the image becomes smaller than 40 pixels, we stop degrading the resolution. As a result, we produced a dataset of 64,000 images for 189 galaxies spanning a wide range of resolution and image depths. This dataset is available on Zenodo\footnote{\href{https://zenodo.org/records/17585609}{Dataset DOI: 10.5281/zenodo.17585608}}, and can be well-suited for other analyses or for training deep learning models.

We then measure the quantitative morphologies described in Sec. \ref{sec:morphology} using \texttt{statmorph} \citep{statmorph} by passing the image, the error array as a weight map, the new PSF, and the segmentation map. In addition to the \texttt{statmorph} analysis, we calculate a new isophotal asymmetry ($A_X$; Sec. \ref{sec:aiso}), substructure ($St$; Sec. \ref{sec:substructure}), and run \textsc{Galfit} \citep[Sec. \ref{sec:sersic};][]{galfit} to obtain single-S\'ersic fits. Finally, we compute an effective signal-to-noise per pixel in the augmented image.


\section{Analysis of morphology metrics} \label{sec:morphology}

In this section, we discuss the morphological parameters measured in our test suite by \texttt{statmorph}, \textsc{Galfit}, and our custom algorithms. Table \ref{tab:toc} lists all the parameters we studied in this work, and Fig. \ref{fig:summary} provides a visual summary of our results -- we encourage the readers to use these to find the parameters they are most interested in.

We measure each metric $p$ on all augmented images, and compare its value to a tentative ``baseline'' value measured on images with resolution $\mathcal{R}=100$ pc/px and surface brightness limit $\mu_0=23.5$ mag/arcsec$^2$\footnote{We define a surface brightness limit $\mu_0$ as the effective limit after accounting for cosmological dimming, $\mu_0 \equiv \mu - 7.5 \log (1+z)$, where $\mu$ is the 1$\sigma$ surface brightness limit measured directly from the image. The factor of 7.5 arises from cosmological dimming and an AB magnitude system K-correction as described in \cite{Hogg2002}. Since our sources are within 200 Mpc, $\mu$ and $\mu_0$ are almost interchangeable.}. We use this ``baseline'' as a comparison, since it is impossible to measure a ``true'' value of a morphological parameter. Moreover, for parameters that do not converge and depend intrinsically on resolution and/or depth, the true value may not even be defined. Therefore, we opted to define a ``baseline'' which is the measurement you get when the source is reasonably well-resolved and faint features are visible.

For each metric, we consider that its measured value can depend on three variables: the baseline value $p_{\rm{base}}$, resolution, and the signal-to-noise \new{ratio}. We do not consider that some metric may depend on a different metric (such as a centroid) for simplicity. The resolution dependence might be two-fold: a metric might depend on our ability to resolve some specific features with a given size (hence on $\mathcal{R}$ in pc/px) or on our ability to resolve some feature relative to a galaxy's size. To distinguish these two cases, we also measure the effective resolution, $\mathcal{R}_{\rm{eff}}$, defined as Petrosian radius (Sec. \ref{sec:rpet}) divided by the pixel scale, or the number of resolution elements sampling the galaxy\footnote{Note that since the pixel size is always half the PSF FWHM, this is also the effective resolution relative to the PSF times two.}. Similarly, while some metrics depend on detecting low surface brightness features and hence $\mu_0$, others may depend more on the total flux and hence a signal-to-noise ratio, so we also measure a\new{n average} signal-to-noise per pixel \new{(\avgsnr{})} within 1.5$\times$Petrosian radius\footnote{The Petrosian radius used here is measured on the baseline image.}. So in our modelling approach, $p$ may depend on all of these factors, i.e. $p = p(p_{\rm{base}}, \mu_0, \avgsnrm{}, \mathcal{R}, \mathcal{R}_{\rm{eff}}$), but not on any other morphological parameters.

In practice, since $\mathcal{R}$ and $\mathcal{R}_{\rm{eff}}$ as well as $\mu_0$ and \avgsnr{} are degenerate with one another, it is more useful to find which factors drive the measurement most. To do this, we use a mutual information score from \textsc{Scikit-Learn} \citep{sklearn}. A mutual information score compares the joint distribution of two variables to their marginal distributions, and quantifies how much knowing one variable informs the other. The score is 0 if the two variables are independent, and 1 if one completely depends on the other. First, we compute the mutual information scores between $p$ and each of $p_{\rm{base}}$, $\mu_0$, \avgsnr{}, $\mathcal{R}$, $\mathcal{R}_{\rm{eff}}$. Then, we choose one variable between ($\mathcal{R}$, $\mathcal{R}_{\rm{eff}}$), and one between ($\mu_0$, \avgsnr{}) with the higher score as the more important one. In most cases, $\mathcal{R}_{\rm{eff}}$ and \avgsnr{} are more important than $\mathcal{R}$ and $\mu_0$. \new{The feature importance plots for each parameter are available in the online materials.}

We then choose the final ``baseline" value for each galaxy. In cases where $\mathcal{R}_{\rm{eff}}$ is more important to the measurements, it would be incorrect to choose the baseline as some arbitrary resolution in parsecs, since the \textit{effective} resolution will change between larger and smaller galaxies at the same pixel scale. This is also true for \avgsnr{}. Therefore, if these parameters are preferred, instead we selected the baseline based on $\mathcal{R}_{\rm{eff}}$ and \avgsnr{}. We chose the baseline of $\mathcal{R}_{\rm{eff}}$ to be the median of the entire sample when $\mathcal{R} = 100$ pc/px, which was $\mathcal{R}_{\rm{eff}} = 42$. Similarly, our baseline  $\avgsnrm{} = 5.3$.  We then reassigned the baseline augments for each galaxy.


Finally, for convenience, we provide fits to the function $p(p_{\rm{base}}, \mathcal{R}, \mu, \mathcal{R}_{\rm{eff}}, \avgsnrm{})$ obtained using the symbolic regression package \texttt{SymbolicRegression.jl} \citep{Cranmer2023}. The details of the symbolic regression implementation are described in Appendix \ref{app:pysr}. These convenience functions enable observers to convert the measurement they obtained on a given image to the expected values for an image of a different quality, and they are listed in Tab. \ref{tab:pysr}. We caution that these fits are purely empirical rather than universal relationships. For most measurements, as the image quality decreases, the scatter in $p$ increases, so using these equations to find the expected $p$ for a deeper or a better resolved image will introduce significant uncertainties in the estimate, and so should be done with care. Moreover, we assumed a Moffat PSF with a two-pixel FWHM in our augmentations. For a different PSF, the fits will be different, and so it is best to recalibrate the correction based on a known PSF. We would also not advise users to extrapolate our models beyond the fitted regime.

For each parameter $p$, we provide several analysis products. First, a mutual information score for each parameter and each variable, shown in the online materials\footnote{\href{https://github.com/astro-nova/statmorph-lsst}{https://github.com/astro-nova/statmorph-lsst}}. Second, a grid showing the variation in the parameter compared to the baseline as a function of resolution and depth. Third, a \texttt{SymbolicRegression} fit to the relationship between $p$ and $p_{\rm{base}}$ (Tab. \ref{tab:pysr}), and a distribution of corrected parameters. Finally, we also show how each parameter varies with depth and resolution for two example galaxies. Since there are a lot of figures, we only provide the most informative plots in this paper, and the remaining plots are available online. We explain the plots in detail in the following section describing the centroid (Sec. \ref{sec:centre}); the other sections adopt the same format. 

We group the parameters roughly by their intended purpose: first, we discuss the simplest geometric measurements (image moments and ellipticity; Sec. \ref{sec:moments}), then radii and S\'ersic modelling (\ref{sec:radii}), then concentration or ``bulge strength'' measurements ($C$, $G$, $M_{20}$, Sec. \ref{sec:bulge}), then asymmetry parameters (Sec. \ref{sec:disturbance}), and finally other metrics that do not fit well in these categories (smoothness, substructure, multimode-intensity-deviaiton; Sec. \ref{sec:other}). We additionally define two new parameters: isophotal asymmetry (Sec. \ref{sec:aiso}) and substructure (Sec. \ref{sec:substructure}).

\begin{figure*}
    \includegraphics[width=\linewidth]{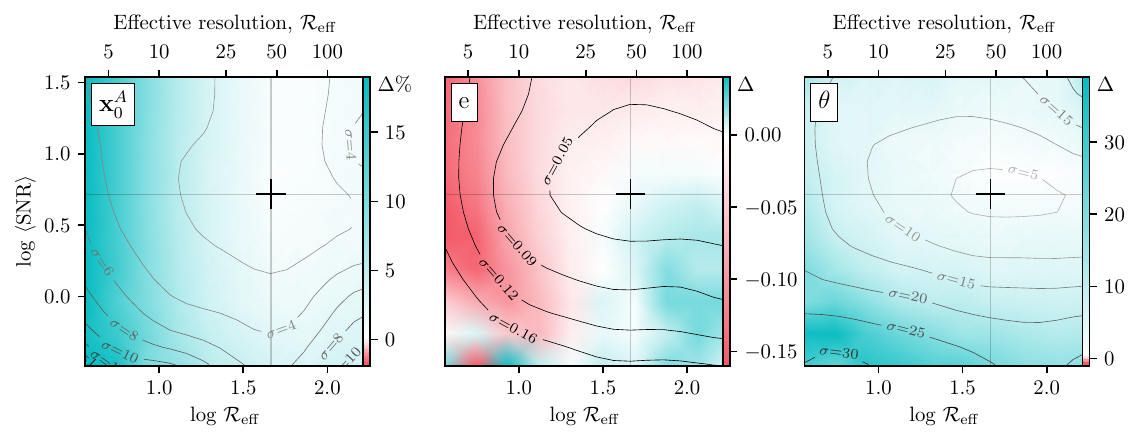}
    \caption{The measurement error compared to the baseline for the asymmetry centre ($\mathbf{x}_0^A$, \textbf{left}), ellipticity ($e$, \textbf{middle}), and orientation ($\theta$, \textbf{right}), as a function of the average signal-to-noise per pixel \avgsnr{} and the effective resolution \reseff{}. For each galaxy and each image, the error is calculated as the difference in measurement compared to the baseline. $\mathbf{x}_0^A$ is normalized by the Petrosian radius (Sec. \ref{sec:rpet}). The colored distribution shows the median error in each bin, while the gray contours show 1$\sigma$ scatter. Both $\mathbf{x}_0^A$ and $e$ are fairly robust to noise, with an increased uncertainty at \avgsnr{}<1 but no systematic offsets. They are however biased at low resolution, with typical error in $\mathbf{x}_0^A$ reaching 20\%. $\theta$ is less reliable in the low signal-to-noise regime, but the error is mostly driven by circular sources, where $\theta$ is poorly constrained.}
    \label{fig:moments_grid}
\end{figure*}

\subsection{Moments of light}\label{sec:moments}

The simplest approach to quantifying a light distribution is to measure the moments of the image. For a 2D image, the moment is calculated independently in $x$ and $y$ directions, and so the moment of order $i,j$ is

\begin{align*}
    M_{ij} &= \int_{-\infty}^{\infty} \int_{-\infty}^{\infty}  x^i y^j I(x,y) dx dy
           &= \sum_x \sum_y x^i y^j I(x,y).
\end{align*}

The 0\ts{th} moment is the total flux in the image. The centroid, i.e. a weighted average location of light, is the ratio of the first order moments and the total flux:

\begin{equation*}
    \mathbf{x}_0 = (x_0, y_0) = \left( \frac{M_{10}}{M_{00}}, \frac{M_{01}}{M_{00}}\right).
\end{equation*}

Finally, the second-order moments can be used to construct a covariance matrix for the light distribution,

\begin{equation*}
    \textrm{cov}(I) = \frac{1}{M_{00}} 
    \begin{pmatrix}
        M_{20} & M_{11} \\ M_{11} & M_{02}
    \end{pmatrix},
\end{equation*}

\noindent which defines a Gaussian distribution with the same moments as the image. The eigenvalues of this covariance matrix give the semi-major and semi-minor lengths of the equivalent Gaussian, and so can be used to estimate the axis lengths, ellipticity, and orientation of the source. The size estimate is the standard deviation of a Gaussian and so isn't particularly physically meaningful; however, we can use the axis ratio and the orientation measurements.

\subsubsection{Centre}\label{sec:centre}

To calculate the covariance matrix, we first need to center the image, i.e. subtract the centroid position from the pixel coordinates. The easiest choice is to use the centroid calculated from the first moments. However, as noted by \cite{statmorph}, this often does not correspond to the brightest region of the galaxy, as asymmetries in the light distribution can shift the centroid away from the physical centre. We can get a more robust measurement of the centre using the pixel about which asymmetry is minimized (Sec. \ref{sec:asymmetry}). \texttt{statmorph} calculates both of these centers, and therefore outputs elongation, ellipticity, and orientation measured about both coordinates, with suffixes \texttt{\_centroid} and \texttt{\_asymmetry}. For all other measurements the asymmetry centre is used by default. This centre differs from the centre \new{provided by source detection catalogs}, and while most surveys quote their astrometric precision, there are no studies of how well can we measure the centre in morphological pipelines.

The left panel of Fig. \ref{fig:moments_grid} shows how the asymmetry center, $\mathbf{x}_0^A$, depends on \avgsnr{} and \reseff{}. We calculated the centroid for each image and then found the distance between the measured and the baseline centers. We expressed the distance as a percentage of the baseline Petrosian radius (Sec. \ref{sec:rpet}). We binned the images into \new{11} groups of matched \avgsnr{} and \reseff{} \new{spanning the 5\textsuperscript{th}-95\textsuperscript{th} percentile range of the baseline sample,} and calculated the median offset as well as its 16\ts{th} and 84\ts{th} percentiles. The colored plot shows a smoothed histogram of the median percentage offset ($\Delta \%$). The gray contours show the scatter in the offset, calculated as half the range between the 84\ts{th} and 16\ts{th} percentiles. This scatter can be interpreted as the expected error in an individual measurement. For example, at $\log$ \avgsnr{} $=0.75$ and $\log$ \reseff{} $ = 0.5$, $\mathbf{x}_0^A$ is offset on average by 0.12$R_p$ with a \new{4}\% spread about this value. 

$\mathbf{x}_0^A$ is robust to changes in the signal-to-noise \new{ratio}: we don't find any systematic variations in the centre even in the lowest \avgsnr{} bins, although the expected error increases to $>$0.04$R_p$ when \avgsnr{}$<1$. However, the estimate of the centre becomes less reliable as the resolution degrades, with up to $0.20 \pm 0.07 R_p$ offsets when there are fewer than 5 pixels spanning $R_p$. This result is unsurprising: at the lowest resolutions, one pixel spans larger distances, and so subpixel precision is required to estimate the centre well. Small errors in the subpixel centre estimate lead to significant offsets. This is especially true for the asymmetry center, since minimizing asymmetry with subpixel precision leads to interpolation errors when the image is rotated about this centre \citep{Sazonova2024}. 

For brevity, we do not show the results for the moment-based centroid, as they are almost identical, but these results are available in the online materials. The errors as the resolution degrades are comparable to those we found with the asymmetry center, but the centroid also depends on the signal-to-noise \new{ratio}, with up to $0.05\pm 0.07 R_p$ offsets at fixed resolution and $\avgsnrm < 1$. In most cases, the two centre estimates agree well, so it is reasonable to use either one, but the asymmetry-based centre is more robust to extended tidal features \citep{statmorph}.

\subsubsection{Ellipticity}

Ellipticity quantifies how elongated or circular the source is by finding the ratio of the semi-major and the semi-minor axes ($a$ and $b$). There are several different definitions of this ratio in literature: the ellipticity, $e = 1 - b/a$, elongation, $a/b$, or the axis ratio, $q = b/a$. They are perfectly degenerate with one another, so here we only discuss the ellipticity $e$. Ellipticity of 0 means a perfectly circular source, while ellipticity close to 1 means an elongated, or a cigar-shaped source. \texttt{statmorph} calculates ellipticity about the asymmetry centre and the centroid; here we only discuss the measurement relative to the asymmetry centre as the two are consistent in most cases.

The middle panel of Fig. \ref{fig:moments_grid} shows how $e$ varies with \avgsnr{} and \reseff{}. $e$ is largely insensitive to the signal-to-noise \new{ratio}, although the uncertainty in measurement increases to $\pm$0.12 when \avgsnr{}$<1$. The offset in $e$ is driven by resolution, with an up to $0.15 \pm 0.12$ underestimate when there are fewer than 5 pixels spanning the galaxy. This is likely because \new{the non-parametric ellipticity is calculated} from the raw image moments without accounting for the PSF, which circularizes the incoming flux and reduces ellipticity. The issue is even more pronounced in Source\new{-}Extractor, which first smooths the image with a small kernel, and then computes ellipticity. The discretization can also be an effect in the most extreme cases, where the entire source is spanned by a few pixels. In our set-up, the effect of the PSF is small, but in ground-based observations where the PSF is large relatively to the pixel scale, the effect may be more pronounced. This dependence on the PSF or pixel size also means that in a given observation, there is a \textit{lower bound} on measured ellipticity, so extremely edge-on galaxies will appear more circularized \citep[e.g.,][]{vanderWel2014}. \new{This can be remedied by a forward modelling approach, such as fitting a S\'ersic profile (Sec. \ref{sec:sersic})}

We provide our best fit to correct for the resolution offset \new{in} Appendix \ref{app:pysr}, however we must caution that since the error is driven primarily by the PSF, the correction is only applicable in cases where the PSF FWHM is roughly 2 pixels, and it is better to derive a correction based on your PSF. Alternatively, the best approach is to derive ellipticity from a parametric fit that includes a PSF model, e.g. through a S\'ersic fit (Sec. \ref{sec:sersic}). This is often done for studies of high-redshift galaxies \citep[e.g.,][]{VegaFerrero2024,Pandya2024}.

\begin{figure*}
    \centering
    \includegraphics[width=\linewidth]{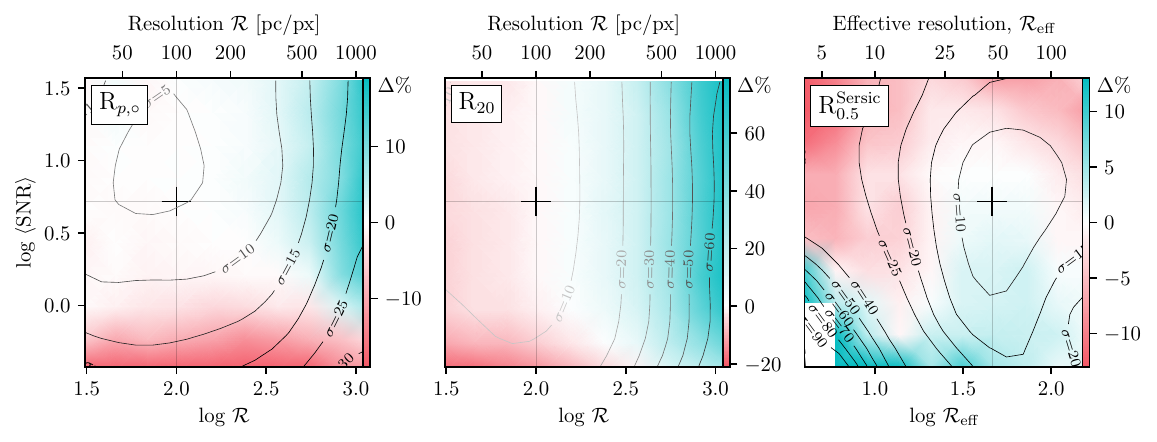}
    \caption{Same as Fig. \ref{fig:moments_grid}, for three radius metrics: $R_{p,\circ}$ (\textbf{left}), $R_{20}$ (\textbf{middle}), and S\'ersic $R_{0.5}$ (\textbf{right}, discussed in Sec. \ref{sec:sersic}). The bias and scatter are measured as a fraction of the baseline, rather than an absolute offset. $R_{p,\circ}$ and $R_{20}$ are more correlated to the pixel scale $\mathcal{R}$ while $R_{0.5}^{\rm{Sersic}}$ depends on the effective resolution $\mathcal{R}_{\rm{eff}}$. The non-parametric radii are overestimated at low resolutions, up to 20\% and 80\% for $R_{p,\circ}$ and $R_{20}$ respectively, but $R_{0.5}^{\rm{Sersic}}$ is slightly underestimated (up to 10\%) compared to the baseline. However, the uncertainty in $R_{0.5}^{\rm{Sersic}}$ is larger than that for non-parametric radii.}
    \label{fig:r_grid}
\end{figure*}

\subsubsection{Orientation}

The final parameter that uses the image moments is the orientation $\theta$, i.e. the angle of inclination relative to the chosen centre. It is measured starting from the vertical, going counter-clockwise, spanning from $-\pi$ to $+\pi$. In this work, we converted orientation from radians to degrees. Fig. \ref{fig:moments_grid} shows the orientation behavior in the right panel. The uncertainty in the orientation angle is over $10\degree$ for most resolutions and depths when compared to the baseline, increasing to 20$\degree$ in low-resolution images and to \new{30}$\degree$ in noisy images. The error in orientation reaches 20$\degree$ when $\avgsnrm{} \sim 1$ and 5$\degree$ when $\reseffm \sim 5$. However, most of this error comes from the intrinsic uncertainties in calculating a position error of round sources. \new{For more edge-on galaxies, the error is reduced -- we discuss this further in Appendix \ref{app:theta}.} When restricting our sample to only galaxies with $e > 0.3$, the error in orientation is $2 \pm 5\degree$ in all but the lowest \avgsnr{} and \reseff{} bins. \new{Appendix \ref{app:theta} shows the average errors for other choices of $\theta$.}

\subsection{Radius measurements}\label{sec:radii}

Here we describe three galactic radii computed by \texttt{statmorph} and \textsc{Galfit}: Petrosian, isophotal, and S\'ersic. There are, however, many other ways to measure a galaxy's size, some more physicall\new{y} motivated than others -- such as looking at the break in the exponential disk profile \citep[e.g.,][]{Hunter2006}, a stellar mass cut-off \citep{Chamba2022}, or the H\textsc{i} mass threshold \citep[e.g.,][]{Wang2016}. While it is beyond the scope of our work to discuss these other radii here, we intend to analyze them in the future and upload the results to the online repository.

\subsubsection{Petrosian radius}\label{sec:rpet}

The \cite{Petrosian1976} radius $R_p$ is where the mean flux in an annulus centered on $R_p$ is a fraction $\eta$ of the mean flux up to $R_p$:

\begin{equation}
    \frac{1}{4\pi R_p \delta r} \int_{R_p-\delta r}^{R_p+\delta r} I(r)dr = \eta \times \frac{1}{\pi R_p^2}\int_0^R I(r) dr,
\end{equation}

\noindent where $I(r)$ is galaxy flux and $\delta r$ is some small annulus width (1 pixel in \texttt{statmorph}). Most often, $\eta$ is chosen to be 0.2\footnote{This definition of $\eta$ is actually the inverse of the original definition \citep{Petrosian1976,Graham2005}. We adopt the current definition since it is used by \texttt{statmorph} and SDSS.}. \texttt{statmorph} calculates $R_p$ in circular and elliptical isophotes ($R_{p,\circ}$ and $R_{p,e}$), using the asymmetry centre (Sec. \ref{sec:centre}) and the ellipticity found using image moments (Sec. \ref{sec:moments}). Petrosian radius is largely insensitive to the signal-to-noise ratio or the surface brightness limit \citep[e.g.,][]{Petrosian1976,Lotz2004} and so \new{is often used} in high-redshift science \citep[e.g.,][]{Shen2003,PerezGonzalez2023}, where the cosmological surface brightness dimming is important.

Fig. \ref{fig:r_grid} shows the dependence of $R_{p,\circ}$ in kpc on \avgsnr{} and \reseff{}. $R_p$ is a robust measurement in a low signal-to-noise regime, resulting in consistent radius estimates as long as \avgsnr $\geq 1$, and an up to $30\pm20$\% underestimate in noisier images. There are only two other studies that investigate the dependence of $R_p$ on the signal-to-noise \new{ratio}: \cite{Lotz2004} give a conservative estimate that $R_p$ is reliable up to \avgsnr $\sim$5, while \cite{Ren2024} find no dependence of $R_p$ on image depth. Our results agree with \cite{Ren2024} in that on average, $R_p$ is only underestimated in extremely low signal-to-noise images. However, the scatter in low-\avgsnr{} bins reaches 20\%, which means that measured $R_p$ on noisy images carries about a 20\% uncertainty, and should be treated with caution on an individual galaxy basis.


Similarly to \cite{Lotz2004}, \cite{Ren2024} and \cite{Yu2023}, we found that $R_{p}$ is sensitive to resolution. It is stable at resolutions better than 500 pc/px (or, equivalently, 1 kpc-wide PSF). At lower resolutions, $R_{p,\circ}$ is overestimated by $15\%$ with a $30\%$ scatter. Our results are in complete agreement with \cite{Lotz2004}, who also found that $R_p$ is robust at resolutions below 500 pc/px and then begins to increase, albeit using a much smaller sample of galaxies.  \cite{Ren2024} show that the dependence of $R_p$ on resolution arises from the PSF blurring the central flux to larger radii. \cite{Yu2023} also find that the error in $R_p$ is always 0.77$\times$PSF FWHM. We cannot directly test the effect of the PSF separately from the pixel scale with our set-up. However, in Sec. \ref{sec:sersic} we show that the S\'ersic effective radius, which is computed by accounting for the PSF, does not suffer from the same resolution effects, so we can indirectly conclude that this error is infact driven by the PSF smearing.

\begin{SCfigure*}[0.45]
\begin{wide}
    \centering
    \includegraphics[width=0.65\textwidth]{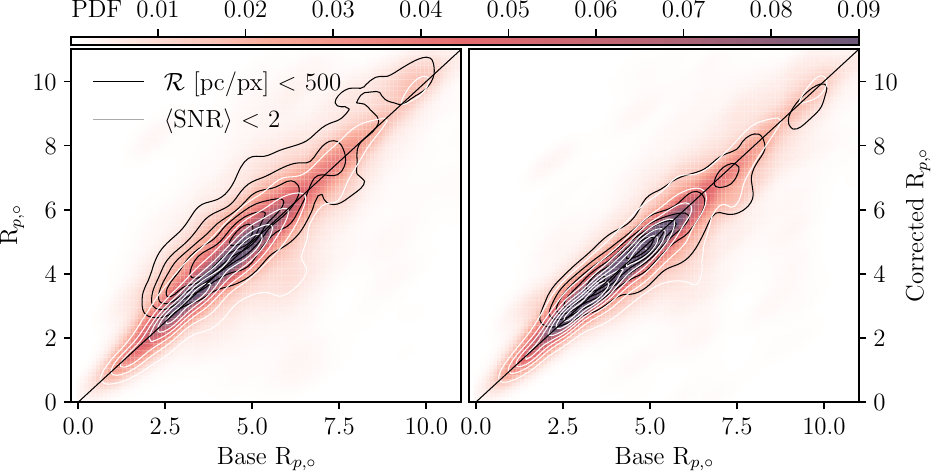}
    \caption{The distribution of $R_{p,\circ}$ measurements compared to the baseline before (\textbf{left}) and after (\textbf{right}) a signal-to-noise and resolution correction from Eq. \ref{eq:rpet}. Before the correction, $R_{p,\circ}$ is overestimated when resolution is worse than 500 pc/px (black contours), and slightly underestimated when \avgsnr{} is below 2 (white contours). After applying the correction, we can recover the baseline $R_{p,\circ}$ for a wide range of resolutions and depths.}
    \label{fig:rpet_corr}
\end{wide}
\end{SCfigure*}

The relationship between $R_p$ and $R_p^{\rm{base}}$ we obtained with \texttt{SymbolicRegression} is:
\vspace{-3mm}

\begin{equation}\label{eq:rpet}
    R_p = R_p^{\rm{base}} \tanh (3.2 \avgsnrm ) + \left( \frac{\mathcal{R}}{1000} \right)^{1.86}.
\end{equation}

\noindent It is then possible to measure $R_p$ and invert this relationship to estimate $R_p$, which we call ``corrected $R_p$''. This correction is different to the correction found in \cite{Yu2023}: we have a signal-to-noise term, which depends on the baseline $R_p$ and thus affects larger galaxies more, and a resolution term, which is flat with respect to the galaxy's intrinsic size. However, the resolution term is not linear like in \cite{Yu2023}. Fig. \ref{fig:rpet_corr} shows the distribution of the measured $R_p$ compared to the baseline on the left, and the corrected $R_p$ on the right. The kernel density estimate of the entire dataset is shown in color, and we overplotted the distribution of \avgsnr{}$<1$ measurements in white and $\mathcal{R} > 500$ pc ones in black. On the left panel, we see the same effect as in Fig. \ref{fig:r_grid}: $R_p$ is overestimated at low resolutions. The correction allows us to match $R_p$ to the baseline value well in both the low-\avgsnr{} and low-\reseff{} regimes. Our results contradict \cite{Ren2024} and \cite{Yu2023}, who find that the bias in $R_p$ is flat with respect to the intrinsic radius. \cite{Ren2024} find that the error is flat overall, while \cite{Yu2023} derive a correction that depends linearly on resolution. We find that the best correction includes all three of the input parameters, and can accurately recover the baseline $R_p$ even in the lowest signal-to-noise or resolution regimes. However, since the resolution dependence likely arises from the PSF rather than discretization, it is best to define $\mathcal{R}$ as the width of the PSF rather than the pixel scale.

The behaviour we see when measuring $R_p$ in elliptical rather than circular apertures is virtually identical, and we show the plots made for $R_{p,e}$ online. $R_{p,e}$ has the same noise dependence as $R_{p,\circ}$, unsurprisingly, since both the ellipticity and orientation are robust to noise (Sec. \ref{sec:moments}). In terms of resolution, $R_{p,e}$ is slightly better-behaved, with a $10\pm25$\% error. The \texttt{SymbolicRegression} fit is also similar to the one we obtained for $R_{p,\circ}$, although the functional form is not exactly the same, highlighting the arbitrary nature of the fit: while we tried to set up the run in a physically motivated manner, there are infinitely many functional combinations that may provide similarly good fits to our data. All of the fit results for each parameter as shown in the Appendix \ref{app:pysr}.

The final important consideration to note is that $R_p$ is used by \texttt{statmorph} to define the outer edge of the galaxy and the total flux. This edge is then used downstream, e.g. in isophotal radii or asymmetry calculations. The outer edge is given by the \texttt{petro\_extent\_cas} parameter times $R_p$, which by default is $1.5$. However, \cite{Graham2005a} showed that for early-type galaxies, as much as 20\% of the total flux is contained outside $R_p$, and so it is advisable to increase \texttt{petro\_extent\_cas} if the foreground contaminants are well-masked in the image. We quantify the magnitude of this missed flux as a function of S\'ersic index $n$ in Appendix~\ref{sec:appendix_g}.

\subsubsection[R50, R20, and R80]{R$_{50}$, R$_{20}$, and R$_{80}$}\label{sec:r20}

The half-light radius $R_{50}$, as well as $R_{20}$, and $R_{80}$, are isophotal radii measuring regions containing 50\%, 20\%, and 80\% of the galaxy's light respectively. Isophotes are calculated about the asymmetry centre ($\mathbf{x}_0^A$, Sec. \ref{sec:centre}). Calculating a percentile of a galaxy's light requires knowing the extent of the \textit{total} flux as well, so \texttt{statmorph} defines the total flux as the light contained within \texttt{petro\_extent\_cas} of $R_{p,\circ}$. \new{In this work, we used the default value of 1.5.} $R_{50}$ is commonly used as a half-light radius, while $R_{20}$ and $R_{80}$ are uncommon on their own and are mainly used to calculate the Concentration parameter (Sec. \ref{sec:concentration}).

For readability of this paper, we do not show all fit results of these radii or all of the correction functions, which are available online. Here we show the error in $R_{20}$ in the middle panel of Fig. \ref{fig:r_grid} -- qualitatively, the other parameters behave the same way but with a weaker bias. Perhaps unsurprisingly, since it depends on $R_p$, $R_{20}$ is extremely sensitive to resolution. While $R_p$ is overestimated by at most 30\%, $R_{20}$ is overestimated by more than 100\% when there are fewer than 5 resolution elements. The measurement appears to slowly converge in well-resolved imaging although not entirely; there is an approximately $5\%$ bias between our baseline images and images at higher resolutions. The behavior of $R_{50}$ and $R_{80}$ is qualitatively the same, with a bias reaching 50\% and 25\% at \reseff{}$<5$ respectively.

The only other study discussing $R_{20}$ is \cite{Bershady2000}, who also note that smaller percentile isophotes are more sensitive to resolution, but find that even $R_{20}$ is stable at \reseff{}$\sim$10. However, \cite{Bershady2000} simply downsampled their images to investigate the resolution dependence, rather than also including the effect of the PSF. The stark difference between our results and those of \cite{Bershady2000} strongly indicate that the bias is caused by the PSF, and could be potentially remedied if the effect of the PSF was accounted for (as it is in the S\'ersic fit, see below). 

The \texttt{SymbolicRegression} fit to our data allowed us to correct for the resolution well. However, since the bias appears to stem from the PSF, our fit would only work well in the identical set-ups, where the PSF is a 2-pixel Moffat. Therefore, we discourage using these fits in images with a different PSF without testing.

\subsubsection{S\'ersic fit \& Effective radius}\label{sec:sersic}

The S\'ersic fit is one of the most widely used parametric ways to quantify galaxy structure. \cite{Sersic1963} generalized the \cite{deVaucouleurs1948} empirical profile to obtain a good fit to many different galaxies along a Hubble sequence:

\begin{equation}
    I(r) = I_{0.5} \exp \left\{ -b_n \left[ \left( \frac{r}{R_{0.5}^{\rm{Sersic}}}\right)^{1/n} -1 \right] \right\}
\end{equation}

\noindent where $I_{0.5}$ is the light intensity at the half-light radius $R_{0.5}^{\rm{Sersic}}$, $n$ is the S\'ersic index, and $b_n \approx 2n - 1/3$ and can be calculated exactly via Gamma functions \citep[e.g.,][]{Graham2005}. The S\'ersic profile describes an exponential light profile when $n=1$ and a \cite{deVaucouleurs1948} profile of bright early-type galaxies when $n=4$. In practice, in many applications the S\'ersic profile is fit to the 2D light distribution, and so the fit has four additional free parameters: the centre $(x_0, y_0)$ of the galaxy, the (constant) axis ratio $q$ or ellipticity $e$, defined as $q \equiv b/a \equiv 1-e$, and the orientation or position angle $\theta$. These are the same parameters as those obtained from the moments (Sec. \ref{sec:moments}), except in this case a galaxy is modelled by a more appropriate profile than a Gaussian. One advantage of the S\'ersic fit is that the effect of the PSF can be incorporated into the model by defining a model that is convolved with an input PSF, allowing to remedy some of the resolution effects.

There are several popular codes that \new{fit} a 2D S\'ersic profile: \textsc{Galfit} \citep{galfit}, \textsc{GalfitM} \citep{galfitm}, \textsc{Galight} \citep{galight}, which is a wrapper for \textsc{Lenstronomy} \citep{lenstronomy}, Bayesian fitter \texttt{pysersic} \citep{Pasha2023}, \textsc{SourceXtractor++} \citep{sextractorplus}, \textsc{Astrophot} \citep{astrophot}, \textsc{Galfit-Corsair} for fitting core-S\'ersic models \citep{Bonfini2014}, and more. An exhaustive comparison of these implementations is beyond our scope, \new{but we refer the reader to \cite{EuclidMorph2023}}, who find a good agreement between different fitters. One caveat we would like to highlight is that all of the software mentioned above fit the data to a \textit{pixelized} model, which accounts for the binning of the profile flux onto square pixels, while the default Astropy (and thus \texttt{statmorph}) fitter does not do this, leading to systematically lower S\'ersic indices for peaked profiles (a difference of $\Delta n \sim 1$ at $n \sim 4$). For this reason, we run our tests using \textsc{Galfit} rather than \texttt{statmorph}. \new{The new version of \texttt{statmorph-lsst} corrects this issue by subsampling high-$n$ galaxies, similar to \textsc{Galfit}.}

The S\'ersic model is non-linear and so the fit may converge to a local minimum. Therefore, a good initial guess for the parameters is important. Moreover, it is useful to keep the number of free parameters to a minimum, especially when the data are not sufficient to constrain more complex fits. On one hand, \cite{Peng2010} showed that galaxies are not well-described by a S\'ersic profile when they are sufficiently resolved: nuclear structures, bars, spiral arms, \new{accreted stellar haloes,} and other features mean that a single S\'ersic fit is often insufficient for nearby objects. Codes such as \textsc{Galfit} \citep{Peng2010} allow the inclusion of any number of arbitrary models as well as Fourier perturbations. However, when the resolution is low and these features are smoothed over, as is the case for high-redshift objects, a single S\'ersic fit is often better as additional modes are poorly constrained, and the derived parameters are highly degenerate. In our tests, we fit a single S\'ersic profile convolved with a PSF for all galaxies to minimize the degeneracies between parameters. We use the \texttt{statmorph} estimates of the S\'ersic parameters as initial guesses \citep[as was done in][]{Sazonova2021}.


The dependence of $R_{0.5}^{\rm{Sersic}}$ on image properties is plotted in Fig. \ref{fig:r_grid}, and the advantages of accounting for the PSF become clear. While both $R_p$ and $R_{20}$ overestimate the true radius at low resolutions, $R_{0.5}^{\rm{Sersic}}$ is, on average, much closer to the baseline, with a $\sim 15\%$ difference between highest and lowest \avgsnr{}, slightly overestimating the radius in shallow imaging. The resolution dependence is weak. This acts as an indirect evidence that our results agree with \cite{Ren2024}, who found that the error in radius measurements stems from the PSF smearing out the flux, not the pixelization itself. When the PSF is accounted for, we get much closer radius estimates to the baseline. However, $R_{0.5}^{\rm{Sersic}}$ is a much more uncertain measurement: at \avgsnr{} $\sim1$ the scatter depends on the resolution and ranges from 50\% to 15\%, while it stays at 10\% for $R_{20}$ and 15\% for $R_p$. This stems from the fact that $R_{0.5}^{\rm{Sersic}}$ requires fitting for other S\'ersic parameters simultaneously, and there may not \new{be} enough information to constrain the fit. 

\begin{figure*}
    \centering
    \includegraphics[width=\linewidth]{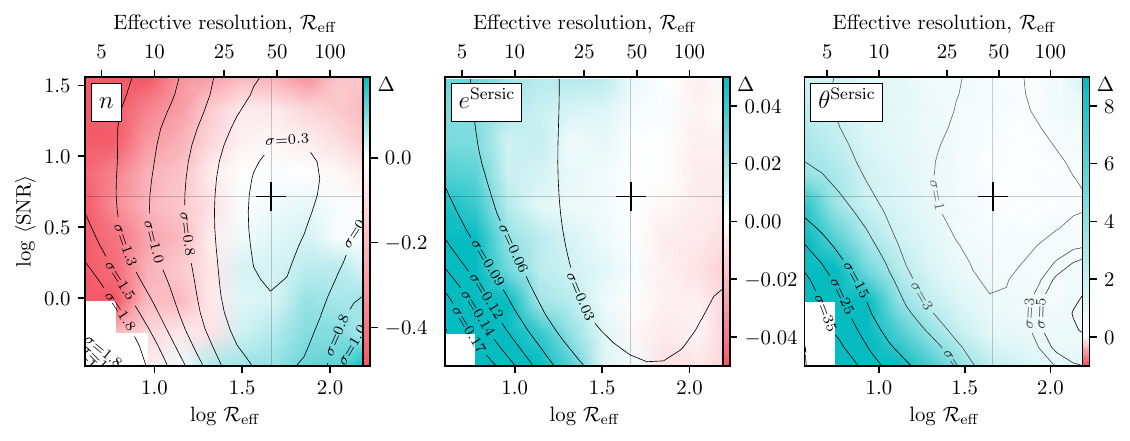}
    \caption{Same as Fig. \ref{fig:moments_grid}, for three S\'ersic parameters: $n$ (\textbf{left}), ellipticity (\textbf{middle}), and orientation (\textbf{right}). $n$ is underestimated by up to 0.5 when there are fewer than 10 resolution elements per object, and it is uncertain, with $\pm1.3$ scatter at lowest resolution. However, $n$ is robust to noise on average, with a $\pm0.5$ scatter at lowest \avgsnr{}. Both $e^{\rm{Sersic}}$ and $\theta^{\rm{Sersic}}$ are, on average, much more reliable than their non-parametric counterparts (Fig. \ref{fig:moments_grid}), with negligible offsets from the baseline and increasing scatter in the lowest resolution and signal-to-noise regimes. }
    \label{fig:sersic_grid}
\end{figure*}

Fig. \ref{fig:sersic_grid} shows the behaviour of other S\'ersic parameters: $n$ (left), $e$ (middle), and $\theta$ (right). Ellipticity and orientation are robust at all resolution and noise levels, with a negligible ($-0.04 \pm 0.15$) bias in $e$ with fewer than 5 resolution elements. While $R_{0.5}^{\rm{Sersic}}$ is weakly affected by resolution, $n$ has a higher resolution dependence, with a systematic bias of $\Delta n = -0.4$ on average in the least resolved galaxies. This bias is quite small given the typical range of $n$ is between $n=1$ for late-type and $n\gtrapprox4$ for early-type galaxies, but \new{the bias can be important when classifying galaxies into Hubble types based on a chosen S\'ersic index value. For example, if the distinction is drawn at $n=2$, some ``early-type'' may be misclassified as ``late-type''.} \new{Our derived correction for the bias in $n$ is:}

\begin{equation}
    n = n_{\rm{base}} \tanh (0.2 \mathcal{R}_{\rm{eff}} ).
\end{equation}

\new{This implies that the bias in $n$ depends on the intrinsic $n$ value snd is larger for more bulge-dominated galaxies. Their S\'ersic indices will hence be systematically more underestimated, making the overall distribution of S\'ersic indices flatter. At \reseff{}$<$5, the average observed S\'ersic index of $n_{\rm{base}}=4$ galaxies is reduced to 2.5, consistent with what \cite{Ferreira2023} and \cite{Martorano2025} found at $z\sim2$. An example of the dependance of $n$ on $n_{\rm{base}}$ can be found in supplementary materials. Correcting for this effect is challenging, since the scatter in $n$ is on the order of $\sim$1.5, much larger than the bias -- so on individual galaxy basis, the uncertainty in the measurement will dominate the correction and increase the scatter.}

Part of the reason for the scatter in $n$ and $R_{0.5}^{\rm{Sersic}}$ is that these two parameters are degenerate with one another \citep[e.g.,][]{Graham1996,Trujillo2001}, so varying them together can result in similarly well fitting profiles. \textsc{Galfit} estimates the uncertainty in fitted parameters from the error image, but does not give uncertainties based on this degeneracy, so the errors are always underestimated. Bayesian codes, such as \texttt{pysersic} \new{\citep{Pasha2023}} provide more robust error estimates, but are slower due to the need to run Monte Carlo chains. An alternative approach to quantify this degeneracy was proposed in \cite{Trujillo2001}. We can define a ``true'' S\'ersic profile, a ``fitted'' one, and a $\chi^2$ between the two. Then, by fixing $n$, we can find the values of $R_{0.5}^{\rm{Sersic}}$ and $I_{0.5}$ that minimize the $\chi^2$ given a fixed $n$. While \cite{Trujillo2001} showed this relationship for $n_{\mathrm{fit}}=4$, we used a similar approach to analytically solve for the degeneracy between $R_{0.5}^{\rm{Sersic}}$, $n$, and $I_{0.5}$ for arbitrary values of $n$ in Appendix \ref{app:sersic}. 

We find that the degeneracy has the form

\begin{equation}
    \frac{R_2}{R_1} = \exp 1.223 \left(\Delta n / n_1^{\sqrt{\mathcal{R}_{\rm{min}}}}\right)
    \label{eq:sersic_degen}
\end{equation}

\begin{figure*}
    \centering
    \includegraphics[width=0.7\linewidth]{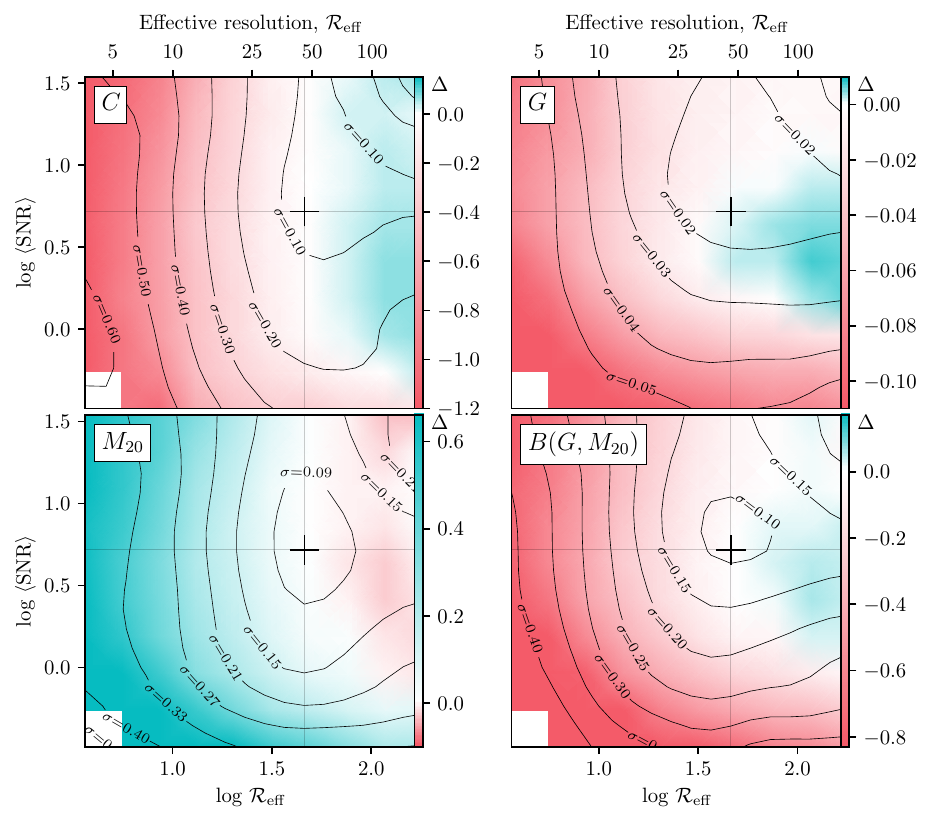}
    \caption{same as Fig. \ref{fig:moments_grid}, for bulge strength metrics concentration ($C$, first panel), Gini index ($G$, second panel), $M_{20}$ (third panel) and the Gini-$M_{20}$ bulge strength (last panel). All four metrics are biased with decreasing resolution, leading to lower bulge strength measurements with less \reseff{}. Since $B(G, M_{20})$ is constructed from two resolution-dependent metrics, it is even more sensitive resolution, leading to a bias of $\sim 1$ at $\reseffm < 10$. While $M_{20}$ and $C$ are fairly insensitive to \avgsnr{}, $G$ and hence $B(G, M_{20})$ are also underestimated at $\avgsnrm < 3$.
    }
    \label{fig:bulge_strength_grid}
\end{figure*}

\noindent where $n_1$ and $R_1$ are the ``true'' parameters, and $n_2$, $R_2$ are the ``fitted'' ones. Here $\Delta n \equiv n_2 - n_1$. $\mathcal{R}_{\rm{min}}$ roughly corresponds to the image resolution, where we defined it as half the pixel scale. 

This equation can be used to find the uncertainties on the fitted parameters, accounting for the degeneracies between them. If the best-fitting values are $n_1, R_1$, then the \textit{next}-best fitting $\Delta n, \Delta R$ can be computed with Eq. \ref{eq:sersic_degen}. These next-best-fits, of course, will result in a higher $\chi^2$. We can then set a limit on an acceptable $\chi^2$ to find the range of parameters that all produce acceptable fits -- this is an uncertainty intrinsic to the degeneracies of the S\'ersic profile, given the image variance. In our analysis, we accept the fits with $p(\chi^2) > 0.05$, i.e. we calculate a 2$\sigma$ intrinsic uncertainty on $n$ and $R_{0.5}^{\rm{Sersic}}$.

In Appendix \ref{app:sersic}, we computed the $\chi^2$ for these ``next best fits''. We find that a typical uncertainty on $R_{0.5}^{\rm{Sersic}}$ that is \textit{intrinsic} to the S\'ersic profile is 20\% at $\avgsnrm = 3$ and $\reseffm = 10$, comparable to the empirically measured scatter in Fig. \ref{fig:r_grid}. Similarly, we found that the intrinsic uncertainty in $n$ ranges from $0.2$ in late-type (true n$=$1) galaxies to $0.6$ in early-type (true n$=$4) galaxies. This is a significantly higher error than what is usually reported by \textsc{Galfit}, but still is an underestimate of the observed scatter in Fig. \ref{fig:sersic_grid}. Running a Monte Carlo-based fit will provide more robust uncertainty estimates specific to a given image, but this approach can still be used to estimate the joint uncertainties on $n$, $R_{0.5}^{\rm{Sersic}}$, and $I_{0.5}$ much more realistically than a typical least-squares fit does.

As $\mathcal{R}_{\rm{min}}$ increases, the degeneracy between $n$ and $R_{0.5}^{\rm{Sersic}}$ becomes stronger, and $n$ is essentially impossible to constrain. \new{There are several consequences of this. First, in cases with poor resolution, the central flux is averaged over a large pixel area, and $\mathcal{R}_{\rm{min}}$ effectively becomes the pixel size. For large $\mathcal{R}_{\rm{min}}$ the degeneracy is larger, and so telling apart different S\'ersic models becomes more difficult. This is especially important to consider for large $n$, e.g. where $n>6$, where the fitted uncertainties may not fully capture this degeneracy.}

\new{Another important consequence} is that two or more component S\'ersic fits, such as bulge-disk decomposition, lead to degenerate parameters when $n$ is allowed to vary. During decomposition, the central flux is dominated by the bulge component, while the outer flux is typically modelled as a disk. However, for both models, most of the constraining power lies in the inner regions where the signal-to-noise \new{ratio} is highest. This means that the inner component is better-constrained, but for the disk component, $\mathcal{R}_{\rm{min}}$ is effectively set by the radius where the inner component is no longer dominant, rather than the resolution. 

\new{A solution with the least degeneracy would be to restrict the} bulge-disk decomposition to a conservative $n_{\rm{bulge}} = 4$ and $n_{\rm{disk}} = 1$ choice \citep[e.g.,][]{Mendel2014,Bottrell2019,Nedkova2024}. \new{This is likely the best approach for decomposition of high-redshift, marginally resolved galaxies \citep[e.g.,][]{Shuntov2025,Genin2025}}. \new{However, if the resolution is sufficiently good, forcing $n=4$ may result in poor fits since the central region has the most weight in determining $n$ and $R_{\rm{eff}}$. The value of $n_{\rm{bulge}}$ can also distinguish between bulges and pseudobulges and thus provide insight into the formation mechanism of a galaxy, so keeping $n_{\rm{bulge}}$ free provides valuable information. Varying the bulge's S\'ersic index while keeping the disk fixed at $n=1$ is a good alternative} \citep[e.g.,][]{Andredakis1995,Graham2001a,Allen2006,Gadotti2009,Simard2011,Dimauro2018,Bottrell2019,LimaDias2024,Quilley2025a}. Allowing both components to have a free value of $n$ \citep[e.g.,][]{Lange2016,galfitm} will suffer from the greatest joint uncertainties of the components, since the slope of the disk-like component cannot be easily constrained in the center, which carries the most weight during the fit. \cite{MacArthur2003} investigated the reliability of the two-parameter fit using mock profiles and found 20\% systemic errors in $n$ for the free bulge component and only 5\% for the disk, but did not discuss the degeneracy between the two components. A solution could be to model the components with a Bayesian approach, which accounts for the correlation between variables \citep[e.g.,][]{Robotham2017,Argyle2018,Casura2022,Cook2025}.  \new{Fully testing the degeneracies between S\'ersic parameters is beyond the scope of this paper, but to summarize, we reiterate \cite{galfit} recommendation that researchers take a data-informed decision and choose a model as flexible as the data permits, keeping in mind that spatial resolution is crucial in breaking parameter degeneracies.} 

\begin{SCfigure*}[0.45]
\begin{wide}

    \centering
    \includegraphics[width=0.65\textwidth]{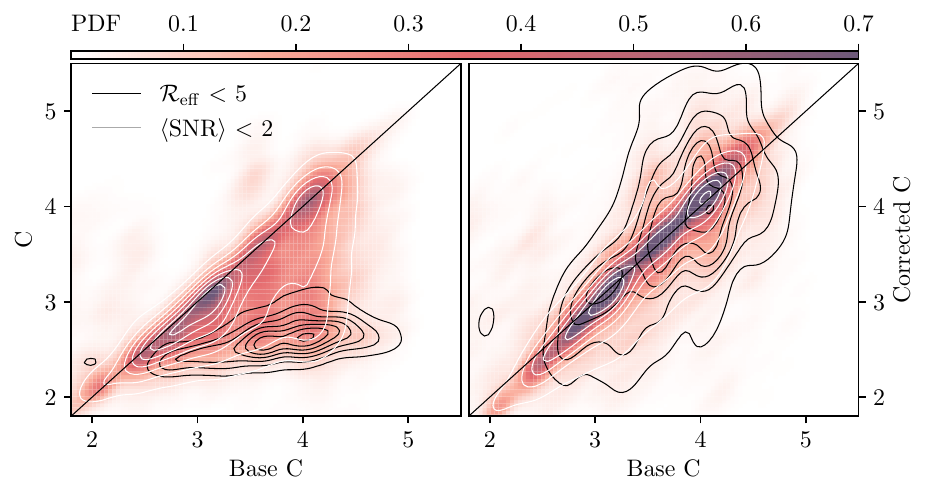}
    \caption{Same as Fig. \ref{fig:rpet_corr}, showing the distribution of the measured (\textbf{left}) and corrected (\textbf{right}) $C$ compared to the baseline. There is a bias where $C$ is underestimated more for early-type galaxies at low \reseff{}, and the scatter is larger for low \avgsnr{}. The \avgsnr{} effect is mitigated by the correction. The \reseff{} bias is also alleviated, but the measurement uncertainties are amplified by the correction, leading to a larger scatter.}
    \label{fig:c_corr}

\end{wide}
\end{SCfigure*}



\subsection{Bulge strength measurements}\label{sec:bulge}

\subsubsection{Concentration}\label{sec:concentration}

The concentration parameter $C$ quantifies the relative brightness of the centre of a galaxy relative to its outskirts. $C$ is measured by comparing two circular isophotal radii, $R_{\rm{inner}}$ and $R_{\rm{outer}}$ \citep{Vaucouleurs1977a,Okamura1984,Kent1985,Bershady2000,Conselice2003}. Most commonly, the inner and outer radii are taken as the 20\textsuperscript{th} and the 80\textsuperscript{th} percentile isophotes, although different choices are possible: e.g., using a surface brightness threshold for the outer isophote as in \cite{Kent1985} or a ratio of effective S\'ersic radii as in \cite{Trujillo2001}. \cite{Bershady2000} showed that the 20\textsuperscript{th} and 80\textsuperscript{th} isophotes are stable to changes in resolution and the signal-to-noise \new{ratio}, \new{but did not account for the PSF effects, while we find that $R_{20}$ is very sensitive to resolution (\ref{sec:r20})} Still, the most common definition of concentration, and one adopted by \texttt{statmorph}, is

\begin{equation}
    C = 5 \log \frac{R_{80}}{R_{20}}
\end{equation}

\noindent where the isophotes are found as described in \ref{sec:r20}. 

The first panel in Fig. \ref{fig:bulge_strength_grid} shows the dependence of $C$ on resolution and the signal-to-noise \new{ratio}. $C$ is robust to changes in \avgsnr{}: there is no systematic offset in concentration until $\log$ \avgsnr{} reaches 0.2. This is expected since both $R_{20}$ and $R_{80}$ are not strongly affected by noise. Below this \avgsnr{}, $C$ is slightly underestimated by $\sim 0.2$. The scatter in $C$ slightly increases as \avgsnr{} decreases, reaching an uncertainty of $\pm$0.2 at low \avgsnr{} at a fixed resolution. This is consistent with what \cite{Conselice2000}, \cite{Lotz2004}, and \cite{Ren2024} found. \new{\cite{Graham2001} and \cite{Trujillo2001} note that $C$ requires knowing the total flux and hence a maximal radius, which can lead to a \avgsnr{} dependence. Our total flux is measured in a $1.5R_p$ aperture, which is stable even at low \avgsnr{}, and this makes $C$ robust in the low signal-to-noise regime. In an a very high $n$ case, however, 1.5$R_p$ does not fully enclose the total flux \citep{Graham2005}, leading to a systematic underestimate of the \textit{real} concentration -- but this error does not depend on resolution or depth.}


Fig. \ref{fig:c_corr} (left) shows the relationship between $C$ and base $C$ for the entire sample, the low-\avgsnr{} low-\reseff{} regimes in red, white, and black respectively. The bias in $C$ for poorly resolved galaxies depends strongly on the intrinsic concentration, with a higher bias for more bulge-dominated galaxies, reaching $\Delta C \sim 1.5$ when $\reseffm < 5$. Similarly, when \avgsnr{} is low, early-type galaxies suffer a larger bias and more scatter in their $C$ measurements, consistent with \cite{Wang2024}, who found a bias of $\Delta C = 1$ for bulge-dominated sources. Therefore, at low resolution or poor \avgsnr{}, all galaxies will appear less bulge-dominated, and the distribution of bulge strengths will appear biased lower in agreement with previous studies \citep{Yeom2017,Thorp2021,Yu2023,Ren2024,Wang2024}\footnote{\cite{Conselice2000}, \cite{LanyonFoster2012}, and \cite{Yao2023} found no difference in $C$ when artifically redshifting their sources. Given a wide range of studies that agree with our results, as well as the intrinsic dependence of $C$ on isophotal radii which are resolution-dependent, we are confident that $C$ does in fact depend on resolution.}. The dependence of $\Delta C$ on intrinsic concentration has implications for studies of the galaxy morphology bimodality in the early Universe, where the resolution is poorer, as the fraction of centrally concentrated galaxies may appear to be lower due to this effect and can explain the decreasing $C$ trends seen in recent JWST analyses (see Sec. \ref{sec:discussion}).

The relationship between $C$ and $C_{\rm{base}}$ we obtained with \texttt{SymbolicRegression} is

\begin{equation}
    C = C_{\rm{base}}  \tanh(\log(0.4\reseffm{})) + \frac{3.3}{\sqrt{\reseffm-0.76}},
\end{equation}

\noindent which reflects the dependence on intrinsic $C_{\rm{base}}$. As in the case of $R_p$, we can then correct $C$ measurements by calculating $C_{\rm{base}}$ from this equation. We are able to correct well for the bias in the low signal-to-noise galaxies (Fig. \ref{fig:c_corr}, right-hand panel). Our correction function also brings the measurements closer to $C_{\rm{base}}$, although with a significant scatter introduced by the correction. Since at $\reseffm < 5$ the distribution of the raw $C$ measurements (left) is so flat, any uncertainty in the measurements gets amplified, leading to a larger scatter in the corrected $C$ (right). Moreover, since the correction mostly attempts to mitigate the effect of the PSF, our correction will be effective in images where the PSF FWHM is two pixels, and ideally a new correction should be derived for a range of different PSF sizes.

\subsubsection{Gini index}\label{sec:gini}

The Gini index, $G$, originally proposed as a measure of wealth inequality, has been widely used to characterize the concentration of a galaxy's light since its introduction to astronomy in \cite{Abraham2003} and \cite{Lotz2004}. The Gini index is defined as 

\begin{equation}
    G \equiv \frac{1}{2 \bar{X} n(n-1)}\sum_i^n \sum_j^n \left| X_i - X_j \right|,
\end{equation}

\noindent where $n$ is the number of elements ${X_i}$ and $\bar{X}$ is their mean. $G = 1$ means that all flux is concentrated in a single pixel, while $G=0$ means that it is evenly distributed throughout the entire galaxy. In practice, this can be simplified to a single sum by first sorting ${X_i}$ in ascending order, and then calculating $G$ as 

\begin{equation}
    G = \frac{ \sum_i^n (2i - n - 1) X_i}{n(n-1) \bar{X}}
\end{equation}

By construction, the Gini index is meant to be used with strictly positive quantities -- e.g., income, or a measure of impurity in machine learning applications \citep[e.g.,][]{sklearn}. If $X_i$ is allowed to be negative, $G$ is no longer bounded by the $(0, 1)$ range and interpreting the results becomes difficult. In astronomy, after the sky subtraction, the flux values are often negative, so to remedy this, $G$ is most often calculated on the \textit{absolute} values of flux:

\begin{equation}
    G = \frac{ \sum_i^n (2i - n - 1) |X_i|}{n(n-1) \bar{|X|}}.
\end{equation}

Because of this absolute-valued definition, an interesting edge case is a Gini index of a white noise distribution, where $X_i \sim \mathcal{N}(0, \sigma^2)$. This is equivalent to the Gini index of a blank sky without any source flux. After the absolute value is taken, such a patch follows a half-normal distribution with mean $\mu = \sigma \sqrt{2/\pi}$. For a strictly positive distribution with known cumulative density function $F(x)$, the Gini index can be computed analytically, and for a half-normal distribution it is 

\begin{equation}
    G \equiv \frac{1}{\mu} \int_0 ^\infty F(x) \left(1 - F(x)\right) dx = \sqrt{2}-1 \approx 0.41.
\end{equation}

\cite{Lisker2008} empirically found that as the signal-to-noise \new{ratio} decreases, $G$ approaches a constant value of $\sim 0.4$. This is precisely due to approaching this limit of the Gini index of absolute-valued white noise. 

In general, as more and more sky pixels are included in the aperture where $G$ is calculated, the sum will become dominated by the sky, and so $G$ will approach this value of 0.41. Therefore, $G$ is not only sensitive to the signal-to-noise \new{ratio}, but also to the aperture size \citep{Lotz2004, Lisker2008, statmorph}. In practice, $G$ is therefore calculated within an aperture defined by $R_p$, which is robust to changes in the signal-to-noise \new{ratio} (Sec. \ref{sec:rpet}) and thus cosmological surface brightness dimming. In \texttt{statmorph}, an image is first smoothed by a filter with a size $R_p/5$, and then a new segmentation map is computed using the average SB at $R_p$ as a threshold. $G$ is then calculated only using the (un-smoothed) pixels included in this map. In their tests, \cite{Lisker2008} confirm that using an aperture with a Petrosian radius has a sufficiently small contamination from sky pixels that $G$ does not yet approach the noise-limited value of $0.41$.

In Fig. \ref{fig:bulge_strength_grid} (\new{top right} panel), we show that our results are in agreement with previous discussion in \cite{Lotz2004}, \cite{Holwerda2011} and \cite{Lisker2008}, where $G$ is robust up to $\avgsnrm \sim 2$. At lower \avgsnr{}, $G$ begins to decrease towards the noise-limited value of $0.41$. As we showed in Sec. \ref{sec:rpet}, $R_p$ is robust to changes in \avgsnr{}, only decreasing by $\sim 10 \%$ as $\avgsnrm \rightarrow 1$, and so as the signal-to-noise \new{ratio} decreases and $R_p$ stays roughly constant, the pixels within $R_p$ become more noise-dominated, leading to decreasing values of $G$. This explains the results in \cite{Petty2014}, who use \textit{HST} images to calculate $G$ and find a flat distribution of $G \approx 0.45$ at high redshifts -- their images were likely already in this noise-dominated regime, and caution should be taken when measuring $G$ on noisy images. However, at medium signal-to-noise, the measurement is very reliable.

Similarly to concentration, $G$ is much more sensitive to resolution than the signal-to-noise \new{ratio}: at more than 10 resolution elements per galaxy, $G$ is stable, but with fewer than 5 it can be underestimated by up to $0.1$ with a $\pm 0.05$ scatter, which is significant given a small dynamic range of $G$ values. This is consistent with the tests of $G$ and resolution in \cite{Yeom2017} and \cite{Wang2024}. In particular, \cite{Wang2024} -- with their slightly different definition of \reseff{} -- also find an increasing bias with resolution that flattens at $\reseffm > 6$, at slightly lower but similar resolutions than in our tests. We computed the Gini index for some simple analytic S\'ersic profiles with a changing pixel scale and PSF size independently, and found that while both affect $G$, the PSF has a much stronger impact. As the PSF redistributes flux from the bright point sources outwards, it naturally decreases the Gini index. The correction derived with \texttt{SymbolicRegression} is:


\begin{equation}
    G = G_{\rm{base}} \left( 0.83-\frac{0.59}{\reseffm} \right)+ 0.1 \tanh \avgsnrm 
\end{equation}

The offset depends on both \avgsnr{} and \reseff{}, but the fit applies a flat correction to the noise-dependence. Therefore, comparing the \textit{relative} Gini indices of galaxies at the same redshift observed with the same instrument is still possible provided the resolution is stable and the signal-to-noise is not so poor that $G$ becomes noise-dominated. However, care must be taken when comparing galaxies with different effective resolutions.

\subsubsection[M20]{$M_{20}$}\label{sec:m20}

$M_{20}$ was first defined in \cite{Lotz2004} as an alternative measure of a galaxy's concentration, and is often used in conjunction with the Gini index to find bulge-dominated or merging galaxies. It is based on the second moment of a galaxy's light distribution. In general, we can find the second moment of the $X$ brightest galaxy pixels by first sorting all the pixels by flux, and then finding the sum

\begin{equation}
    m_X = \sum_i^X f_i \left( (x_i - x_0)^2 + (y_i - y_0)^2 \right).
\end{equation}

$M_{20}$ is then defined as the moment of the brightest 20\% of pixels divided by the total moment:

\begin{equation}
    M_{20} \equiv \frac{m_{20\%}}{m_{100\%}}.
\end{equation}

In practice, the total light cannot be known, and so $m_{100\%}$ is defined within some aperture. \texttt{statmorph} uses the Gini segmentation map, constructed with the average SB at $R_p$ as a threshold, and so roughly corresponds to flux within $R_p$.

$M_{20}$ is a purely photometric quantity similarly to $C$, and so is robust to changes in \avgsnr{}, as shown in the third panel of Fig. \ref{fig:bulge_strength_grid}. As \avgsnr{} decreases, the scatter in $M_{20}$ increases as photometry becomes more uncertain, but there are no systematic biases in shallow imaging. This was also concluded in \cite{Ren2024}, while \cite{Lotz2004} and \cite{Wang2024} found slightly higher values of $M_{20}$ at very low signal-to-noise. The difference in our results is likely due to implementation; if the Petrosian-based segmentation map is used to compute $M_{20}$ like in \texttt{statmorph}, the values are stable even at $\avgsnrm < 1$.

However, $M_{20}$ is biased at low resolutions: as \reseff{} decreases, $M_{20}$ can increase by up to $0.6$, which could lead a galaxy to be misclassified from bulge-dominated to disk-dominated. This is a natural consequence of the PSF: it distributes the flux from bright point sources, such as the central bulge and any star clusters, to large distances; therefore naturally increasing $M_{20}$ by its definition. This was also the conclusion in \cite{Ren2024}. The effect of the PSF is therefore important to keep in mind and correct for. The correction function we derived is

\begin{equation}
    M_{20} = \left[0.74 - \frac{1}{\reseffm} \right] M_{20, \rm{base}} - 0.5.
\end{equation}

\noindent This correction depends only on \reseff{} as expected, and reduces the bias introduced by the PSF. However, since the scatter and hence the uncertainty in $M_{20}$ at low resolutions is high, the correction will also amplify these uncertainties; moreover, it depends on the exact shape of the PSF and so ideally it should be re-derived with a PSF significantly different from a 2-pixel Moffat used in this work.

Finally, several studies that artificially redshifted their galaxies from $z \approx 0.5$ to $z > 2.5$ found no change in their $M_{20}$ measurement \citep{Petty2014,Yao2023,Salvador2024}. We expect this stems from the fact that at $z > 1$, the decrease in angular size with proper distance is small, and so galaxies with similar intrinsic sizes will have similar apparent sizes. On the other hand, cosmological surface brightness dimming does not strongly affect $M_{20}$ as it is insensitive to the signal-to-noise \new{ratio}. However, observers should still be careful when comparing $M_{20}$ measurements of galaxies with different \textit{intrinsic} sizes, as it is the \textit{effective} resolution that introduces the bias in $M_{20}$. As galaxies at higher redshifts are intrinsically smaller, their effective resolution is still lower and thus measured $M_{20}$ is likely overestimated -- see Sec. \ref{sec:discussion} for further discussion.

\subsubsection[Gini-M20 bulge strength]{Gini-$M_{20}$ bulge strength}\label{sec:bgm20}

Using subtle differences in otherwise similar metrics can be a powerful way to probe more detailed galaxy structure. \cite{Lotz2004} use $G$ and $M_{20}$ in tandem in this way. Both metrics probe how concentrated a galaxy light is, but $G$ is agnostic to the centre of the galaxy, and measures bright pixels anywhere in the image. On the other hand, $M_{20}$ is heavily weighted to the distance of these bright pixels from the centre of flux, making these two similar measurements actually quite different. 

Regular early-type and late-type galaxies therefore
form a ``main sequence'' on the $G-M_{20}$ plane. Bulge-dominated galaxies have bright pixels in the center, leading to high $G$ and low $M_{20}$ values. Star-forming disks have many bright knots of star formation, leading to low $G$ and high $M_{20}$ values instead, thus forming this ``main sequence''. But there are rarer galaxies off the axis of this main sequence: for example, during a merger, when two galaxies have already a shared envelope but the nuclei have not yet coalesced, there are two bright nuclei -- leading to high $G$ -- that are both removed from the common centre of light, leading to high $M_{20}$. Therefore, the $G-M_{20}$ plane can be used to both classify galaxies along the Hubble sequence, and identify a particular stage of mergers. \cite{Snyder2015b} and \cite{statmorph} define a bulge strength and a merger statistic based on the distance along the main sequence and the offset away from it. Here we discuss the bulge strength statistic, computed in \texttt{statmorph} as 

\begin{equation}
    B(G, M_{20}) \equiv -0.693M_{20} + 4.95G - 3.96.
\end{equation}

\noindent We note that this is simply an empirical relation derived using local galaxies in \cite{Snyder2015b}. \cite{Sazonova2020} and \cite{EspejoSalcedo2025} show that at $z>1$, galaxies still follow a ``main sequence'' in the $G-M_{20}$ plane, but the slope and the equation of this main sequence are different. Therefore it may be meaningful to redefine this statistic for each individual study, using a representative sample of early- and late-type galaxies at the same redshift with comparable imaging quality. 

Perhaps unsurprisingly, we find that $B(G, M_{20})$ depends on \avgsnr{} the same way as $G$ does, with decreasing bulge strength at low signal-to-noise (Fig. \ref{fig:bulge_strength_grid}). The dependence of $B(G, M_{20})$ on resolution is however more striking. $G$ decreases in low-resolution imaging, leading to galaxies being classified as more disk-like. $M_{20}$ increases at low resolutions, also producing more disk-like morphologies. Therefore, $B(G, M_{20})$ absorbs both of these effects -- as resolution decreases, $B(G, M_{20})$ decreases by up to $0.8$. The dynamic range of this statistic on the Pan-STARRS sample used in \cite{statmorph} is approximately $(-1, 1)$, and so a bias of $-0.8$ is a significant decrease. While the bias in $G$ does not depend on $G_{\rm{base}}$, the bias in $M_{20}$ does. Therefore, even re-defining a new ``$G-M_{20}$ main sequence'' in imaging with a different resolution will not fully mitigate this resolution dependence of $B(G, M_{20})$. 

Similarly to $C$, a correction for $B(G, M_{20})$ can be derived directly from $G$ and $M_{20}$. Alternatively, the best fit provided by \texttt{SymbolicRegression} is shown in Appendix \ref{app:pysr}. For our PSF, this provides a good correction on average, but the scatter at $\reseffm < 5$ after the correction is  high, reaching $\pm 1$ -- see the figure in the online materials.

\subsection{Disturbance measurements }\label{sec:disturbance}

\begin{SCfigure*}[0.35]
\begin{wide}
    \includegraphics[width=0.7\textwidth]{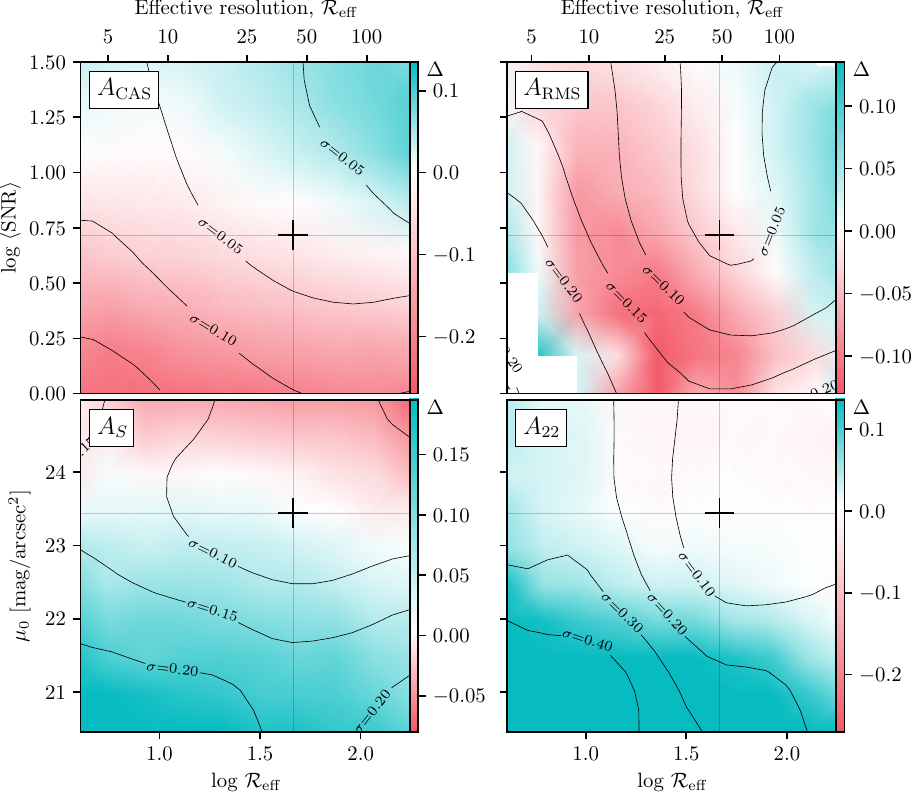}
    \caption{
    same as Fig. \ref{fig:moments_grid}, for disturbance metrics: $A_{\rm{CAS}}$ \citep[first panel;][]{Conselice2003} (first), $A_{\rm{RMS}}$ \citep[second panel;][]{Sazonova2024}, shape asymmetry $A_S$ \citep[third panel;][]{Pawlik2016}, and $A_{22}$, asymmetry of the 22 mag/arcsec$^2$ isophote (last panel; this work).  All asymmetry metrics except $A_{\rm{RMS}}$ are biased at decreasing \avgsnr{}, although $A_{22}$ is robust as long as $\mu_0 > 22$. $A_{\rm{RMS}}$, on the other hand, is much more sensitive to resolution than the other metrics.
    }
    \label{fig:asym_grid}
\end{wide}
\end{SCfigure*}

\subsubsection[Gini-M20 disturbance]{Gini-M$_{20}$ disturbance}\label{sec:sgm20}

Following the Gini-M$_{20}$ bulge strength, the disturbance in $G$-M$_{20}$ plane is calculated as a vertical offset from the ``main sequence'' and corresponds to galaxies with concentrated but off-centered peaks, such as double nuclei. \texttt{statmorph} defines it as 

\begin{equation}
    S(G, M_{20}) = 0.139 M_{20} + 0.99G - 0.327.
\end{equation} 

It is important to note that while $S(G, M_{20}) > 0$ indicates a deviation from the ``main sequence'' towards more unusual morphologies, high values being associated with mergers, a negative value of $S$ does \textit{not} mean that the galaxy is ``less disturbed'', so we caution against over-interpreting negative values of this parameter.

The behaviour of $S(G, M_{20})$ at different image qualities is shown in the online materials for brevity. Interestingly, since $S(G, M_{20})$ measures the \textit{perpendicular} distance in the $G$-$M_{20}$ plane, the resolution dependencies of the two parameters mostly cancel out, so $S$ is a stable measurement for the majority of galaxies in our sample, with only a $\pm 0.02$ average deviation and a $0.1$ average scatter across the entire parameter plane. However, there are only 20 galaxies in the sample with $S(G, M_{20}) > 1$. For these objects, there is a systematic decrease of roughly $-0.1$ at low resolutions, although this trend suffers from low-number statistics. Finally, if $B(G, M_{20})$ is re-defined at higher redshifts, then $S(G, M_{20})$ must also be re-defined accordingly \citep{Sazonova2020,EspejoSalcedo2025}.

\subsubsection{CAS Asymmetry}\label{sec:asymmetry}

Asymmetry ($A_{\rm{CAS}}$) is a part of the concentration-asymmetry-smoothness (CAS) system from \cite{Conselice2003} and is one of the most widely used measurements of the irregularity of a galaxy. It is conceptually simple: an image is rotated by 180$\degree$ about its centre, and then the rotated image is subtracted from the original. The asymmetry measurement is the sum of the residual normalized by the total flux of the galaxy.

In practice, this calculation involves many hidden nuances. The centre of rotation has a strong impact on the final value: while the original definition in \cite{Schade1995} used the centroid of the galaxy, \cite{Conselice2003} find a centre which \textit{minimizes} asymmetry. Moreover, the sky background itself has a non-zero asymmetry, both due to random fluctuations in flux and potential sky gradients, and so must be subtracted. A modern definition of $A_{\rm{CAS}}$, and that adopted by \texttt{statmorph}, is:

\begin{equation}\label{eq:acas}
    A_{\rm{CAS}} = \min_{\mathbf{x}_0} \frac{\sum |I_{ij} - I_{ij}^{180}|}{|I_{ij}|} - \langle A_{\rm{CAS, bg}}\rangle
\end{equation}

\noindent where $I_{ij}^{180}$ is the flux rotated about $\mathbf{x}_0$, and $\langle A_{\rm{CAS, bg}} \rangle$ is the average \new{asymmetry of} an empty patch of the sky. However, while this definition is used in most codes, the exact implementation (e.g., between \texttt{statmorph} and \textsc{Morfometryka}) differs and so produces different measurements on the same set of images.

We have discussed this measurement in great detail in \cite{Sazonova2024}, and so we keep the discussion in this work brief, and suggest the reader refers to our previous work, where we the impact of the sky subtraction, centering, and interpolation needed when the image is rotated on $A_{\rm{CAS}}$. The primary conclusion in \cite{Sazonova2024} is that the absolute-valued definition makes a perfect sky subtraction impossible, and $\langle A_{\rm{CAS, bg}}\rangle$ always overestimates the true contribution of the background to asymmetry. Therefore, this definition will \textit{over-}subtract the background asymmetry, leading to systematically lower $A_{\rm{CAS}}$ values. 

The dependence of $A_{\rm{CAS}}$ on resolution and depth in our current series of tests is shown in the left panel of Fig. \ref{fig:asym_grid}, agreeing with our previous conclusions: asymmetry is systematically underestimated at low \avgsnr{}. Our results are qualitatively consistent with previous works \citep[e.g.,][]{Bottrell2019,Thorp2021}. Similarly to \cite{Thorp2021}, we also find a weak dependence on resolution, although it is not as significant as the dependence on \avgsnr{}.

There are several approaches to mitigate the severe dependence of $A_{\rm{CAS}}$ on the signal-to-noise \new{ratio}. While they are discussed at length in \cite{Sazonova2024}, for consistency with other parameters, we also provide a \texttt{SymbolicRegression} correction in Appendix \ref{app:pysr}. We also test an \textit{outer} asymmetry, $A_o$, defined as a\new{n} asymmetry where the inner pixels up to $R_{50}$ are masked -- these results are in the online materials and Appendix \ref{app:pysr}. 
Again, we caution that while this correction achieves consistent measurements with the baseline \textit{on average}, it introduces a large scatter for low-\avgsnr{} objects and so it is better to either approach it with caution, or employ other metrics when \avgsnr{} is variable or low (e.g., $A_{\rm{RMS}}$, see below).

\subsubsection{RMS Asymmetry}\label{sec:arms}

Root-mean-squared asymmetry ($A_{\rm{RMS}}$) is an alternative definition of the same metric, where the residual flux is squared rather than absolute-valued. This metric was also proposed in \cite{Conselice2003} but has historically gained less traction. It provides an important advantage: the sky contribution to asymmetry is separable from the object asymmetry since the square is better-behaved than an absolute value \citep{Sazonova2024}. Therefore, it is, by definition, much more robust to noise. $A_{\rm{RMS}}$ is defined as:

\begin{equation}
    A^2_{\rm{RMS}} = \min_{\mathbf{x}_0} \frac{\sum (I_{ij} - I_{ij}^{180})^2}{I^2_{ij}} - \langle A^2_{\rm{RMS, bg}}\rangle
\end{equation}

\noindent We have also discussed this metric in detail in \cite{Sazonova2024}, but we still show the resolution and signal-to-nose plot in the second panel of Fig. \ref{fig:asym_grid}. There is almost no dependence on signal to noise, aside from a small $-0.1$ average underestimate at $\avgsnrm \sim 1$, and a small amount of scatter as the signal-to-noise \new{ratio} decreases. However, the metric is sensitive to resolution, with a $\pm 0.2$ spread at different resolution levels. Since the PSF spreads out the flux, the sum of the squared fluxes becomes smaller, and so the measurement decreases when the PSF is large. Galaxies with higher intrinsic asymmetries are more affected by this. Our derived correction is:

\begin{equation}
    A_{\rm{RMS}} = A_{\rm{rms, base}} \times 0.5 \log \reseffm.
\end{equation}

\noindent While this correction provides a good fit to our data on average, it significantly overestimates $A_{\rm{RMS}}$ at low resolutions, $\reseffm < 5$ (see the online materials). Moreover, since the impact of the PSF depends on the shape of the PSF itself, we advice to derive a new correction if the user's PSF is significantly different from a 2-pixel Moffat used in this work, or explore deconvolution methods as described in \cite{Sazonova2024}.

\begin{figure*}
    \centering
    \includegraphics[width=\linewidth]{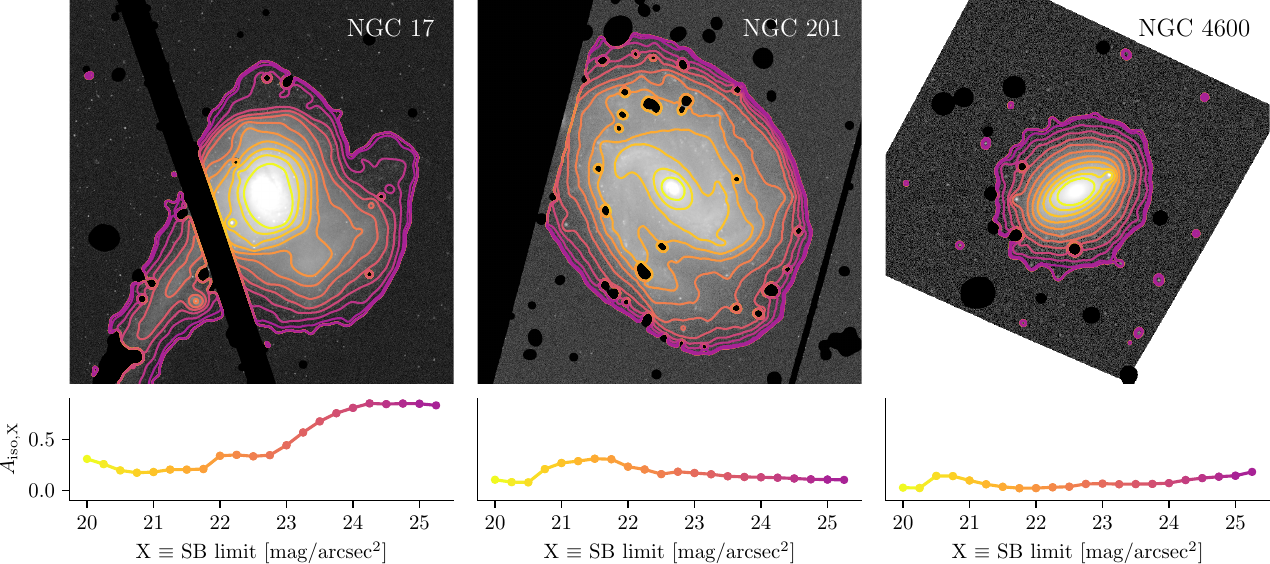}
    \caption{The isophotal asymmetry contours for three galaxies: NGC 17 (\textbf{left}), NGC 201 (\textbf{centre}), and NGC 4600 (\textbf{right}) for a range of surface brightness thresholds from 20 to 25.5. NGC 17 is a post-merger and has strong tidal features, reflected in high $A_{>23}$. NGC 201 is a spiral galaxy with mild asymmetries induced at the medium SB level $A_{21.5}$ by the bar. Finally, NGC 4600 is a lenticular galaxy with low $A_X$ except a small bump at $A_{20.5}$ caused by a star cluster. $A_X$ as a function of $X$ therefore is a powerful tool to trace which features cause which disturbances in a galaxy.}
    \label{fig:a_iso}
\end{figure*}

\subsubsection{Shape asymmetry}\label{sec:ashape}

CAS asymmetry and, to a larger extent, RMS asymmetry are more sensitive to bright disturbances such as star-forming clumps or multiple nuclei, since they are flux-weighted. However, in many cases, such as in tidal tails induced during late- or early-stage mergers, the asymmetric structures have a low surface brightness. Traditional metrics such as asymmetry and Gini-$M_{20}$ disturbance are not necessarily designed to best detect these large but faint features. An alternative approach was proposed in \cite{Pawlik2016}: to measure an asymmetry of a binary detection map of a galaxy, without a flux weighting. This is a measure of \textit{shape} asymmetry, $A_S$.

While there are different ways of constructing this binary detection maps, they are generally similar. \texttt{statmorph} implements $A_S$ by first smoothing an image with a $3\times3$ boxcar filter, and then drawing a mask of all connected pixels above a $1\sigma$ threshold of the background noise. Then Eq. \ref{eq:acas} is used to calculate the asymmetry using the same centre as that in $A_{\rm{CAS}}$. Since the background pixels have a value of 0, a background term is not necessary. 

Fig. \ref{fig:asym_grid} (third panel) shows the behaviour of $A_S$. While this metric is fairly robust to changes in resolution, it is sensitive to the surface brightness limit, $\mu_0$, leading to a systematic underestimate of $\Delta A_S \approx 0.15$ and a large scatter at $\mu_0 < 22$ compared to the baseline $\mu_0 = 23.5$. While this metric is often used to detect and time mergers or other disturbances \citep[e.g.,][]{Pawlik2016,Snyder2015a,Foster2025,Yanagawa2025}, there are no studies showing this behaviour which is crucial in detecting mergers in different surveys or at different redshifts. The intrinsic uncertainty in $A_S$ compared to the baseline is too large compared to the bias to get an effective correction, but we provide the error statistics in Appendix \ref{app:pysr}.

This dependence on noise is a natural consequence of the definition of $A_S$. Since the segmentation map used to calculate it is defined using a $1\sigma$ background noise threshold, then the features used to calculate the asymmetry are the features that correspond to the surface brightness limit of the image. While a deep image, such as those expected from the LSST, will probe tidal features of nearby galaxies at $\mu_0 \approx 30$, an SDSS-like imaging of the same objects will probe the asymmetry of internal structures, such as the outskirts of the spiral arms. Therefore, $A_S$ measured in these two cases will quantify the asymmetry of \textit{different physical features}. Instead, we propose a similar, but a more physically motivated approach to quantifying a ``shape asymmetry'' discussed below.

\subsubsection{Isophotal asymmetry}\label{sec:aiso}

The goal of the new isophotal asymmetry ($A_X$) metrics is to alleviate the dependence of $A_S$ on the background, instead providing features at a specific surface brightness or luminosity (and by proxy, mass) surface density.

It is defined in the same way as $A_S$, except we use a different threshold to compute the binary map. We define $A_X$ as the asymmetry of a galaxy's surface brightness $\rm{SB} = X$ mag/arcsesc$^2$ isophotes rather than the background noise level. This can be an even more powerful approach to trace disturbances in both the internal structure and the low surface brightness features of a galaxy. To measure an isophotal asymmetry, we simply create a mask of all pixels brighter than a particular SB threshold $X$. We then rotate this mask about the $A_{\rm{CAS}}$ centre and calculate the residual, divided by the size of the detection mask. Unlike for shape asymmetry, we do not require the regions to be connected, since at a high surface brightness threshold this measurement probes disconnected clumps. This approach requires a good masking routine, ideally masking out any foreground stars. To make sure this measurement is physically meaningful, it should also be corrected for any cosmological dimming effects -- in this work, we define rest-frame surface brightness $\mu_0$ as $\mu_0 \equiv \mu - 2.5 \log (1+z)^3$, which includes the $(1+z)^4$ dimming and a K-correction arising bandpass shift in the AB magnitude system \citep{Hogg2002}.

Fig. \ref{fig:a_iso} shows example isophotal contours for three galaxies: NGC 17, a recent merger, NGC 201, a spiral galaxy, and NGC 4600, a lenticular galaxy in the Virgo cluster. Each image is scaled to 25 parsec/pixel and has a surface brightness limit of 24 mag/arcsec$^2$. We constructed isophotes with a range of SB $= 20 \sim 25.5$ (light yellow to dark purple). Isophotal asymmetry measured using these different contours is shown below. The post-merger galaxy has a strong asymmetry at a low surface brightness levels -- $A_{22}$ and deeper -- tracing the large tidal feature to the southwest. On the other hand, the spiral galaxy has symmetric faint isophotes and a large bright $A_{21}$, which traces the barred structure in the centre. Finally, the lenticular galaxy has low isophotal asymmetry at all surface brightnesses, slightly bumped up with a star cluster at $X\approx 20.5$. This shows the power of our proposed metric: a surface brightness, after correcting for cosmological dimming, maps directly to the luminosity density of the galaxy. Therefore measuring $A_{21}$ or $A_{24}$ probes the asymmetry of different features with different light, and hence mass surface density, and can help differentiate between tidal tails, double nuclei, knots of star formation or star clusters, and other structures. The profile of the galaxy where $A_X$ is plotted against $X$ has a potential to distinguish between different types of disturbances. It is also possible to define an integrated measure, such as a flux-weighted asymmetry of different isophotes, similar to the ``twistiness'' in \cite{Ryden1999}. We plan to continue this exploration in an upcoming work. 

\begin{SCfigure*}[0.5]
\begin{wide}
    \centering
    \includegraphics[width=0.63\textwidth]{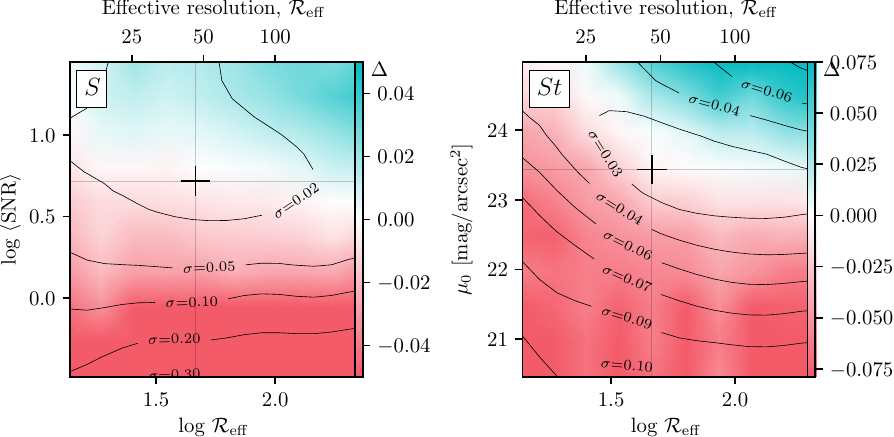}
    \caption{same as Fig. \ref{fig:moments_grid}, for smoothness \citep[$S$;][\textbf{left}]{Conselice2003} and substructure ($St$; Sec. \ref{sec:substructure}, \textbf{right}). Both decrease at low \avgsnr{} as clumps become more difficult to distinguish from the noise. The scatter in $S$ however dominates the typical measurements of $S$ at $\avgsnrm < 3$, so this metric is noise-dominated in most cases. The scatter in $St$ is smaller than its typical dynamic range, making it more reliable, but it has a resolution dependence absent in $S$.}
    \label{fig:smoothness}
\end{wide}
\end{SCfigure*}
\new{The power of this metric is not in measuring A$_X$ at a \textit{single} threshold, but at a series of thresholds spanning physically meaningful features -- from the highest to the lowest surface brightness levels, and then studying the galaxy's $A_X$ profile. The \texttt{statmorph-lsst} code has a new input parameter, \texttt{asymmetry\_isophotes}, where a user can provide a list of levels in image units to calculate $A_X$ in. We strongly recommend first deciding on a list of surface brightness thresholds in mag/arcsec$^2$, and then converting it to image units, for a physically motivated measurement -- however it is difficult for us to implement this as default without knowing the user image's zeropoint and units.}


The last panel in Fig. \ref{fig:asym_grid} shows the performance of $A_{22}$, the asymmetry measured using a 22 mag/arcsec$^2$ isophote. The measurement is not affected by the limiting depth by definition, as long as the depth exceeds the measured surface brightness. In practice, $A_X$ is reliable when $X \lesssim \mu_0 - 0.5$, i.e. even if the sky surface brightness limit is shallower by up to $0.5$ mag/arcsec$^2$ than the desired contour, $A_X$ is still a good measurement. At very low resolutions, $A_{22}$ is slightly underestimated. This is due to the loss of small, bright clumps as well as the PSF smearing effect as resolution decreases. High surface brightness measurements, such as $A_{19}$, will be affected more since they are more sensitive to individual clumps; and low surface brightness measurements are more robust to changes in resolution. In particular, in the low surface brightness regime probing the faintest tidal features, low resolution is \textit{beneficial} since it increases the effective image depth while retaining these large features.

\subsection{Other metrics}\label{sec:other}
 
\subsubsection{Smoothness}\label{sec:smoothness}

$S$, often referred to as either ``smoothness'' or ``clumpiness'' was originally defined in \cite{Conselice2003} as part of the $CAS$ system to quantify how uneven the light distribution is. 

It is measured by convolving the image with a smoothing kernel, and then subtracting the smoothed profile from the original:

\begin{equation}
    S = \frac{\sum \max(0, I_{ij} - I^S_{ij})}{\sum I_{ij}} - \langle S_{bg} \rangle,
\end{equation}

\noindent where $I^S_{ij}$ is the smoothed image, and $\langle S_{bg} \rangle$ is the average clumpiness of the background. Since the nucleus of the galaxy is bright, any smoothing kernel would lead to a significant difference between smoothed and unsmoothed images in the centre, so the centre is excised from the calculation. \texttt{statmorph} therefore calculates $S$ in an aperture between $0.25 R_p$ to $1.5 R_P$, using a uniform kernel of size $0.25 R_p$. Pixels where the smoothed flux is \textit{larger} than the original, so the residual is negative, are set to zero and thus ignored. 

A high value of $S$ typically corresponds to a galaxy with many bright star-forming regions, which disappear when smoothed over, giving rise to a large residual. However other small-scale structures such as bars and spiral arms will also contribute to the residual, so $S$ measures the amount of substructure rather than simply clumps. Sometimes $A$ and $S$ are used in tandem to detect true mergers: \cite{Conselice2003} suggests selecting $A>0.3$ and $A>S$ to eliminate galaxies that are asymmetric due to their distribution of star-forming clumps. However, late-stage mergers such as those with embedded multiple nuclei will also have large values of $S$, since only one of the nuclei is excised, so this threshold should be used with caution. 

The first caveat of this measurement is that for small galaxies, with $R_p \approx 4$ pixels, the filter radius is 1 pixel or less, and so the image is not smoothed, leading to $S=0$. This becomes important in large catalogs that include either low-mass or high-redshift sources with a small angular size. We recommend only using $S$ measurements when $R_p > 12$ pixels, where the filtering step is reliable\footnote{One could also set \texttt{petro\_fraction\_cas} to a lower value, but this would affect the size of the excised region, which should ideally be well beyond the nucleus of the galaxy.}.

The second caveat is that the background contribution to $S$ can be large since the negative residual pixels are set to zero. Consider a Gaussian noise distribution with $I_{bg} \sim \mathcal{N} (0, \sigma)$. This noise is then passed through a uniform filter with area $A$ to produce $I^S_{bg}$. Accounting for the covariance between the two, the residual then follows a distribution
\begin{equation}
    I_{bg} - I^S_{bg} \sim \mathcal{N} \left(0, \sigma \sqrt{1-\frac{1}{A}}\right).
\end{equation}

\noindent Half of the residual pixels are set to zero and do not contribute to the expectation of the residual. The remaining ones follow a half-normal distribution, so the expectation of the residual becomes

\begin{equation}
    \langle S_{\rm{sky}} \rangle = \frac{\sigma}{2} \sqrt{\frac{2}{\pi} \left(1-\frac{1}{A}\right)}.
\end{equation}

While this background term is separable and can be subtracted, unlike in the CAS asymmetry case (Sec. \ref{sec:asymmetry}), the magnitude of the background term in practice is comparable to the magnitude of the residuals in the signal. Therefore, the random variance about the background median often dominates the real flux in the residual, and the measurement is noisy and background-dominated. In large surveys without exquisite depth, the background contribution is so significant that the smoothness values tend to vary randomly about 0 without a strong correlation to the underlying morphology of the source \citep[e.g.,][]{statmorph}. We show the behavior of $S$ with respect to the resolution and the signal-to-noise \new{ratio} in the left panel of Fig. \ref{fig:smoothness} -- there is a clear trend where $S$ decreases, on average, in lower signal-to-noise regimes. But more importantly, the scatter increases as \avgsnr{} drops and becomes more dominant than the systematic bias. This is due to the fact that the background subtraction noise dominates the measurement.

\begin{figure*}
    \centering
    \includegraphics[width=\linewidth]{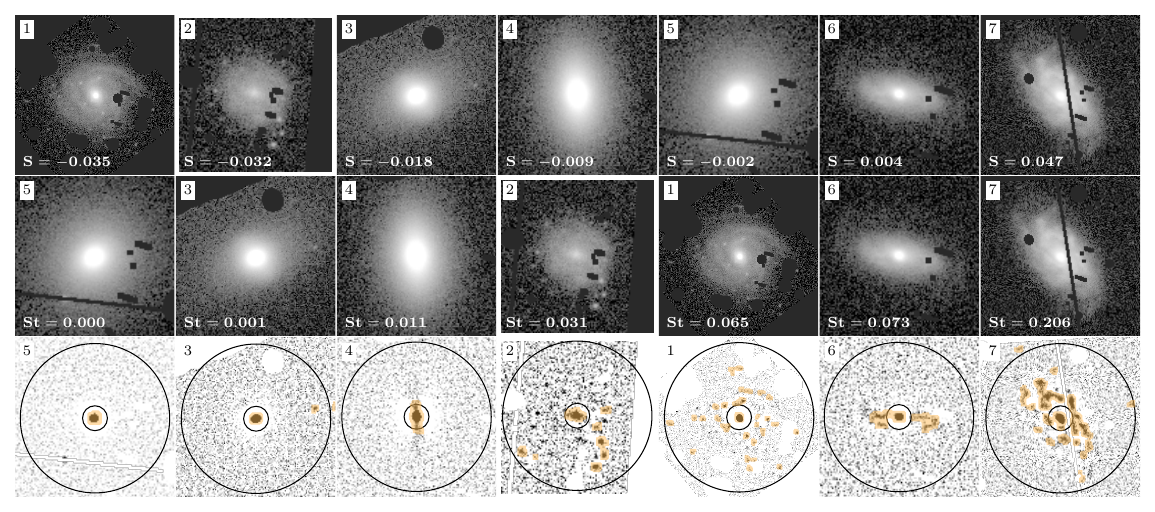}
    \caption{Smoothness ($S$, \textbf{top}) and substructure ($St$, \textbf{middle}) measurements for 10 galaxies along with the unsharp residuals (\textbf{bottom}). All pixels in the residual are used to calculate $S$, and only contiguous regions (orange) are used in $St$. Galaxies are sorted by increasing $S$ in the top row an $St$ in the bottom two rows. $St$ correlates better with a visual Hubble sequence while $S$ varies a lot due to noise.}
    \label{fig:substructure}
\end{figure*}

\subsubsection{Smoothness alternative: Substructure}\label{sec:substructure}

To address this issue, we propose a new metric, which we call \textit{Substructure} ($St$). The motivation behind this metric is the same as for the original smoothness. By applying a smoothing kernel to an image and looking at the residual, we are essentially performing an unsharp mask highlighting small-scale features -- substructures -- in the residual. Therefore, we repeat this process exactly like we did for smoothness, producing the same residual image.

The main difference is the treatment of the residual. Galaxy structures, such as clumps, spiral arms, bars, and others, are typically extended regions of several connected pixels. On the other hand, while the background also has spikes in the residual, they are disconnected as they arise from random noise. Therefore, our goal is to detect coherent residual structures that correspond to true galaxy features. We run \texttt{detect\_sources} to identify any connected regions above $1\sigma$ of the residual noise. We use connectivity of 4 to exclude spurious connections in the noise, and choose a minimum area of 10 pixels as it is statistically unlikely to get 10 nearby pixels all above the noise threshold. We then compute the total flux contained in these detected clumps, normalized by the total flux in an aperture. Note that we use the image rather than the residual to calculate fluxes, unlike in the smoothness calculation. The background correction is similar: we compute the total flux in detected clumps in the background (if any), normalize by the total flux in the sky aperture, and then rescale the mean background substructure to the source aperture and subtract it. In most cases, due to our stringent detection criteria, the background substructure is 0. 

Fig. \ref{fig:substructure} shows an example of 8 galaxies in our sample, all imaged at 100 pc/px with $\mu_0 \sim 22.5$ mag/arcsec$^2$. The galaxies are sorted by increasing smoothness in the top row, and by increasing substructure in the bottom two rows. The bottom row shows the residual used in calculating both quantities, and orange regions show detected ``clumps'' that contribute to the substructure measurement. The inner aperture is excised. First, it is clear in the images that the coherent regions detected have a range of morphologies: some are star-forming clumps, but some are spiral arms or the central bar which are also picked up by the unsharp mask -- hence we opted to call the metric ``substructure'' rather than ``clumpiness'' to reflect the range of structures that contribute to it. Second, the substructure metric maps better to the diversity of galaxy types. When sorting galaxies by substructure (the middle row of Fig. \ref{fig:substructure}), we see a clear Hubble sequence, starting from two early-type galaxies, then a barred early-type, and then a series of disks with clumps and spiral arms. On the other hand, smoothness is dominated by noise variability for all but the most clumpy objects and thus the sorting is less meaningful.

\new{Our method of source detection on a filtered image} is the same approach as what is used to detect clumps at high redshifts \citep[e.g.,][]{Shibuya2016} \new{and so our metric can be a useful diagnostic for studies of clumpy galaxies in the early Universe. At high redshifts, we expect it to work well to detect large clumps or satellites, although distinguishing between the two is challenging, and perhaps may be done with colour information. However, at high spatial resolutions, as seen in Fig. 11, the substructure metric also picks up other small-scale features\footnote{\new{``Small-scale'' here refers to two-dimensional features where one of scales is small, such as in the case of spiral arms or bars.}} -- spiral arms and bars -- and so the metric evaluates the flux fraction in \textit{all small-scale features} rather than clumps alone. One potential solution to distinguish between these two modes is to also measure the axis ratio distribution or the spatial distribution of these clumps, but this is beyond the scope of the current work. }

The right panel of Fig. \ref{fig:smoothness} shows the dependence of substructure on depth and resolution. Similarly to smoothness, the parameter is very depth-dependent. This is to be expected: the calculation involves source detection above a noise level, and so the more noise there is, the fewer dim sources are detected, decreasing the overall flux contained in substructures. Similarly, as resolution decreases, substructures are more challenging to resolve. This is a natural consequence of what a metric aims to measure -- however, for consistency across inhomogeneous surveys, one could adapt it to look for clumps that are brighter than a given limit, or train deep learning methods to ensure a homogeneous detection of clumps.

The substructure metric naturally traces bright star-forming clumps \citep[e.g.,][]{Shibuya2016}, as well as spiral arms, bars, and other sharp, high-contrast features. This approach has many potential applications from detecting clumps and globular clusters to characterizing their flux ratios, colours, and relative distribution. This metric can be especially powerful when combined with integral field spectroscopy, which would trace the differences in ionization mechanisms, stellar populations, and metallicity within these clumps. In a future work, we plan to use the substructure metric in tandem with the visual classifications of galaxies observed with the MaNGA survey \citep{VazquezMata2025} to go beyond the Hubble sequence presented in Fig. \ref{fig:substructure}.

\subsubsection{Multimode}\label{sec:multimode}

The multimode ($M$), intensity ($I$), and deviation ($D$) -- together, MID -- statistics were developed by \cite{Freeman2013} as an alternative way to probe disturbed and clumpy flux distributions. \texttt{statmorph} uses a slightly different definition of the statistics from \cite{Peth2016}, which we adopt here as well.

First, a new segmentation map is generated to calculate the MID parameters. This map is defined similarly to the Petrosian radius: a surface brightness threshold $q_{\rm{SB}}$ for source detection is chosen such that the average surface brightness in the detected region is $1/\eta \times q_{\rm{SB}}$. As for $R_p$, the segmentation process is repeated iteratively until the optimal $q_{SB}$ is found such that $\eta = 0.2$. 

Multimode finds two brightest peaks inside the segmentation map and compares their relative areas. First, we define some intensity quantile $q_x$. Then, construct a binary map -- similar to the shape asymmetry map -- where all pixels brighter than $q_x$ are set to 1 and the rest to 0. If $q_x$ is low (e.g., $x=0\%$), this selects all pixels inside the segmentation map and creates a single continuous region. If $q_x$ is high (e.g., $x=99\%$), a small region in the nucleus is created. For a simple analytic S\'ersic profile with a single peak, regardless of $q_x$, there will only be one region. However, for multimodal sources such as mergers or even spiral galaxies, there exists a value of $q_x$ where the map contains several disconnected regions. We then sort these regions by their areas and calculate the ratio between the areas of the largest and second largest regions:

\begin{equation*}
    R(x) = \frac{A_2 (x)}{A_1(x)} \times A_2 (x).
\end{equation*}

If there is only one region on the binary map, $R(x) = 0$. This ratio is calculated iteratively until an optimal $\hat{x}$ is found such that $R(\hat{x})$ is maximized. Multimode is then defined as\footnote{Note that this is the definition from \cite{Peth2016} where $M$ is a dimensionless quantity. In \cite{Freeman2013}, $M \equiv R(\hat{x})$.}

\begin{equation*}
    M \equiv A_2 (\hat{x}) / A_1 (\hat{x}).
\end{equation*}

This is a difficult optimization problem since it involves two steps: finding the optimal segmentation map and then $\hat{x}$. $R(x)$ can have many local maxima, so \texttt{statmorph} implements a brute-force method, increasing $x$ by 2\%, followed by a basin-hopping optimization step. 

For a galaxy with two equally bright nuclei, $M = 1$; and an early-type galaxy will have $M \approx 0$ -- not necessarily 0 due to possible noise spikes. 

\begin{figure}
    \centering
    \includegraphics[width=\linewidth]{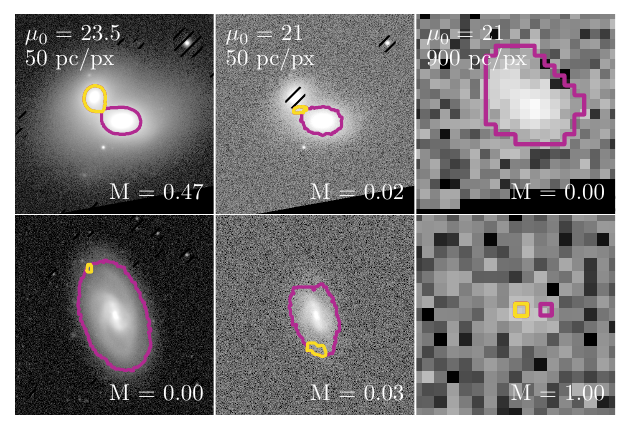}
    \caption{Measurements of multimode \citep[$M$;][]{Freeman2013,Peth2016} for a merger (\textbf{top}) and a spiral galaxy (\textbf{bottom)} in three regimes: deep and high-resolution (\textbf{left}), shallow low-resolution (\textbf{middle}), and shallow \new{high}-resolution (\textbf{right}). In the best imaging, $M$ distinguishes the merger from the normal galaxy. In shallow high-resolution imaging, the second nucleus of the merger is deblended, leading to \new{$M\approx0$}. In low-\avgsnr{} imaging, a noise spike in the spiral galaxy leads to a spurious $M=1$.}
    \label{fig:multimode}
\end{figure}

In practice, for the majority of images in our test series, $M$ is low ($M < 0.1$ for 77\% of the sample and $M = 0$ for 20\%). While in theory $M$ could pick up star-forming clumps, they are much smaller and fainter than the central region of the galaxy, leading to low $M$. Since low-$M$ images dominate our dataset, we do not provide a resolution-depth grid for $M$ here, but it is available in the online materials. Instead, we discuss the general behaviour of the metric.

Fig. \ref{fig:multimode} shows an example multimode calculation for two galaxies: a merger with two nuclei in the top row and a spiral galaxy in the bottom, for a range of resolution and surface brightness limits. The two regions $A_2$ and $A_1$ for optimized $\hat{x}$ are in purple and yellow contours, and the masked foreground sources are hatched. We show the calculation in three regimes: deep and high-resolution, shallow high-resolution, and shallow poor resolution. In the first regime, the metric behaves as expected, with the merger having high $M$ and the spiral galaxy essentially 0. The middle and third columns show edge cases: where $M$ drops to 0 for a clearly disturbed galaxy, and where $M$ increases to 1 for an undetected source.

In general, in sources with double nuclei where $M$ is high, the measurement is robust to changes in the signal-to-noise \new{ratio} and resolution up to a certain limit. This limit depends on the intrinsic properties of a galaxy. For example, in the top left panel in Fig. \ref{fig:multimode}, the two nuclei are enveloped in a single halo, leading to a high $M$. As the signal-to-noise \new{ratio} decreases, eventually the halo flux becomes lower than the detection threshold used in the main segmentation algorithm, and the two nuclei are deblended; so $M$ drops from 0.47 to 0. In our example, this happens at $\mu_0 = 21$ mag/arcsec$^2$, but the exact threshold depends on the flux ratio between the nuclei and the halo, and so will change from one galaxy to another. Similarly, as resolution decreases, the two peaks become more difficult to resolve, and eventually they blend into a single peak with $M \approx 0$. For the galaxy in Fig. \ref{fig:multimode} this happens at $\mathcal{R} \sim 900$ pc/px. However, the exact resolution where this happens depends on the surface brightness of the galaxy, the flux ratios of the two peaks, and their physical separation, and so will also vary for different mergers. Until these thresholds in \avgsnr{} and \reseff{} are reached, the measurement is quite robust to changes in image properties -- but the problem is that these thresholds are impossible to define for all galaxies simultaneously. One way to reframe the multimode metric is that it measures a particular combination of merger mass ratios and times where the two nuclei are bright, separable, but not deblendable.

Finally, for extremely noisy images, where $\avgsnrm < 1$, the optimizer tends to find a high value of $\hat{x}$ which leaves several single-pixel noise spikes in the binary map, in addition to a (possible) real peak in the nucleus. Then the optimizer returns $M = 1/A_1$. In the most noise-dominated cases, \textit{all} detected regions, including $A_1$, are simply due to noise, which typically gives $M = 1, 0.5$, or $1/3$. These fractional $M$ values should be treated carefully, especially if \avgsnr{} is low. An example of this is shown in the bottom right panel of Fig. \ref{fig:multimode}, where only two pixels are identified in the binary map, resulting in $M=1$. One improvement to the algorithm could be setting a minimum area for the contiguous regions in the binary map, making it more robust to spurious noise detections; however, this is only an issue in the extremely low signal-to-noise regime.

\begin{figure}
    \centering
    \includegraphics[width=\linewidth]{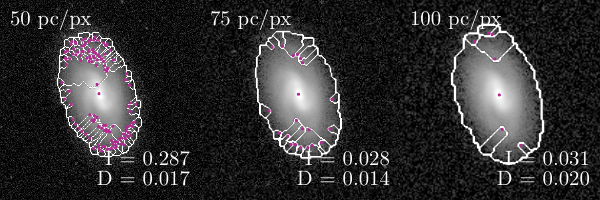}
    \caption{The intensity and deviation \citep[$I$ and $D$;][]{Freeman2013} watershed map for an example spiral galaxy at different resolutions: 50, 75, and 100 pc/px on the left, centre, and right, respectively. While the changes in resolution are relatively small and visually the same features are present in the galaxy, the watershed maps are markedly different.  $D$ is stable across all three images despite the differences in the watershed maps.}
    \label{fig:intensity}
\end{figure}

\subsubsection{Intensity}\label{sec:intensity}

Intensity ($I$) is similar to multimode in that it measures the relative strengths of two peaks, but it is weighted by flux rather than area of the two regions. First, an image is smoothed by a Gaussian kernel (with $\sigma = 1$ px in \texttt{statmorph}). Then we find local maxima in the image (within the MID segmentation map) and create a watershed map for each local peak. In practice, there are many ways to identify local maxima, and the original algorithms described in \cite{Freeman2013} and \cite{Peth2016} are not explicit in how these maxima are defined, so $I$ measurements produced by different codes likely differ. \texttt{statmorph} uses the \textsc{Scikit-Image} library to find local maxima with \texttt{peak\_local\_max} and create the watershed map with \texttt{watershed}. Fluxes in each segment are then summed. Then intensity statistic is defined as 

\begin{equation*}
    I \equiv I_2 / I_1,
\end{equation*}
\begin{figure}
    \centering
    \includegraphics[width=\linewidth]{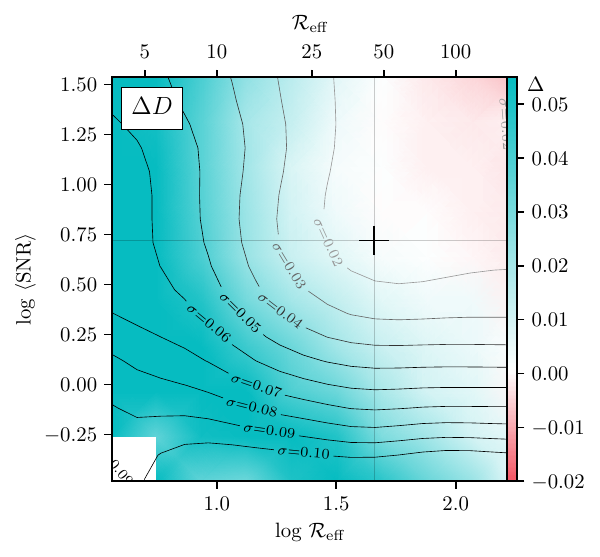}
    \caption{same as Fig. \ref{fig:moments_grid}, for deviation \citep[$D$;][]{Freeman2013}. There is a small bias in $D$ at \avgsnr{}$<$2 and \reseff{}$<$10, but the bias is lower than the intrinsic scatter or the typical values expected of disturbed sources ($D>0.1$), making $D$ a reliable metric overall.}
    \label{fig:deviation}
\end{figure}
where $I_1$ and $I_2$ are the fluxes of the brightest and the second brightest segments, respectively.

In practice, this metric is extremely sensitive to resolution and its output does not necessarily have the intended meaning, for example for late-type galaxies. A large fraction of images in our sample ($>50\%$) have intensity of 0, so we do not provide a resolution-depth grid, although it is available online. Fig. \ref{fig:intensity} shows an example galaxy imaged at 50, 75, and 100 pc/px, with the intensity peaks and watershed contours outlined. At these resolution levels, visually, the same features in the galaxy are present. However in the 50 pc/px case, the watershed algorithm identifies secondary peaks in the central region, while in the other cases it does not, leading to different intensity values despite a relatively small change in resolution. The peaks identified by the algorithm also do not correspond to any visually obvious peaks. The number of tertiary peaks also depends strongly on the resolution of each image, although these do not contribute to the overall intensity calculation.

The crux of the issue here is that \textsc{Scikit-Image} simply applies a 3$\times$3 maximum filter to the image and labels any pixel where the input and the max-filtered arrays are equal as a peak. This means, for example, that in a noise-dominated array every 3$\times$3 box will contain a peak. While this issue is partly mitigated by the MID segmentation map, the algorithm still identifies many spurious peaks in the low signal-to-noise regions, and the robustness of peaks found in the centre is hard to determine. \textsc{Scikit-Learn} provides additional parameters that can enforce a minimum flux ratio and/or distance between adjacent peaks. Setting these parameters to physically meaningful quantities could help alleviate some of the stochasticity of the peak finder, but this investigation is beyond the scope of our paper, and they are not currently used in \texttt{statmorph}.

Another, more subtle issue is that watershed segmentation is not the best choice for calculating the total fluxes associated with each component. As seen in Fig. \ref{fig:intensity}, the watershed algorithm essentially draws a boundary between two peaks that is equidistant from both of them. Even a small clump -- not visible by eye in our example -- sufficiently near the centre of the galaxy forces the watershed to split the source into two halves, leading to a high $I \approx 0.4$. However, for a more physical meaning, most of the flux should be attributed to the main galaxy halo, not the clump. This requires using more sophisticated decomposition methods, such as doing it in frequency space \citep[e.g.,][]{Sattari2023,Kalita2025b}, using the smoothness/substructure (Sec. \ref{sec:substructure}), or modelling the clumps alongside the main S\'ersic profile \citep[e.g.,][]{Kalita2025a}. With the current approach, $I$ cannot distinguish between a real bimodality in the light distribution -- e.g., two nuclei -- and a small clump causing a bifurcation in the watershed map. Adding in the information from other metrics such as multimode can help break this degeneracy.

\subsubsection{Deviation}\label{sec:deviation}

Deviation ($D$) is the final statistic of the trio, derived directly from the intensity measurement. It is simply the distance between the image centroid and the brightest peak $I_1$ found earlier, normalized by the ``radius'' of the MID segmentaion map of area $N$ defined as $\sqrt{\pi/N}$. The peak $I_1$ typically corresponds to the brightest pixel (nucleus) of a galaxy, while the centroid may be offset due to asymmetries in the light distribution, so $D$ is essentially a measure of asymmetry, similar to $S(G,M_{20})$.

While the intensity watershed map may be unstable, the brightest peak is always identified well in it, so the position of $I_1$ is reliable. The centroid is also a robust metric (see Sec. \ref{sec:centre}), so $D$ is robust to changes in the signal-to-noise \new{ratio} and resolution as shown in Fig. \ref{fig:deviation}. There is a small positive bias in $D$ measured on low-resolution or shallow images, however, this bias is comparable to intrinsic scatter, and is lower than typical $D$ values expected for disturbed objects ($D > 0.1$).  The slight increase in $D$ is due to poorer deblending of foreground sources at low resolutions. For true double nuclei or mergers, $D$ actually decreases at low resolutions where the two brightness peaks blend in and are no longer identified as separate intensity peaks.

\section{Discussion \& Implications} \label{sec:discussion}

\begin{figure*}[ht]
\centering
\begin{minipage}[t]{0.25\linewidth}
    \vspace{0pt}
    \centering
    \begin{tabular}{lll}
        \toprule
        \midrule
        \multicolumn{3}{l}{Geometric} \\
        \midrule
        $\mathbf{x}_0$ & Centroid & \ref{sec:centre} \\
        $e$ & Ellipticity & \ref{sec:moments} \\
        $e_{\rm{Sersic}}$ & S\'ersic $e$ & \ref{sec:sersic} \\
        $\theta$ & Orientation & \ref{sec:moments} \\
        $\theta_{\rm{Sersic}}$ & S\'ersic $\theta$ & \ref{sec:sersic} \\

        $R_p$ & Petrosian $R$ & \ref{sec:rpet} \\
        $R_{20,50,80}$ & Isophotal $R$ & \ref{sec:r20} \\
        $R_{0.5}^{\rm{Sersic}}$ & S\'ersic $R$ & \ref{sec:sersic} \\
        \midrule
        \multicolumn{3}{l}{Bulge strength} \\
        \midrule
        $n$ & S\'ersic index & \ref{sec:sersic} \\
        C & Concentration & \ref{sec:concentration} \\
        G & Gini index & \ref{sec:gini} \\
        M$_{20}$ & M$_{20}$ & \ref{sec:m20} \\
        B(G,M$_{20}$) & Bulge strength & \ref{sec:bgm20} \\
        \midrule
        \multicolumn{3}{l}{Disturbance} \\
        \midrule
        S(G,M$_{20}$) & Disturbance & \ref{sec:sgm20} \\
        A & Asymmetry & \ref{sec:asymmetry} \\
        A$_O$ & Outer $A$ & \ref{sec:asymmetry} \\
        $A_S$ & Shape $A$ & \ref{sec:ashape} \\
        $A_{\rm{RMS}}$ & RMS $A$ & \ref{sec:arms} \\
        S & Smoothness & \ref{sec:smoothness} \\
        M & Multimode & \ref{sec:multimode} \\
        I & Intensity & \ref{sec:intensity} \\
        D & Deviation & \ref{sec:deviation} \\
        \midrule
        \multicolumn{3}{l}{New in this work} \\
        \midrule
        $A_{X}$ & Isophotal $A$ & \ref{sec:aiso} \\
        St & Substructure & \ref{sec:substructure} \\
        \bottomrule
    \end{tabular}
\end{minipage}
\begin{minipage}[t]{0.72\linewidth}
    \vspace{0.5cm}
    \centering
    \includegraphics[width=\linewidth]{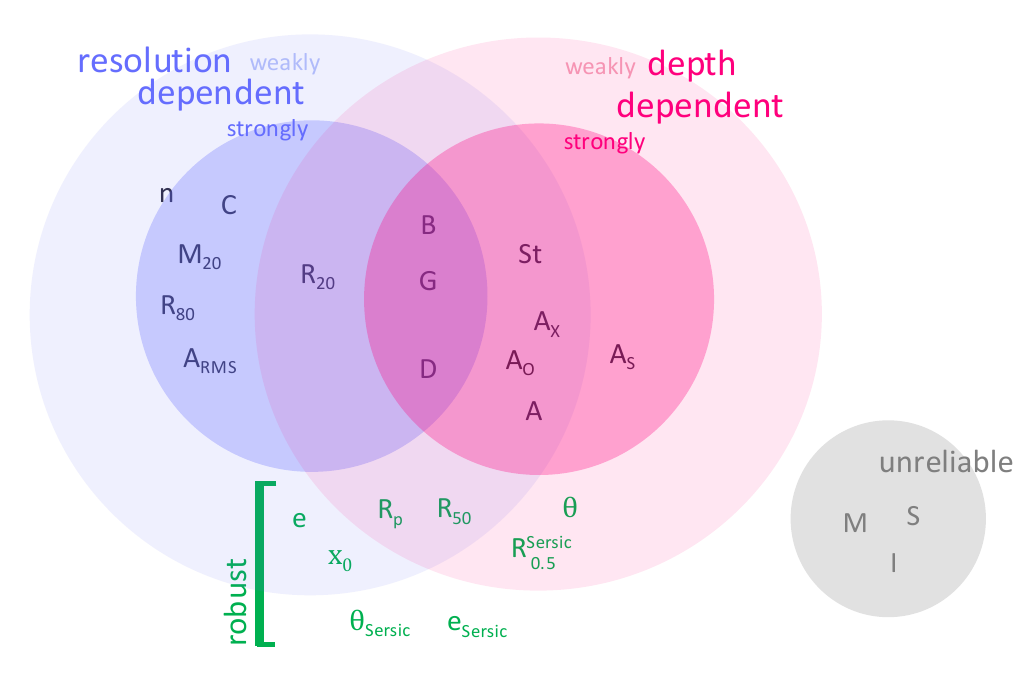}
\end{minipage}
    \caption{A summary of the parameters studies in this \and   work, roughly broken down into four regions: resolution-dependent, depth-dependent, robust, and unreliable. The table on the left provides references to the sections where each parameter is discussed.}
    \label{fig:summary}
\end{figure*}
In this work, we analyzed all morphological metrics measured by the \texttt{statmorph} package in terms of their dependence on resolution and depth. In Fig. \ref{fig:summary}, we offer a concise review of our findings: some metrics depend on resolution, some on the signal-to-noise \new{ratio}, some on both, and some metrics are robust to changes in either. It is important to consider which metrics you are using based on your science needs, which we roughly split into three broad areas.

\textbf{Large surveys:} large surveys, such as the SDSS, LSST, Roman, UNIONS, or Euclid typically have a homogeneous resolution and depth, so morphological catalogs should be consistent across the survey area \new{within a data release}. However, \new{in surveys such as LSST, the images are stacked with each new release and so the the effective depth will change over time, so} some parameters -- such as $A$, $A_S$ -- will not be consistent between different data releases. \new{Because of this, one should not expect the same morphological measurements of the same galaxies in LSST DR1 and DR10.}

\textbf{Individual / archived programs:} archival observations, obtained with different science goals, by construction have  inhomogeneous exposure times and therefore image depth. So parameters such as $A$ and $S$ would not be consistent across observations. On the other hand, in archives of space-based imaging where the resolution is consistent, metrics such as $C$, $G$, $M_{20}$, and $A_{\rm{RMS}}$ will be robust from one image to another.

\begin{figure*}
    \centering
    \includegraphics[width=\linewidth]{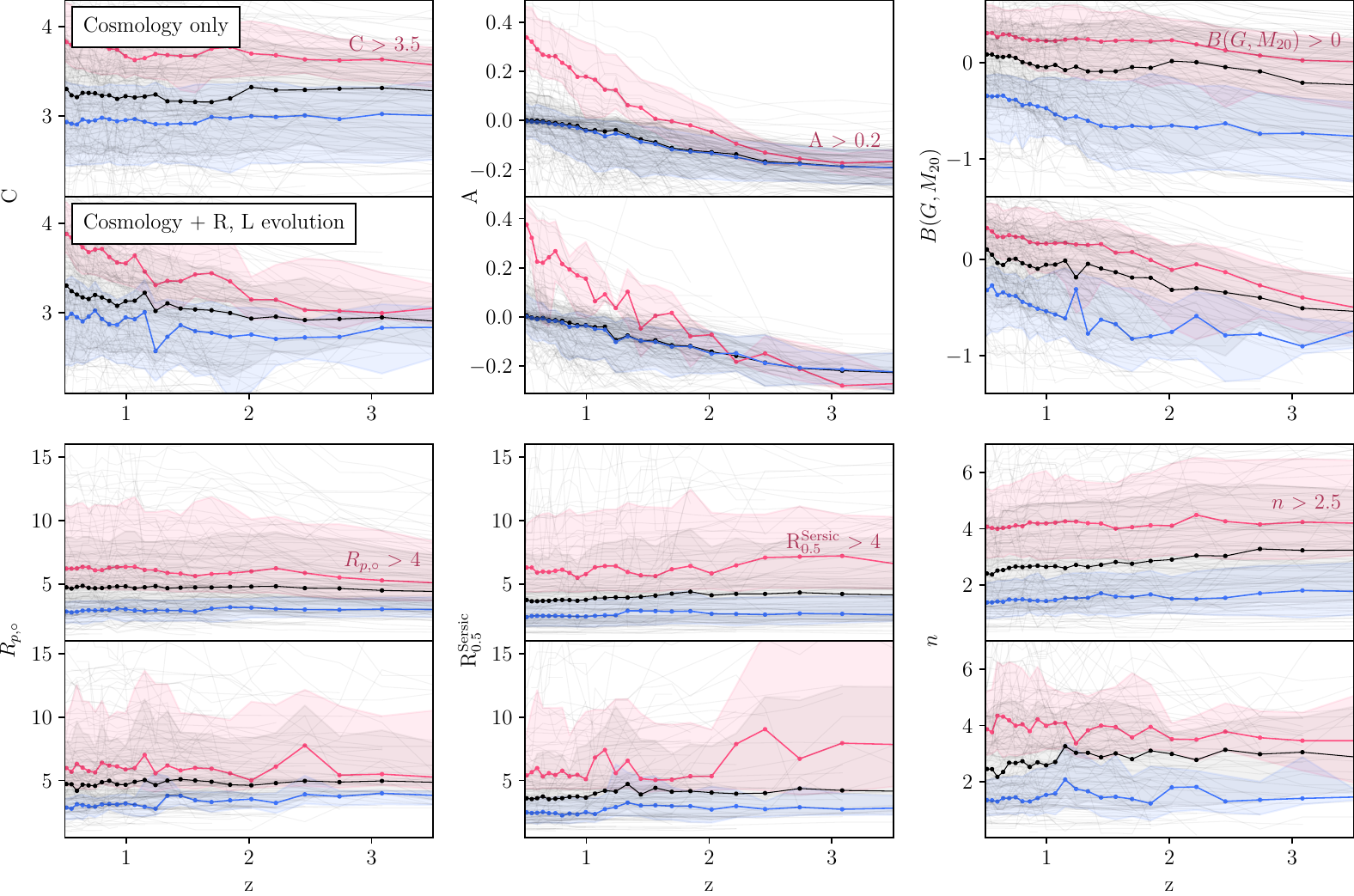}
    \caption{The change in six common morphological metrics as a result of moving the same galaxies to a higher redshift (\textbf{top}), and also accounting for a size and luminosity evolution of galaxies with redshift (\textbf{bottom}). The black, pink, and blue lines show the median of the entire sample, and high-$p$ and low-$p$ subsamples for each parameter $p$, respectively. $C$, $A$ and $B(G, M_{20}$) parameters suffer a systematic bias that needs to be corrected for when looking at structural evolution across redshifts. $R_p$, $R_{0.5}^{Sersic}$, and $n$ are largely unbiased but the uncertainty increases as the effective resolution and depth degrade. We assume a size evolution $R \sim R_0 (1+z)^{-0.71}$ \citep{Ormerod2024} and luminosity evolution $L \sim L_0 (1+z)$ \citep{Yu2023}.}
    \label{fig:evol}
\end{figure*}

\subsection{Evolution of bias from $0.5 < z < 3.5$}

In both of these modes, the main difficulty arises when studying objects at different cosmic times -- whether found in the same survey footprint or in archival observations. Even with a consistent angular scale, the \textit{effective} resolution in terms of galactic radii changes as objects further away appear smaller. Similarly, the same surface brightness limit probes different physical surface brightnesses in $z=0$ and $z=2$ objects due to cosmological dimming. For lower-redshift objects ($z\lessapprox 0.2$), the surface brightness dimming is negligible while the resolution change is significant, so parameters such as $C$, $G$, $M_{20}$ will be affected more. For high-redshift objects ($z>1$) the inverse is true as the angular diameter dependence on distance flattens. However, if we account for changes in intrinsic luminosity and radii of galaxies at a fixed mass as was done in \cite{Whitney2021} and \cite{Yu2023}, due to the rapid size evolution, the effective resolution \reseff{} does change as a function of redshift even at $z>1$ despite the almost constant angular scale. 

In Fig. \ref{fig:evol}, we show the observed changes in six commonly used morphological parameters: $C$, $A$, $B(G, M_{20})$, $R_p$, $R_{0.5}^{\mathrm{Sersic}}$, and $n$ due to biases caused by the losses in depth and resolution if the same source is redshifted from $z=0.5$ to $z=3$. We do not consider changes in wavelength for this toy example as we assume the filter is changed appropriately to probe the same rest-frame wavelength. Therefore, our analysis in Fig. \ref{fig:evol} aims to show the differences in morphological measurements arising \textit{purely from biases}. The evolution of intrinsic properties -- stellar structure, dust content, star forming regions, and others -- would be seen as trends on top of the ones caused by biases.

For each parameter, we show the changes due to cosmological dimming and angular size change in the top panel. In the bottom panel, we also include an intrinsic evolution in size and luminosity: $R \sim R_0 (1+z)^{-0.71}$ \citep{Ormerod2024} and $L \sim L_0 (1+z)$ \citep{Yu2023}. As galaxies of a given mass are intrinsically smaller at higher redshifts, the physical resolution \res{} in parsecs per pixel does not change much, but the effective resolution in \reseff{} does. For each galaxy in our sample, we assume that the baseline measurement is taken with JWST NIRCam at $z=0.5$ with resolution 200 pc/px and depth $\mu_0 = 23.5$ mag/arcsec$^2$. Then, for a given redshift, we calculated the change in resolution (\res{}, top or \reseff{}, bottom) and depth ($\mu_0$, top or \avgsnr{}, bottom) arising from cosmological effects (top and bottom) and intrinsic size/luminosity evolution (bottom only). For each galaxy and redshift, we then found a run from our suite \new{with a closest resolution and depth}. Fig. \ref{fig:evol} shows how the median and $\pm1\sigma$ quantiles of the entire sample change with redshift in black. In red/blue we show the median and $\pm1\sigma$ quantiles for galaxies with high or low baseline values of a given parameter, respectively, using the following thresholds: $C=3.5, A=0.2, B=0, R_p=4, R_{0.5}=4, n=2.5$.

As seen in Fig. \ref{fig:evol}, when size evolution is accounted for, there is significant evolution in $C$, $A$, and $B(G, M_{20})$ that arises due to observational biases rather than a real evolution in these parameters. $C$ and $B$ will decrease by $1$ each from $z=0.5$ to $z=3.5$, while for late-type $C$ remains constant on average and $B$ decreases by 0.5. $A$ decreases by 0.5 on average for disturbed galaxies and $0.1$ for undisturbed ones. Galaxies at higher redshifts appear less bulge-dominated and less disturbed than they would at lower redshift due to insufficient resolution and depth. On the other hand, both radii measurements and $n$ are fairly robust across this redshift range, although there is large scatter at high $z$, which reflects the increasing uncertainty in these measurements. 

These biases might explain some of the puzzling results from morphological analyses in the past. On one hand, we expect that major mergers are more frequent at earlier times from the hierarchical structure formation \citep[e.g.,][]{RodriguezGomez2015}. On another, repeated analyses of HST and JWST data found that asymmetry is, on average, low and constant with redshift \citep{Ferreira2023,Yao2023,Ren2024,Costantin2025}, while \cite{Mortlock2013} and \cite{Ferreira2022b,Ferreira2023} find that up to 40\% of galaxies are visually peculiar. Considering the strong bias in asymmetry we see in Fig. \ref{fig:evol}, we expect that the \textit{true} asymmetry increases significantly with redshift, but this signal is washed away in observational effects. Accounting for the loss in image quality for distant galaxies, Sazonova et al. (in prep.) find a strong increase in non-parametric disturbance between z$=$0.5 and z$=$3.

Similarly, several studies find that concentration and bulge strength are lower at high $z$ \citep{Yao2023,Ren2024,Costantin2025}, while the S\'ersic index does not evolve significantly \citep{Ferreira2023,Sun2024,Martorano2025}. Concentration and S\'ersic indices are correlated by construction \citep{Trujillo2001}, so this result appears puzzling at first -- but Fig. \ref{fig:evol} shows that while $n$ is not affected by imaging biases, $C$ is. The decrease in $C$ by $\sim$1 over the 0.5$<$z$<$3 redshift range for early-type galaxies is completely consistent with the findings of \cite{Ren2024}, who showed that $C$ decreases by 1 among the quiescent massive galaxies, but not star-forming, and \cite{Law2012}, who found that $C$ is constant with $z$ in star-forming galaxies. Finally, \cite{Lotz2008} used $B(G,M_{20})$ measured with HST data at 0.2$<$$z$$<$1.2 to split galaxies into early- and late-type, and found that the fraction of bulge-dominated (E, S0, Sa) galaxies increases from 21\% at z$=$1.2 to 44\% at z$=$0.3. However, this redshift range covers a significant change in the effective resolution, which in our tests leads to $\Delta B$$\sim$1. Accounting for this, the evolution of the E/S0/Sa fraction could be much weaker, which agrees with the recent results from \cite{Kartaltepe2023} that 30\% of visually classified galaxies at z$<$5 have a spheroid. Our results indicate that the apparent decline in bulge strength at $z$$>$3 may
largely arise from observational biases, rather than true structural evolution.

\vspace{0.5cm}

One way to study parameters such as CAS, G, and $M_{20}$ across cosmic time is to measure the observed evolution and subtract off the expected bias induced from changing image quality -- either obtained from redshifting sources \citep[e.g.,][]{Whitney2021} or by fitting analytical solutions in terms of \avgsnr{} and \res{} \citep[e.g.,][]{Pawlik2018}. We fitted for this dependence using symbolic regression, and provided correction functions for each parameter (Appendix \ref{app:pysr}). However, our correction assumes that the PSF is a 2 pixel-wide Moffat -- for different PSF sizes, it is best to recalibrate the correction terms. 

Another approach is to artificially degrade the imaging across the sample such that the physical scale and surface brightness limits are matched for each galaxy, regardless of the instrument taken or the redshift of the source. Sazonova et al. (in prep.) took this approach for galaxies in JWST JADES survey, and found a strong evolution between 0.5$<$z$<$3, where high-redshift galaxies have distinctly more disturbed structures. 

Finally, machine learning approaches can be used to remedy the imaging effects -- for example, applying meaningful augmentations that represent the loss of resolution and the signal-to-noise \new{ratio} in the training step can produce a network that is invariant to observational biases \citep[e.g.,][]{VegaFerrero2024}. We plan to use our augmentation dataset to train a deep learning network that is invariant to the changes in the signal-to-noise \new{ratio} and resolution considered in this work, and then train this model on data from different telescopes and observations to be invariant to realistic imaging artifacts.

\subsection{Multi-wavelength morphology}

Throughout this work we focused only on galaxy structure in the rest-frame \textit{I}-band, which traces intermediate stellar populations and dust. In our redshift analysis, we assumed that the observer is able to switch bands to match the rest-frame wavelength. If instead a survey is designed in a single filter, then higher-redshift sources will move further to the rest-frame UV, and the observations will trace younger stellar populations. When observed in the UV, even early-type galaxies can have late-type or clumpy morphology \citep{Windhorst2002,Mager2018}. \cite{Rutkowski2025} in particular found that 15\% of 0.5$<$$z$<$1.5$ quiescent galaxies still have residual star formation and so would appear to have structures of late-type galaxies when viewed in the UV, where the older stellar populations are not visible. It may be possible to correct for these effects with ``morphological k-corrections'' \citep{Windhorst2002}, but this would require \textit{assuming} some relationship between UV and optical structures, and so is not advisable for populations of high-redshift or irregular galaxies where this relationship is not known a priori.

A more common approach is to design high-redshift surveys using imaging from multiple bands, such that each band traces the same rest-frame filter in a given redshift range. This should also be done carefully, since one would need to account both for the change in the central wavelength and the bandwidth, so that redder (tracing higher-redshift) bands are wider than the bluer (tracing lower-redshift) ones. For example, the filters on the \textit{HST} and \textit{JWST} telescopes are designed with this in mind. Finally, when analyzing populations that include quiescent galaxies, it may be more important to not match the central wavelength precisely, but to ensure that the filter does not straddle the 4000\r{A} break since the contribution blueward of the break will be much weaker.

However, while we must be careful to compare like rest-frame images, in general it is valuable to consider the structure of galaxies in different wavelengths, from UV all the way to radio. Different wavelengths of light are produced different emission mechanisms and hence correspond to different galaxy components: UV, optical, and near-IR trace young, intermediate, and old stellar populations respectively. Dust attenuates light to varying extents, so a comparison between these wavelengths can also constrain dust properties. Mid- and far-IR highlight dust emission itself as well as emission from an AGN, if present. Finally, radio observations can trace atomic and molecular hydrogen gas content of the galaxy at different densities, and IFU observations can do so for ionized gas. Comparing how galaxy structure changes across all of these components by measuring morphology can be an extremely powerful tool to tie together various physical processes and their role in redistributing the gas, dust, and stellar content of a galaxy. With this approach we can start to constrain quenching processes more precisely: for example, a galaxy with asymmetries in the UV and IR is likely a post-merger, while disturbances that are only visible in the UV are likely caused by star formation. Mergers also induce changing signatures on the timescales of $\sim$1Gyr, which can be seen as changes on the CAS/G/$M_{20}$ plane \citep{Snyder2015b,TaylorMager2007}.

Given the results of our work, care must be especially taken when comparing structural parameters in different wavelength regimes, since multi-wavelength observations typically have varying resolution and signal-to-noise levels. While we did not explicitly run the tests for varying wavelengths, our results should be applicable to this use case as well, since the main difference between UV and IR imaging is still simply the depth and resolution. When biases arising from imaging quality are accounted for, any changes in the measured structural parameters can be used to detect changes in actual substructures. For example, in \cite{Sazonova2021} we used this approach to detect disturbances in dust substructure in post-starburst galaxies by comparing \textit{HST} \textit{B}, \textit{I}, and \textit{H}-band imaging. \new{There is a significant body of work in the relationship between S\'ersic parameters and wavelength, which allows comparing the spatial distributions of different stellar populations \citep[e.g.,][]{Casura2022,Baes2024,Quilley2025a,Quilley2025}. In particular, }\textsc{GalfitM} \new{and \textsc{SourceXtractor++}} incorporate different wavelength observations to study how the S\'ersic profile changes from one band to another, \new{while \textsc{Buddi} extends S\'ersic fitting to IFU data \citep{buddi}. Given our conclusions on how strongly all these parameters depend on image properties, however, the comparison of morphologies measured in different bands must be done by first carefully matching image quality.}

The final consideration is that extending this analysis to multiple wavelengths will greatly increase the dimensionality of the problem -- for 30 or more morphological parameters and 10 bands, each galaxy would have 300 structural measurements. This goes far beyond diagnostic diagrams traditionally used in morphological studies \citep[e.g.,][]{Conselice2003,Lotz2004}. These large datasets would be perfectly suited for machine learning and dimensionality reduction algorithm, where the entire parameter space is compressed to a few important features which are then used to classify galaxies or probe their physical processes. For example, \cite{Mukundan2024} and \cite{Holwerda2025} used k-nearest neighbours classification on the space of structural measurements to classify galaxies into different Hubble types and irregulars. \cite{AguilarArgueello2025} used XGBoost for the same task, and provided feature importances for each morphological metric they considered. In Sazonova et al. (in prep.), we use the Uniform Manifold Approximation \& Projection (UMAP) technique to reduce the dimensionality of the space of \texttt{statmorph} outputs to two important new variables, and study the evolution of galaxies in the UMAP phase space. These studies still focused on single-band imaging, but they offer a promising way towards tackling a larger-dimensional dataset of multiwavelength observations.

\subsection{Final remarks}


This work is intended to prepare for the morphological analysis of galaxies in the Rubin LSST. The changes to parameters such as smoothness/substructure, S\'ersic index, and shape asymmetry discussed here will be incorporated in a \texttt{statmorph-lsst} public package, which will be used to construct catalogs in the upcoming Rubin data releases. This package is publicly available on a GitHub repository\footnote{\href{https://github.com/astro-nova/statmorph-lsst}{\texttt{statmorph-lsst} GitHub repository}} and contributions from other researchers interested in this effort are welcome. Finally, we also make available the entire suite of 64,000 augmented images we used for these tests on Zenodo, and we developing a tool for researchers to test the performance of custom metrics and construct their own corrections. The tool will be linked to the \texttt{statmorph-lsst} repository.








\section{Summary} \label{sec:summary}

Using a sample of 189 local galaxies, we created a dataset of 64,000 images at varying physical resolutions and surface brightness limits. This dataset can enable a wide range of further analyses and machine learning applications, and it is available online for public use on Zenodo\footnote{\href{https://zenodo.org/records/17585609}{Dataset DOI: 10.5281/zenodo.17585608}}.

We then investigated how every morphological parameter measured by \texttt{statmorph} depends on the the signal-to-noise \new{ratio} and resolution of the galaxy. 
\new{The numerical values of the average bias and scatter in our sample, as a function of the signal-to-noise \new{ratio} and resolution, are available in Appendix \ref{app:pysr}. Our sample spans a range of galaxy morphologies, masses, and star formation rates, so if one needs a bias estimate for a particular galaxy population -- for example, late type galaxies -- it is best to re-derive it using the full catalog of morphological measurements we provided on Zenodo.} Our results are broadly summarized in Fig. \ref{fig:summary}: 

\begin{enumerate}
    \item Geometrical parameters -- ellipticity, orientation, Petrosian radius $R_p$, and $R_{0.5}^{\rm{Sersic}}$ --- are robust to changes in image properties and can be used consistently across different surveys and redshifts (Sec. \ref{sec:moments} and \ref{sec:radii}).
    \item Bulge strength parameters -- Gini, $M_{20}$, $C$ -- depend on the effective resolution \reseff{}, defined as the ratio of $R_p$ to pixel scale. These will be biased when resolution changes, either from the changing image quality, or from cosmological effects  (Sec. \ref{sec:bulge}). This means that high-resolution imaging, such as that from \textit{Euclid} or JWST, is required to accurately measure these parameters at z$>$1.
    
    \item Disturbance parameters -- $A$, $A_S$,  substructure $St$ -- depend on the signal-to-noise \new{ratio} and thus will be inconsistent across different surveys, or for objects at different redshifts due to the cosmological surface brightness dimming. $A_{\rm{RMS}}$, introduced in \cite{Sazonova2024}, does not depend on the signal-to-noise \new{ratio} but is sensitive to resolution. (Sec. \ref{sec:disturbance} and \ref{sec:other}).
    
    \item Smoothness and intensity are unreliable parameters: first due to the strong noise contribution, and second due to its watershed definition that bifurcates the galaxy flux. Multimode is robust to changes in \avgsnr{} and \reseff{} up to a point and detects double nuclei well, but the threshold where it becomes unreliable varies from one galaxy to another and so is difficult to generalize. Deviation, although defined using Intensity, is a robust metric for a wide range of \avgsnr{} and \reseff{} (Sec. \ref{sec:other})
    
    \item \textsc{Galfit} S\'ersic fitting results generally do not suffer from a strong bias. However, we quantify an intrinsic uncertainty in S\'ersic parameters caused by the degeneracies in the fit to be on the order of 20\% for $R_{0.5}^{\rm{Sersic}}$ and up to 40\% for $n$ (Appendix \ref{app:sersic}). The default \texttt{statmorph} implementation of the S\'ersic fit will also lead to an underestimate in $n$ compared to other libraries \new{-- this issue is addressed in \texttt{statmorph-lsst} with a subsampling approach similar to \textsc{Galfit}.}
\end{enumerate}

We applied our analysis to see how would the morphology of galaxies appear to evolve from $z=0.5$ to $z=3.5$ due to observational biases alone, and found that the observed decrease in the numbers of centrally concentrated galaxies can be explained by the bias in the concentration index. We also found that the true number of disturbed or asymmetric galaxies is likely even higher than what is currently reported.

We implemented several changes in the new \texttt{statmorph-lsst} package, which will be used in the analysis of Rubin LSST data:

\begin{enumerate}
    \item Introduced a new isophotal asymmetry, $A_X$, similar to $A_S$ from \cite{Pawlik2016}. This is an asymmetry of the $X$ mag/arcsec$^2$ isophote. Since it is defined as a function of surface brightness, it is not dependent noise as long as the isophote is above the surface brightness limit,  and has a mild resolution dependence when $X$ is low. One advantage of $A_X$ is that it probes physical features of a particular luminosity (and by proxy, mass) surface density. \new{A user can calculate $A_X$ by passing a list of isophotes, in image units, as the \texttt{asymmetry\_isophotes} argument. We recommend defining isophotes as surface brightness levels as they carry physical meaning about light density, and then converting these to image units.}
    
    \item Introduced a new substructure metric $St$, defined similarly to $S$, but removing the noise contribution by requiring the substructure clumps to be connected. This metric allows detecting high-contrast features such as clumps, spiral regions, and bars, and together with multi-wavelength data can be used to further dissect properties of star-forming regions -- their flux ratios, colours, metallicities, and more.
    
    \item Added the functionality to first degrade the image to the desired physical resolution and depth, and then compute the parameters to ensure consistency across different masses and redshifts.
\end{enumerate}

\new{The repository includes diagnostic plots for all parameters stated in the paper, and a notebook with an example script to reproduce these plots for our catalog, or for a new parameter calculated by the user. The lists of plots is:}

\new{
\begin{enumerate}
    \item \textbf{Feature importances:} which parameter out of the $(\avgsnrm, \mu_0)$ and $(\reseffm, \mathcal{R})$ pairs determines the bias, calculated with the mutual information criterion;
    \item \textbf{Bias grids:} figures equivalent to Fig. \ref{fig:moments_grid}, used in this paper, showing the average bias and scatter in the parameter as a function of resolution and noise;
    \item \textbf{Typical examples:} examples of how a parameter changes as a function of resolution or noise for two galaxies which span two extremes of that parameter's distribution;
    \item \textbf{Correction:} a plot showing the distribution of $p$ and $p_{\rm{base}}$ before and after a correction derived with symbolic regression.
\end{enumerate}
}

This \texttt{statmorph-lsst} package is available on GitHub and contributions are welcome. In addition to this package, we published our suite of galaxy augmentations on Zenodo, and will release a web-app that allows testing new, custom functions before they are implemented in \texttt{statmorph-lsst}. We encourage other morphology experts to try our suite and contribute to our package!

\section*{Acknowledgements}

\new{We are very grateful to the anonymous referee for a thorough reading of the paper and their helpful suggestions.} We thank the University of Waterloo for their support of the Canada Rubin Fellowship program. ES thanks Roan Haggar, Ana Ennis, and the members of the unofficial ``galaxies office'' for many fruitful scientific discussions throughout this project. CRM acknowledges support from an Ontario Graduate Scholarship. D.D acknowledges support from the NAWA (Bekker grant BPN /BEK/2024/1/00029/DEC/1). HMHT acknowledges financial support from DGAPA-PAPIIT project AG101725 and CONAHCYT project CF-2023-G-1052. W.J.P. has been supported by the Polish National Science Center project UMO-2023/51/D/ST9/00147. RAW acknowledges support from NASA JWST Interdisciplinary Scientist grants NAG5-12460, NNX14AN10G and 80NSSC18K0200 from GSFC.

\vspace{5mm}
\textit{Data Access:} \href{https://zenodo.org/records/17585609}{Zenodo DOI: 10.5281/zenodo.17585608}, \href{https://github.com/astro-nova/statmorph-lsst/tree/master}{GitHub repository: \texttt{statmorph-lsst}}

\vspace{5mm}
\textit{Facilities:} HST (WFC3, ACS, WFPC2), MAST, HLA, NED

\vspace{5mm}
\textit{Software:} \texttt{statmorph} \citep{statmorph}, photutils \citep{photutils}, astropy \citep{astropy,astropy2,astropy3}, \textsc{GalSim} \citep{galsim}, \textsc{Galfit} \citep{galfit,galfit2}, Scikit-Image \citep{skimage}, Scikit-Learn \citep{sklearn}, Matplotlib \citep{matplotlib}, pandas \citep{pandas}, NumPy \citep{numpy}, SciPy \citep{scipy}, Julia, \texttt{SymbolicRegression.jl} \citep{Cranmer2023}

\bibliography{references}{}
\bibliographystyle{mnras_custom}

\null
\newpage
\clearpage
\appendix

\section{The diversity of morphologies in the RNGC/IC sample}\label{app:sample}

Since the NGC and the IC catalogues are some of the earliest large compilations of extragalactic sources, they are biased towards galaxies that are nearby and bright. Because of this, the RNGC/IC catalog does not contain, for example, nearby dwarf galaxies that were not detected at the time. Due to this limitation we cannot analyze the behaviour of morphological metrics on the faintest dwarf galaxies, but we attempt to remedy this bias by inspecting the \textit{HST} tiles for any nearby sources that also have a spectrosopic redshift placing them within 200 Mpc. This adds some lower-mass galaxies to our sample that were not originally in the RNGC/IC.

\begin{figure}
    \centering
    \includegraphics[width=\linewidth]{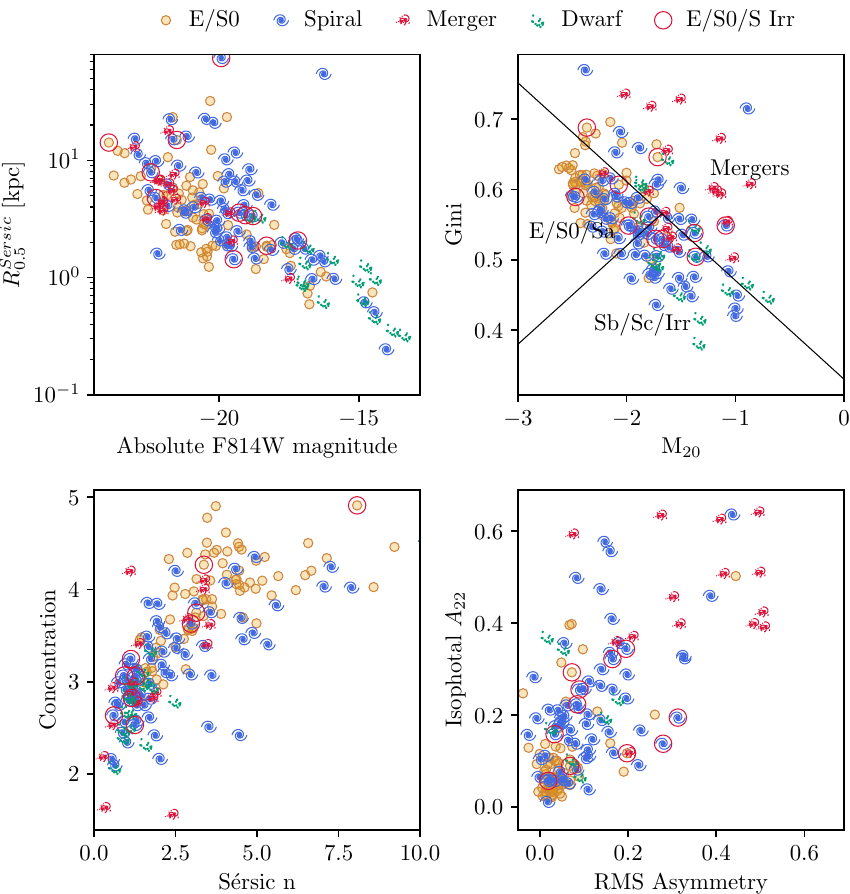}
    \caption{The distribution of several structural measurements in our sample of 189 RNGC/IC and associated galaxies: S\'ersic $R_{0.5}$ versus absolute F814W magnitude (top left), Gini-$M_{20}$ (top right), concentration-S\'ersic $n$ (bottom left), and for two asymmetry metrics (bottom right). The galaxies analyzed in this work sample the distribution of different morphologies along the Hubble sequence well, including both early-type, late-type galaxies and mergers with a range of absolute magnitudes and sizes.}
    \label{fig:sample_dist}
\end{figure}

To demonstrate that we are able to probe a range of morphologies, we show the distributions of some of the parameters studied in this paper, obtained from our ``baseline'' images (100 pc/px and $\mu_0 = 23.5$ mag/arcsec$^2$) in Fig. \ref{fig:sample_dist}. 

In particular, we show the S\'ersic half-light radius as a function of F814W (\textit{I}-band) absolute magnitude (upper left), the Gini-M$_{20}$ plane (upper right), S\'ersic $n$ and concentration (bottom left), and two asymmetry measures (bottom right). \new{We also include Hubble type classifications from the RNGC/IC catalog, with the additional classifications of new sources by the authors. Early-type galaxies are plotted in yellow, spirals in blue, mergers in red, and early- or late- type galaxies with the ``m'' qualifier (i.e. irregular, but not clear mergers) are identified as red circles.}

We use the size and absolute magnitude as a proxy for the size-mass relation since we do not have mass measurements for some galaxy in our sample. However, it is clear that we probe a wide range of galaxy sizes and magnitudes, spanning several orders of absolute magnitudes and physical sizes. In the Gini-$M_{20}$ diagram, the galaxies sample the entire parameter space between early-type and Sa, late-type, and irregular galaxies. We have several merging objects with high asymmetry values, and a good sample of concentration indices. Therefore we are confident that we can appropriately sample the dynamic range of galaxy morphologies, at least outside the dwarf regime. \new{Moreover, the correlation between these morphological indicators and Hubble types is good: mergers are well-separated on asymmetry and Gini-$M_{20}$ diagrams, and early-type galaxies are clearly separated on Concentration-$n$ subplot.}


\section{Background estimation}\label{app:bg}

\begin{figure*}[ht]
    \includegraphics[width=\linewidth]{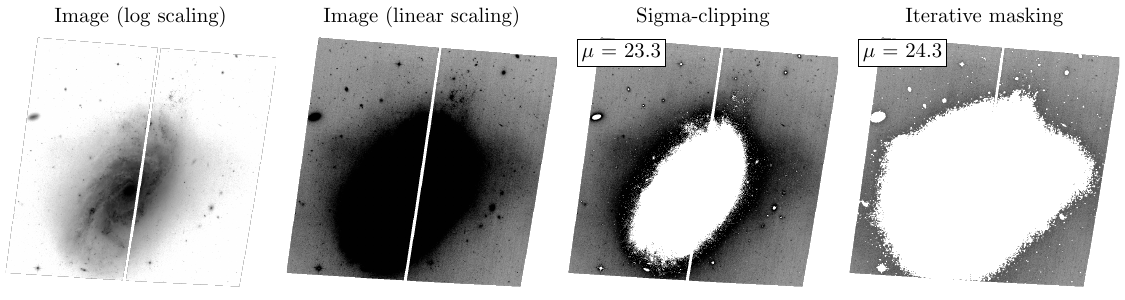}
    \caption{A comparison of background estimation via sigma-clipping (\textbf{third} panel) and our iterative procedure (\textbf{fourth} panel) for an example galaxy shown in log (\textbf{first}) and linear (\textbf{second}) scales. Simple sigma-clipping fails to mask the diffuse light of the source since it dominates the chip. Using iteratations of sigma-clipping and source detection better masks the faint halo, allowing us to get a more accurate measurement of the background noise level and hence the surface brightness limit.}
    \label{fig:sigmaclip}
\end{figure*}

To estimate the background of each image, we use an improved iterative sigma-clipping technique, which we demonstrate in Fig. \ref{fig:sigmaclip} below. An example galaxy NGC 1084, which spans most of the \textit{HST} chip, is shown in a log scaling in the first panel. In the second panel, we show the same galaxy with a linear scale centered on the background median, which highlights the extended low surface brightness envelope of this source.

Sigma-clipping on its own is already iterative, where all pixels above $3\sigma$ are masked, the standard deviation is re-estimated, and so on until convergence. The third panel in Fig. \ref{fig:sigmaclip} shows the pixels that are masked using the default sigma-clipping implementation. However, the source light dominates most of the pixels and so it is not fully clipped. The halo of the galaxy is not fully masked in this way, which leads us to underestimate the surface brightness limit of the image.

Instead, we run source detection to mask out any sources and estimate the background this way. We smooth the image by a 5-pixel Gaussian kernel, and detect any sources above a 1$\sigma$ threshold with a minimum area of 1\% of the image size. For each detection, we expand the source mask by 10\% of its size. Finally, we mask out all the sources and re-run sigma-clipping to get new background statistics. With the new $\sigma$ estimate, we repeat the source detection step, and continue doing so until $\sigma$ converges to within 5\% or 30 iterations have elapsed. The last panel of Fig. \ref{fig:sigmaclip} shows the source mask obtained from our procedure: it captures most of the diffuse light of the galaxy, and provides us with a more accurate $\sigma$ and hence surface brightness limit estimate.

\section{Segmentation}\label{app:segmentation}

\begin{figure*}[ht]
    \includegraphics[width=\linewidth]{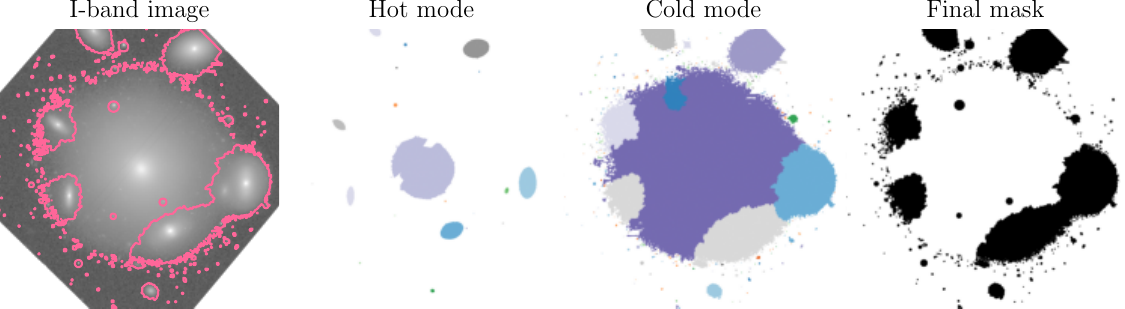}
    \caption{An example of our ``hot-cold'' segmentation approach to the image of NGC 4874, the second brightest galaxy of the Coma cluster (\textbf{first} panel). NGC 4874 is surrounded by other galaxies, which need to be carefully deblended. In the ``hot'' step, all sources with SNR higher than 30 are detected and masked (\textbf{second} panel). Then, we detect all sources above 1.5 SNR in the ``cold'' step and deblend the central galaxy into subcomponents (\textbf{third} panel). Finally we mask any subcomponent with a greater than 20\% overlap with the ``hot'' mask (\textbf{fourth} panel and pink contours in the first panel).}
    \label{fig:segmentation}
\end{figure*}

Since the sources in our sample are extremely local and have large angular extents, we needed to implement a careful segmentation and masking routine to remove as many foreground stars and background galaxies as possible. To do this, we follow a hot-and-cold approach, similar to \cite{Sazonova2021} and \cite{Morgan2024}. Here we describe our routine, which is available in the \texttt{statmorph-lsst/test\_suite} package. \new{The routine is based on the \textsc{Photutils} library and the functions mentioned here refer to its \texttt{segmentation} module.} Fig. \ref{fig:segmentation} shows the steps of the algorithm for one galaxy, where segmentation was particularly challenging. 

To ensure high-quality masks, we perform the segmentation routine iteratively by visually inspecting the image and changing parameters until a satisfactory map is produced. Parameters that can be varied in our routine, with their typical values, are listed in Table \ref{tab:segmentation-params}.

{\renewcommand{\arraystretch}{1.5}
\begin{table*}
\begin{threeparttable}
\caption{Segmentation parameters}
    \begin{tabular}{cccc}
        \toprule
        \toprule
            Param. & Description & 
            Typical value &
            Range \\
        \midrule
        \texttt{hot\_thresh} & Signal-to-noise threshold in the ``hot'' mode & 30 & 5$\sim$1000 \\
        \texttt{cold\_thresh} &  Signal-to-noise threshold in the ``cold'' mode & 1.5 & 1.5$\sim$5 \\
        \texttt{min\_area} & Minimum area of a detected segment, in pixels & 5 & 1$\sim$5 \\
        \texttt{contrast} & Contrast needed to deblend adjacent regions in the central source & 0.001 & 0.001$\sim$1 \\
        \texttt{overlap\_thresh} & Minimum overap between a deblended ``cold'' segment and the ``hot'' map to mask out that segment & 0.2 & 0.05$\sim$0.4\\
    \bottomrule
    \end{tabular}
    \label{tab:segmentation-params}
\end{threeparttable}
\end{table*}
}

First, we load the image and interpolate over any masked or missing pixels other within the detector footprint. This is necessary because the ACS detector has a chip gap in the middle, often passing through the galaxy of interest, and simply masking the chip gap causes the segmentation map to split into two distinct regions, effectively masking out one half of the galaxy. Then, for large images (over 300 pixels), we apply a Gaussian smoothing filter with $\sigma=2$ to average over the noise spikes. For every task involving source detection, we choose 4-connected regions rather than 8 commonly used elsewhere \cite[e.g., in Source\new{-}Extractor][]{Arnouts2007}. We find that a stronger adjacency requirement means fewer spurious noise regions are marked as sources. With this in mind, we proceed with the hot-cold segmentation.


\begin{enumerate}

    \item \textbf{Hot mode:} detect all bright sources above a high \texttt{hot\_threshold} of the background RMS with some \texttt{min\_area}. This step usually selects the central source and other bright sources: foreground stars and other galaxies. We add all sources except the central to a mask\footnote{This could, in principle, also mask out bright star clusters, but we opted to be cautious and mark any unresolved point source as we have no way of distinguishing them from foreground stars.}. We then grow the mask based on the area of each detected ``hot'' source, so that larger detections are grown by a larger amount, since they typically have diffuse light extending further.

    \item \textbf{Cold mode:} next, detect any sources with a lower \texttt{cold\_threshold} on signal-to-noise and the same \texttt{min\_area}. This selects the main source as well as any other sources significantly present in the image. 

    \item \textbf{Deblending:} the central ``cold'' source typically consists of our target galaxy and several embedded contaminants. The brightest peaks of contaminants are already masked in the ``hot'' step, but their diffuse light can still bleed into the main source. We therefore deblend the central source with a high \new{deblending} \texttt{contrast}. We iterate through every deblended segment and calculate the fraction of that segment's area that overlaps with a ``hot'' mask. If there is sufficient overlap, set by \texttt{overlap\_thresh}, then the segment is contaminated and we mask it.
    
    \item \textbf{Mask contaminants:} now grow the constructed mask to ensure the faint light from the contaminants is masked as well. We grow each masked segment proportionally to its size using a single convolution step. By default, \texttt{detect\_sources} returns a map where the pixels corresponding to a source have a value equal to its label. Instead, we assign the pixel values to be the area of each segment. We then convolve this map with a Gaussian filter, using a size that is 5\% of the image. Finally, we assign all pixels with a flat value over 0.05 to a mask. Since different segments have a different area, then after a convolution with a Gaussian, larger segments extent further until they reach this threshold, and therefore the mask grows more. Finally, we return this mask, and the \textbf{cold mode} segmentaion map where only the central source is left.
\end{enumerate}

\section{Symbolic Regression}\label{app:pysr}
We use \texttt{SymbolicRegression.jl} package \citep{Cranmer2023} to estimate the correction functions between our observed and baseline parameters. The advantage of symbolic regression is that it optimizes both the form of the function itself, and any free parameters simultaneously. It works via a ``genetic modification'' scheme: it builds a tree that consists of the input variables, free parameters, and operators. At each iteration, the algorithm can either change the tree (by adding, removing, or swapping nodes) or numerically optimize all free parameters. As the iteration progresses, the tree grows, representing more complex functions. At the end of each iteration, a ``hall of fame" of the functions with the lowest loss is saved and carried over to the next iteration.

{\renewcommand{\arraystretch}{1.5}
\begin{table*}
\centering
\begin{threeparttable}
\caption{\texttt{SymbolicRegression.jl} parameters}
    \begin{tabular*}{\textwidth}{@{\extracolsep{\fill}}lll}
        \toprule
        \toprule
            Parameter & Description & Value \\
        \midrule
            $N$ & \# of galaxies in the training set & 70 \\
            $f$ & Functional form of the fit & $p_{\rm{pred}} = p_{\rm{base}} \times f(x,y) + g(x,y)$ \\
            \texttt{elementwise\_loss} & Loss function during optimization, $\mathcal{L}(p_{\rm{pred}}, p_{\rm{true}})$ & \texttt{L1DistLoss} $\equiv |p_{\rm{pred}}-p_{\rm{true}}|$\\
            \texttt{binary\_operators} & Operators acting on two arguments, $a \otimes b$ & $+, -, \times, \div$\\
            \texttt{unary\_operators} & Operators acting on a single argument, $F(a)$ & $\exp, \tanh, \log_{10}, 1/x, \sqrt{\cdot}$ \\
            \texttt{maxsize} & Maximum model complexity (see text)& 25 \\
            \texttt{complexity\_of\_operators} & See text & 0 for $\times$ and $+$, 1 for the rest \\
            \texttt{niterations} & Number of competition iterations & 1000 \\
            \texttt{ncycles\_per\_iteration} & Number of mutations per iteration & 1000 \\
            \texttt{population\_size} & Number of functions in each iteration & 50 \\
        \bottomrule
    \end{tabular*}
    \label{tab:pysr_params}
\end{threeparttable}
\end{table*}
}
{\renewcommand{\arraystretch}{1.5}
\begin{table*}
\centering
\footnotesize                     
\setlength{\tabcolsep}{2pt}       
\begin{threeparttable}
\caption{\texttt{SymbolicRegression.jl} fit results}
    \begin{tabular}{llllllrrrrrr}
        \toprule
        \toprule
            $p$ & Sec. & $x$ & $y$ & $f(x,y)$ & $g(x,y)$ 
            & \multicolumn{2}{c}{Error [all data]} & \multicolumn{2}{c}{Error [low-res]} 
            & \multicolumn{2}{c}{Error [low-SNR]}\\
            & & & & & & Orig. & Corr. & Orig. & Corr. & Orig. & Corr. \\
        \midrule
            \multicolumn{12}{l}{Geometric measurements (Sec. \ref{sec:moments})}\\
        \midrule
            \footnotemark[1]$\mathbf{x}_0^A$ & 
                    \ref{sec:moments} & \reseff{}& \avgsnr{} & -- & -- &
                    $4.40^{+12.8}_{-3.53}$& --& $16.6^{+11.1}_{-4.47}$& --& $4.34^{+12.3}_{-3.32}$& --\\

            $e$ & 
                \ref{sec:moments} & \reseff{}& \avgsnr{} & $\tanh (\log x)$ &
                0.031 & 
                $-0.02^{+0.09}_{-0.12}$& $-0.00^{+0.12}_{-0.10}$& $-0.13^{+0.18}_{-0.15}$& $-0.00^{+0.31}_{-0.15}$& $-0.02^{+0.14}_{-0.15}$& $-0.00^{+0.16}_{-0.15}$\\
                
            \footnotemark[1]$\theta$ & 
                \ref{sec:moments} & \reseff{}& \avgsnr{} & -- & -- &
                $0.30^{+16.2}_{-14.6}$& -- & $1.20^{+29.6}_{-22.3}$& -- & $0.65^{+30.1}_{-25.2}$ & --\\

            \footnotemark[1]$\theta_{e>0.3}$ & 
                \ref{sec:moments} & \reseff{}& \avgsnr{} & -- & -- &
                $0.08^{+8.44}_{-6.17}$& --& $0.63^{+27.7}_{-20.5}$& 
                --&$0.53^{+21.3}_{-13.8}$&--\\

            \footnotemark[1]$\theta^{\rm{Sersic}}$ & 
                \ref{sec:sersic} & \reseff{}& \avgsnr{} & -- & -- &
                $-0.02^{+4.70}_{-4.82}$& -- & $-0.21^{+25.0}_{-18}$& -- & $-0.02^{+8.71}_{-7.09}$ & --\\

            \footnotemark[1]$\theta^{\rm{Sersic}}_{e>0.3}$ & 
                \ref{sec:sersic} & \reseff{}& \avgsnr{} & -- & -- &
                $-0.04^{+1.98}_{-3.07}$& -- & $-0.50^{+10.8}_{-11.5}$& -- & $-0.07^{+4.00}_{-5.74}$ & --\\
            
            \footnotemark[2]$e_{\rm{Sersic}}$ & 
                \ref{sec:sersic} & \reseff{}& \avgsnr{} & -- & -- &
                $0.00^{+0.06}_{-0.04}$& -- & $0.05^{+0.15}_{-0.10}$& -- & $0.00^{+0.07}_{-0.05}$\\
        
        \midrule
            \multicolumn{12}{l}{Radius measurements (Sec. \ref{sec:radii})}\\
        \midrule
            $R_{p, \circ}$ & 
                \ref{sec:rpet} & \res{}& \avgsnr{} & $\tanh (3.21y)$ & $(x/1000)^{1.86}$ &
                $0.08^{+1.07}_{-0.86}$ & $0.00^{+0.77}_{-0.81}$ & $0.70^{+1.40}_{-1.29}$ & $0.08^{+1.07}_{-1.26}$ & $-0.12^{+1.08}_{-1.68}$ & $-0.04^{+1.18}_{-1.26}$\\
            $R_{p, e}$ & 
                \ref{sec:rpet} & \res{}& \avgsnr{} & $\tanh(y+0.51)$ & $0.19 \log x$ & $0.01^{+1.23}_{-1.36}$ & $-0.02^{+1.83}_{-0.86}$ & $0.47^{+1.81}_{-1.77}$ & $0.43^{+2.41}_{-1.40}$ & $0.01^{+1.23}_{-1.36}$ & $-0.02^{+1.83}_{-0.86}$ \\
            $R_{20}$ & 
                \ref{sec:r20} & \res{}& \avgsnr{} & $0.91$ & $(x/1300)^{1.12}$ & $0.10^{+0.49}_{-0.14}$& $0.01^{+0.16}_{-0.14}$& $0.51^{+0.43}_{-0.28}$& $0.03^{+0.25}_{-0.26}$& $0.05^{+0.48}_{-0.19}$& $-0.00^{+0.19}_{-0.21}$\\
            $R_{50}$ & 
                \ref{sec:r20} & \res{}& \avgsnr{} & $\tanh (3.21y)$ & $(x/1000)^{1.86}$ &
                $0.09^{+0.70}_{-0.24}$& $-0.01^{+0.30}_{-0.38}$& $0.59^{+0.75}_{-0.48}$& $-0.05^{+0.41}_{-0.66}$& $0.03^{+0.68}_{-0.48}$& $0.01^{+0.62}_{-0.44}$\\
            $R_{80}$ & 
                \ref{sec:r20} & \reseff{}& \avgsnr{} & $\tanh (3.35y)$ & $3.04/x$ &
                $0.06^{+0.82}_{-0.48}$& $-0.06^{+0.67}_{-0.38}$& $1.07^{+1.42}_{-0.60}$& $0.24^{+1.35}_{-0.52}$& $-0.02^{+0.82}_{-1.00}$& $-0.04^{+1.04}_{-0.63}$\\
            $R_{\rm{max}}$ & 
                \ref{sec:r20} & \res{}& \avgsnr{} & $\log(y+1.62)$ & $1.08$ &
                $-0.75^{+3.81}_{-5.12}$& $0.05^{+4.27}_{-2.58}$& $-0.47^{+5.19}_{-5.71}$& $0.66^{+5.06}_{-3.03}$& $-4.09^{+2.60}_{-8.44}$& $0.03^{+6.01}_{-4.80}$\\
            \footnotemark[2]$R_{0.5}^{\rm{Sersic}}$ & 
                \ref{sec:sersic} & \res{}& \avgsnr{} & -- & -- & 
                $-0.04^{+0.71}_{-1.13}$& -- & $-0.18^{+0.95}_{-1.32}$& --& $0.04^{+1.48}_{-1.06}$& --\\
        \midrule
            \multicolumn{12}{l}{Bulge strength measurements (Sec. \ref{sec:bulge})}\\
        \midrule

            $n$ & 
                \ref{sec:sersic} & \reseff{}& \avgsnr{} & $\tanh(0.2x)$&
                0 & 
                $-0.07^{+0.54}_{-1.09}$& $-0.02^{+0.71}_{-0.91}$& $-0.81^{+1.30}_{-2.13}$& $0.25^{+2.15}_{-2.10}$& $0.01^{+0.87}_{-0.86}$& $0.03^{+0.99}_{-0.78}$ \\
                
            $C$ & 
                \ref{sec:concentration} & \reseff{}& \avgsnr{} & $\tanh(\log 0.4x))$ &
                $\frac{3.3}{\sqrt{x-0.76}}$ & 
                $-0.22^{+0.29}_{-0.79}$& $0.00^{+0.36}_{-0.38}$& $-1.14^{+0.62}_{-0.56}$& $0.12^{+1.86}_{-1.07}$& $-0.22^{+0.32}_{-0.72}$& $-0.06^{+0.46}_{-0.49}$\\

            $G$ & 
                \ref{sec:gini} & \reseff{}& \avgsnr{} & $0.83 - 0.59/x$ &
                $0.1 \tanh y$ & 
                $-0.03^{+0.04}_{-0.07}$& $-0.00^{+0.05}_{-0.04}$& $-0.08^{+0.08}_{-0.05}$& $0.01^{+0.14}_{-0.06}$& $-0.05^{+0.06}_{-0.07}$& $0.00^{+0.05}_{-0.06}$\\

            $M_{20}$ & 
                \ref{sec:m20} & \reseff{}& \avgsnr{} & $0.74-1/x$ &
                $-0.5$ & 
                $0.15^{+0.49}_{-0.22}$& $0.01^{+0.45}_{-0.29}$& $0.63^{+0.37}_{-0.42}$& $-0.10^{+0.65}_{-0.52}$& $0.20^{+0.52}_{-0.27}$& $0.12^{+0.63}_{-0.35}$\\

            $B(G,M_{20})$ & 
                \ref{sec:bgm20} & \reseff{}& \avgsnr{} & $\tanh x/15$ &
                $-\frac{2.1}{x \tanh y}$ & 
                $-0.25^{+0.30}_{-0.55}$& $-0.02^{+0.32}_{-0.44}$& $-0.82^{+0.50}_{-0.37}$& $0.25^{+1.67}_{-0.84}$& $-0.39^{+0.42}_{-0.60}$& $-0.16^{+0.40}_{-0.52}$\\

        \midrule
            \multicolumn{12}{l}{Disturbance measurements (Sec. \ref{sec:disturbance})}\\
        \midrule
            \footnotemark[2]\footnotemark[3]$S(G,M_{20})$ & 
                \ref{sec:sgm20}\footnotemark[2]\footnotemark[3] & \reseff{}& $\mu_0$ & -- &
                -- & 
                $0.01^{+0.47}_{-0.52}$& -- & $-0.01^{+1.02}_{-0.59}$& -- & $-0.01^{+0.46}_{-0.64}$& --\\
            $A_{\rm{RMS}}$ & 
                \ref{sec:aiso} & \reseff{}\footnotemark[1]& \avgsnr{} & $0.5 \log (x)$ & $0$& 
                $-0.01^{+0.83}_{-1.46}$& $0.39^{+1.52}_{-0.89}$& $0.15^{+1.05}_{-1.67}$& $1.65^{+4.80}_{-1.49}$& $-0.11^{+2.24}_{-2.19}$& $0.23^{+2.50}_{-1.97}$\\
            \footnotemark[4]$A_{\rm{RMS}}$ & 
                \ref{sec:aiso} & \reseff{}\footnotemark[1]& \avgsnr{} & $0.55 \log (x-1.5)$ & $0.2/x$& 
                $-0.01^{+0.83}_{-1.46}$ & $0.07^{+1.17}_{-1.06}$ & $-0.25^{+0.78}_{-2.16}$ & $0.00^{+2.54}_{-1.53}$ & $-0.11^{+2.24}_{-2.19}$ & $-0.08^{+2.28}_{-1.97}$\\
            $A_{\rm{CAS}}$ & 
                \ref{sec:asymmetry} & \reseff{}& \avgsnr{} & $\tanh \left(\frac{xy}{138}\right)$  & $-\frac{0.295}{\exp(0.65y)}$
                & $-0.06^{+0.10}_{-0.21}$ & $0.03^{+0.22}_{-0.26}$ & $-0.03^{+0.08}_{-0.20}$ & $0.00^{+0.59}_{-1.46}$ & $-0.21^{+0.12}_{-0.18}$ & $0.02^{+0.87}_{-1.08}$\\

            \footnotemark[2]\footnotemark[3]$A_{S}$ & 
                \ref{sec:ashape} & \reseff{}& $\mu_0$ & -- &
                -- & $0.03^{+0.20}_{-0.14}$& -- & $0.05^{+0.23}_{-0.17}$& --& $0.12^{+0.20}_{-0.18}$& --\\
                
            $A_{22}$ & 
                \ref{sec:aiso} & \reseff{}& $\mu_0$ & $0.94$& 
                $1/y$ &
                $0.04^{+0.47}_{-0.09}$& $0.01^{+0.50}_{-0.10}$& $0.13^{+0.73}_{-0.13}$& $0.10^{+0.77}_{-0.14}$& $0.17^{+0.61}_{-0.20}$& $0.14^{+0.65}_{-0.20}$\\
            $A_{o}$ & 
                \ref{sec:asymmetry} & \reseff{}& \avgsnr{} & $\log y$ & $0.14(\log y$--0.8) &
                $-0.05^{+0.13}_{-0.20}$& $0.00^{+0.21}_{-0.25}$& $-0.02^{+0.12}_{-0.18}$& $-0.02^{+0.21}_{-0.29}$& $-0.18^{+0.11}_{-0.23}$& $-0.06^{+0.57}_{-0.56}$\\

        \midrule
            \multicolumn{12}{l}{Other measurements (Sec. \ref{sec:other})}\\
        \midrule
            \footnotemark[2]\footnotemark[3]$M$ & 
                \ref{sec:multimode} & \reseff{}& $\mu_0$ & -- &
                -- & 
               $-0.01^{+0.44}_{-0.30}$&-- & $-0.05^{+0.26}_{-0.16}$&--& $0.12^{+1.27}_{-0.38}$& --\\

            \footnotemark[2]$I$ & 
                \ref{sec:intensity}& \reseff{}& $\mu_0$ & -- &
                -- & 
                $-0.00^{+0.08}_{-0.14}$& -- & $0.06^{+0.23}_{-0.19}$& -- & $-0.00^{+0.10}_{-0.17}$ & --\\

            \footnotemark[3]$D$ & 
                \ref{sec:deviation}& \reseff{}& \avgsnr{} & $\tanh(x/14)$-0.12 &
                $\frac{0.53}{x+1.55}$ & 
                $0.17^{+0.81}_{-0.41}$& $0.00^{+1.59}_{-0.76}$& $0.70^{+0.92}_{-0.58}$& $-0.40^{+6.51}_{-3.94}$& $0.30^{+1.06}_{-0.77}$& $0.19^{+2.00}_{-1.00}$\\
                
            \footnotemark[2]$S$ & 
                \ref{sec:smoothness} & \reseff{}& \avgsnr{} & -- &
                -- 
                & $-0.00^{+0.03}_{-0.13}$ & -- & $-0.02^{+0.17}_{-0.35}$& -- & $-0.06^{+0.05}_{-0.45}$ & -- \\

            $St$ & 
                \ref{sec:substructure} & \reseff{}& $\mu_0$ & -- &
                -- & $-0.01^{+0.04}_{-0.11}$& -- &$-0.01^{+0.03}_{-0.07}$& -- &$-0.05^{+0.05}_{-0.13}$&--\\
            
        \bottomrule
    \end{tabular}
    \begin{tablenotes}
        \item \textit{Notes} -- for each parameter, $x$ and $y$ are found using mutual information regression. We then compute the 16\ts{th}, 50\ts{th}, and 84\ts{th} quantiles of the error ($p - p_{\rm{base}}$) distribution for the entire sample, and low-resolution / low-SNR subsamples. We repeat this for ($p_{\rm{corr}} - p_{\rm{base}}$) where a \texttt{SymbolicRegression} correction is fitted.
        \item \footnotemark[1] Here we quantify the offset form the baseline in a random direction, so a correction is not possible.
        \item \footnotemark[2] There resolution/depth bias is significantly smaller than the average scatter, so a bias correction is not needed.
        \item \footnotemark[3] The error quantiles are multiplied by 10.
        \item \footnotemark[4] Better correction for low-resolution images but a similar performance overall.
        
    \end{tablenotes}
    \label{tab:pysr}
\end{threeparttable}
\end{table*}
}

The objective of the \texttt{SymbolicRegression} optimizer is to minimize the loss of the training set, but it also computes the complexity of each proposed function, where the complexity is a number of variables, parameters, and operators used. For example, $f(x) = 3 \times \exp(2\times x)$  would have a complexity 6. The hall of fame stores best-performing functions with different complexities, from lowest to highest, and calculates a \textit{score}: the relative improvement in loss compares to the function with the next lowest complexity. The user then uses the loss and the score in tandem to select the best-fitting function. The user can set up the \texttt{SymbolicRegression} run to fit their science purpose, and many steps of the algorithm can be changed. 

One of the most important steps is the selection of the function space to search. Most morphological parameters are bound between zero and some value by definition, and many converge to the baseline value. Looking at the difference between baseline and observed measurements as a function of resolution or depth, we often found that a sigmoid described the relationship well. To capture these behaviours, we fit a function of the form

\begin{equation}
    p = p_{\rm{base}} \times f(x,y) + g(x,y)
\end{equation}

\noindent where $x$ and $y$ are the measures of resolution and depth selected with the mutual information score. This function captures possible variations in the measurement that depend on the baseline value and any linear offsets. We allowed $f$ and $g$ to consist of basic algebraic operations ($+$, $-$, $\times$, $\div$) as well as $\exp$, $\tanh$, $\sqrt{\cdot}$, and $\log$, to capture sigmoid behaviour. We set up constraints so that $\exp$, $\tanh$, and $\sqrt{\cdot}$ cannot be nested within each other or inside the $\log$. Finally, we set the complexity of $+$ and $\times$ to 0 to allow the code to normalize the resolution or depth metrics with a lower penalty.  With these rules, $f(x) = 3\times \exp(2\times x)$ would have a complexity of 4.

We chose the loss function to be an L1 loss rather than an L2 (root-mean-square) loss, since it is more robust to outliers and so provided us with more stable fits. To train the model, we selected 70 random galaxies from the sample with the most obvious outliers removed. We trained the model for 1000 iterations. Table \ref{tab:pysr_params} shows all of the other parameters of \texttt{SymbolicRegression} needed to reproduce our results. At the end of the training cycle, we tested the highest-complexity model scoring over $10^{-2}$. Note that complexity is highly penalized in the score, so this approach still always preferred low-complexity models. We analyzed the distribution of corrected measurements for the overall sample as well as low-resolution and low-SNR subsets (as in Fig. \ref{fig:rpet_corr}). If the chosen model did not debias these subsamples, we tested the next most complex model, until a satisfactory correction was found for poor image quality. 

Tab. \ref{tab:pysr} shows the derived corrections and the 16/50/84\ts{th} percentiles of the error distribution $p-p_{\rm{base}}$ for each parameter. Note that for some parameters, the intrinsic scatter in the distribution dominated any systematic biases, so we did fit a correction as this would just amplify the scatter.

\section{Robustness of the orientation angle}\label{app:theta}

\new{The orientation angle is a crucial metric for studies of large-scale structure and weak lensing \citep[e.g.,][]{Zuntz2013,Yamamoto2025,Schrabback2025}, so here we show additional tests of $\theta$. As discussed in Sec. \ref{sec:moments}, $\theta$ is not well-constrained for face-on galaxies, and is easier to measure robustly for more edge-on systems. }

\begin{figure}
    \centering
    \includegraphics[width=\linewidth]{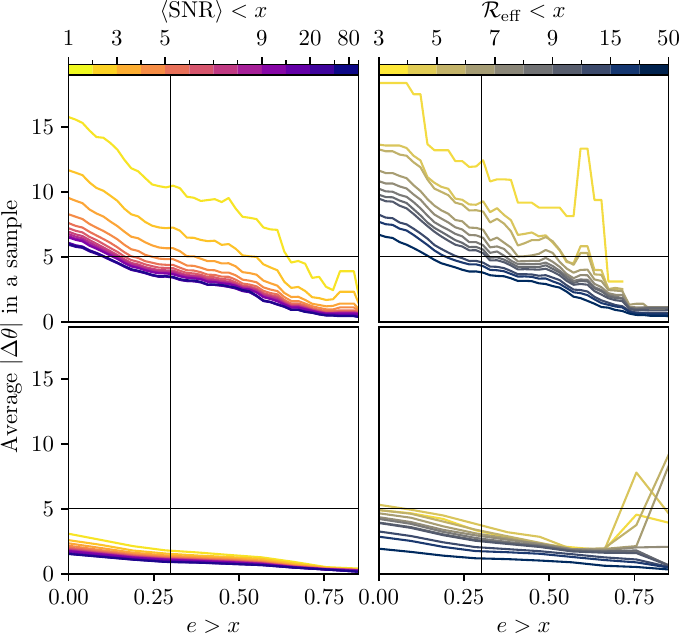}
    \caption{\new{The average error in the orientation angle for subsets of galaxies with ellipticities larger than $x$ as a function of $x$ calculated with image moments (top) or S\'ersic fitting (bottom). Each family of curves shows the error for subsets with \avgsnr{} (left) or \reseff{} (right) better than a given limit. The orientation angle is better constrained for more edge-on systems. S\'ersic-derived orientation has an average error below 5$\degree$ at all resolutions and depths, while the moment-based angle has a higher dependence on image quality.}}.
    \label{fig:theta}
\end{figure}

\new{Fig. \ref{fig:theta} shows the average error in $\theta$, calculated from image moments (top) and with S\'ersic fitting (bottom). For each curve, we chose a subset where \avgsnr{} (left) or \reseff{} (right) is less than a particular threshold, and ellipticity is larger than a given value, and then calculated the average error in $\theta$. }

\new{In general, \textsc{Galfit}-derived $\theta$ values are more robust, and the average error is below 3$\degree$ even for face-on galaxies, at all signal-to-noise levels. This is expected, since S\'ersic fitting also includes a PSF component, making ellipticities and orientations more robust. Another explanation might be that the fitting algorithm in \textsc{Galfit} is more sensitive to the initial guess for $\theta$, and so the fitter does not stray far from the initial guess in face-on systems where $\theta$ is not well-constrained, leading to more consistent values at low ellipticities. The error in $\theta$ at low resolutions is higher but still below 5$\degree$. }

\new{The newest \textit{Euclid} shape and alignment measurements are derived using S\'ersic fits with \textsc{SourceXtractor++} \citep{Quilley2025} and so our results easily transfer to their analysis -- we expect an average error in orientation angles to be within 5$\degree$. Similarly the \textsc{Im3Shape} code \citep{Zuntz2013} used in the DES Year 1 analysis \citep{Samuroff2018} is based on a two-component S\'ersic model, so our results should apply, although we have not tested multiple-component models explicitly. The performance of Gaussian forward-modelling approach \citep[e.g.,][]{Yamamoto2025} is beyond the scope of this paper, but as long as the PSF is accounted for, we expect the uncertainties to be small.}

\section{Degeneracies in the S\'ersic fit}\label{app:sersic}

The fact that the parameters in the S\'ersic model are degenerate with one another has been known for a long time \citep[e.g.,][]{Graham1996,Graham1998,Peng2010,Andrae2011}, however little has been done to \textit{quantify} this degeneracy. An insightful work by \cite{Trujillo2001} does this for one case: an idealized galaxy described by some arbitrary S\'ersic profile with some parameters $n_1$, $R_1$, $I_{01}$ being fit with a fixed $n_2=4$ profile and some free $R_2$ and $I_{02}$. They solve for the degeneracy between the fitted parameters $I_{02}$, $R_2$ as a function of $n_1$, $n_2$, and $R_1$, and show that the degeneracy explains some of the tightness of the correlation between $R$ and $\langle I \rangle$, the two axis of the fundamental plane of elliptical galaxies \citep{Djorgovski1987}. 


We extend this analysis for a general S\'ersic fit, where $n_2$ may also vary. The question we ask is then: given a galaxy well-described by some $(n_1, R_1, I_{01})$, if we try to fit it with some $n_2$, what values for $R_2$ and $I_{02}$ give the best fit? In other words, what is the degeneracy between $\Delta n$, $\Delta R$, and $\Delta I_0$?

To answer this, we want to minimize the total error between the profiles measured between some inner radius $r_{\mathrm{min}}$ and outer radius $r_{\mathrm{max}}$. We define the error as

\begin{equation}
    E = \int_{r_{\mathrm{min}}}^{r_{\mathrm{max}}} \left(I_1(r) - I_2(r)\right)^2 dr
\end{equation}

\noindent where $I_1(r)$ is the true profile and $I_2(r)$ is the profile we are trying to optimize. This is slightly different from the \cite{Trujillo2001} approach, which defines an error somewhat akin to a $\chi^2$ and normalize the square difference by $I_2(r)$, however this leads to numerical instability when $r \rightarrow \infty$. Finally, \cite{Trujillo2001} chose $r_{\mathrm{min}} = 0$ and solve for different $r_{\mathrm{max}}$. While it is true that we cannot know the profile out to infinity from an image, the intensity in the outskirts is negligible compared to the central one, and so $r_{\mathrm{max}}$ does not significantly affect the solution. On the other hand, due to pixelization, we do not know the true central flux, so we choose to keep $r_{\mathrm{min}}$ non-zero in our analysis. Therefore, we want to minimize

\begin{equation}
    E = \int_{r_{\mathrm{min}}}^{\infty} I_1^2(r) + I_2^2(r) - 2I_1(r) I_2(r) dr
\end{equation}

For simplicity of the derivation, we use an alternative (but an equivalent) definition of the S\'ersic profile:

\begin{equation}
    I(r) = I_0 \exp \left[ -b_n \left( \frac{r}{R_e}\right)^{1/n} \right]
\end{equation}

\noindent where $I_0$ is the intensity at the centre of the galaxy rather than at effective radius $R_e$. The two are related to one another as $I(0) = I(R_e) e^{b_n}$.

To find the optimal values of $R$ and $I_0$, we want to solve for $\partial E / \partial R_2 = 0$ and $\partial E / \partial I_{02} = 0$. First, solving for $\partial E / \partial I_{02}$, we get

\begin{equation}
\begin{split}
    \frac{\partial E}{\partial I_{02}} &= 2I_{02} \int_{r_{\mathrm{min}}}^{\infty} \exp \left[ -b_2 \left( \frac{r}{R_2}\right)^{\frac{1}{n_2}} \right] dr \\
    &- 2I_{01} \int_{r_{\mathrm{min}}}^{\infty} \exp \left[ -b_2 \left( \frac{r}{R_2}\right)^{\frac{1}{n_2}} -b_1\left( \frac{r}{R_1}\right)^{\frac{1}{n_1}} \right] dr \\
    &= 0.
\end{split}
\end{equation}

The first term has a closed-form solution. The second term needs to be integrated numerically. For convenience, following \cite{Trujillo2001}, we define this term as 

\begin{equation}
    S(r) \equiv \exp \left[ -b_2 \left( \frac{r}{R_2}\right)^{\frac{1}{n_2}} -b_1\left( \frac{r}{R_1}\right)^{\frac{1}{n_1}} \right],
\end{equation}

\noindent note, however, that our $S(r)$ differs from the one in \cite{Trujillo2001} since both terms in the exponential are always negative, and therefore this exponential always vanishes at infinity; while the one in \cite{Trujillo2001} becomes positive at large $r$ and forces the integral to diverge if $n_2 < n_1$. Substituting in the solution of the first term and rearranging the equation, we get an expression for $I_{02}/I_{01}$:

\begin{equation}
\frac{I_{02}}{I_{01}} 
    = \frac{2^{n_2} b_2^{n_2}}{n_2 R_2} \Gamma^{-1} \left[n_2, 2b_2 \left( \frac{r_{\mathrm{min}}}{R_2}\right)^{\frac{1}{n_2}} \right] \int_{r_{\mathrm{min}}}^\infty S(r) dr
\end{equation}

\noindent where $\Gamma$ is an upper incomplete Gamma function defined as $\Gamma(t, x) \equiv \int_x^\infty t^{s-1} e^{-t} dt$ and $\Gamma^{-1} \equiv 1/\Gamma$. For simplicity, we also define $X \equiv 2b_2 \left( \frac{r_{\mathrm{min}}}{R_2}\right)^{\frac{1}{n_2}}$.

We then follow the same steps using $\partial E / \partial R_2 = 0$. This produces another expression for $I_{02} / I_{01}$:

\begin{equation} 
\begin{split}
\frac{I_{02}}{I_{01}} = \frac{ 2^{n_2+1} b_2^{n_2+1}}{n_2 R_2} &\Gamma^{-1} \left[n_2+1, X \right] \\ 
&\times \int_{r_{\mathrm{min}}}^\infty \left( \frac{r}{R_2} \right)^{\frac{1}{n_2}} S(r) dr.
\end{split}
\end{equation}

\begin{figure*}
    \centering
    \includegraphics[width=\linewidth]{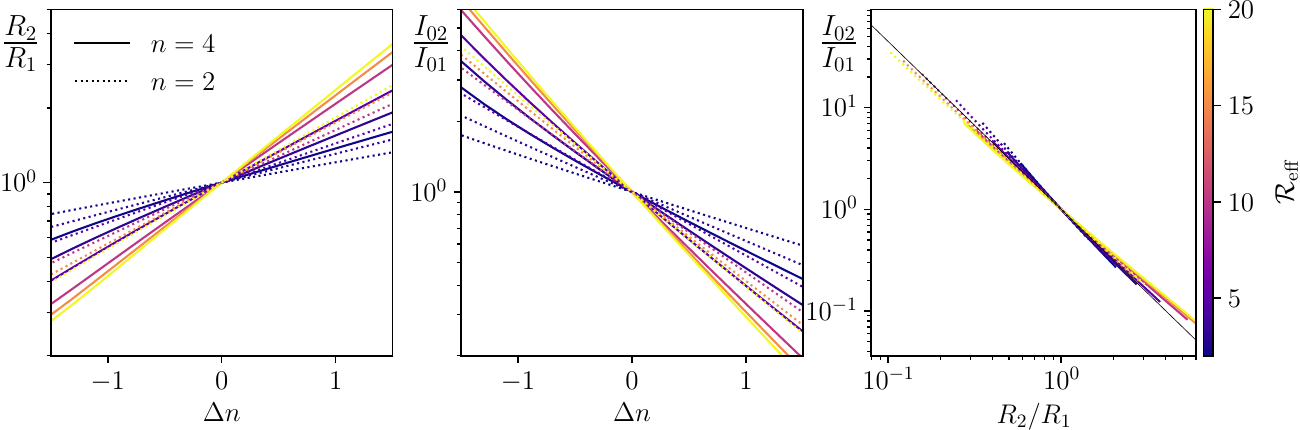}
    \caption{The relationship between a fixed $\Delta n \equiv n_2 - n_1$ and best-fitting $R_2$ (left), $I_{02}$ (center), as well as between $R_2$ and $I_{02}$ (right). The degeneracy in the fit is log-linear and depends on the resolution (i.e. $r_{\textrm{min}}$) and the original S\'ersic index $n_1$. The degeneracy is $R_2 / R_1 = \exp \left[ 1.223 \Delta n / n_1^{r_{\rm{min}}} \right]$ and $I_{02}/I_{01} = (R_2 / R_1)^{-1.65}$. However, while these values of $R_2$ and $I_2$ give the best fit given a fixed $n_2$, the minimized $\chi^2$ of the fit may still be too large for a given $\Delta n$; the largest allowed $\Delta n$ are shown in Fig. \ref{fig:sersic_unc}}.
    \label{fig:trujillo}
\end{figure*}

With these two derivatives, we now have an implicit expression for $R_2$ that is independent from $I_{02}$, and so by solving it numerically, we can evaluate the dependence of the best-fit $R_2$ on $n_2$. Similarly, we can then substitute this $R_2$ to get the dependence of $I_{02}$ on $n_2$. The expression for $R_2$ is:

\begin{equation}
    \int_{r_{\mathrm{min}}}^\infty S(r) \left[ 1 - 2b_2 \frac{\Gamma \left[n_2, X \right]}{\Gamma \left[n_2+1, X \right]} \left( \frac{r}{R_2}\right)^{\frac{1}{n_2}}  \right] dr = 0
\end{equation}

This expression is almost equivalent to the one in \cite{Trujillo2001}, except the functional form of $S(r)$ is different, $n_2$ is free, and the lower bound of the integral is allowed to vary. This integral does not have a closed form solution and has to be solved numerically. We use a \textsc{SciPy} root finder by setting a search bracket of $(r_{\mathrm{min}}, R_1)$ if $n_2 < n_1$ and $(R_1, \infty)$ if $n_2 > n_1$. 

To test this solution in a realistic use-case, we did the following: we first generated a perfect \textsc{Galsim} S\'ersic profiles with a range of chosen $n_1$ and set $R_1 = 10$, $I_{01} = 100$. We chose some pixel scale to discretize the image to, which we now define as $\mathcal{R}_{\mathrm{eff}} \equiv R_0 / \rm{scale}$, similar to the primary tests of this paper. We defined $r_{\rm{min}} \equiv \rm{scale}/2$. Then, we added a varying amount of sky and applied Poisson noise to achieve different levels of \avgsnr{}, calculated in a $2R_1$ aperture. Finally, for a wide range of $n_2$, we found the best-fitting $R_2$ and $I_2$ and calculated the $\chi^2$ of the fit. 

In Fig. \ref{fig:trujillo}, we show how the ratios of best-fitting $R_2/R_1$ and $I_{02}/I_{01}$ change as a function of $\Delta n \equiv n_2 - n_1$ for two test cases: $n_1 = 2$ (dashed) and $n_1 = 4$ (solid). We vary the effective resolution, showed in color from the lowest-resolution (dark) to a relatively high resolution of 20 pixels per $R_1$ (bright). In all cases, the relationship between $R_2/R_1$ or $I_{02}/I_{01}$ and $\Delta n$ is log-linear. In other words, the degeneracy is perfectly described by the functional form

\begin{equation*}
    \frac{R_2}{R_1} = e^{ \alpha \Delta n} \quad \rm{and} \quad \frac{I_{02}}{I_{01}} = e^{ -\beta \Delta n},
\end{equation*}

\noindent where $\alpha$ and $\beta$ depend on $n_1$ and $r_{\mathrm{min}}$. The relationship between $\alpha$ and $\beta$ is independent of $n_2$ and only weakly dependent on the resolution, as seen in the right-most panel of Fig. \ref{fig:trujillo}, and is approximately

\begin{equation*}
    \frac{I_{02}}{I_{01}} \approx \left( \frac{R_2}{R_1} \right)^{-1.65},
\end{equation*}

\noindent plotted as a thin grey line in the third panel of Fig. \ref{fig:trujillo}. As \cite{Trujillo2001} note, the degeneracy between these two quantities partially explains the tightness in the empirical relationship between $R_e$ and $\langle I \rangle$ which makes up one of the axes on the fundamental plane of elliptical galaxies.

\begin{figure*}
    \centering
    \includegraphics[width=\linewidth]{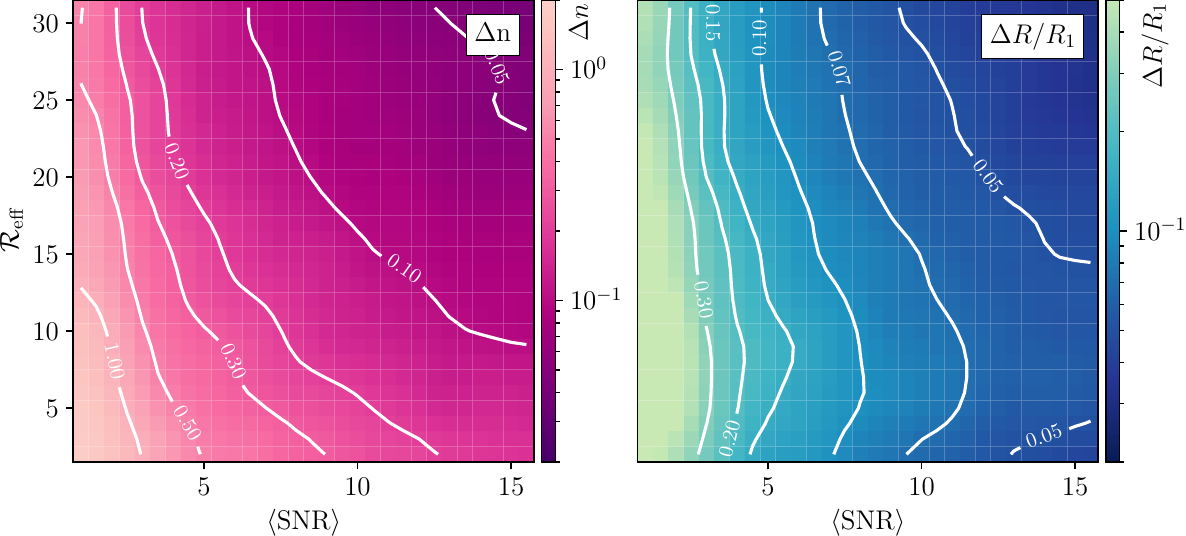}
    \caption{The uncertainty in the S\'ersic index and effective radius arising from the degeneracies in the analytic profile. Shown are the maximum $\Delta n$ (left) and hence $\Delta R / R_1$ (right) allowed by $p(\chi^2) > 0.05$ for an $n_1=4$ profile, as a function of resolution and depth. At low \avgsnr{}, the uncertainty in $\Delta R$ reaches 30\%, which explains the scatter in S\'ersic radius measurements on the observed dataset in Fig. \ref{fig:r_grid}. The uncertainty in $\Delta n$ at medium resolutions is $\pm 0.25 \sim \pm 1$, depending on \avgsnr{}, slightly lower than the scatter in Fig. \ref{fig:sersic_grid}}.
    \label{fig:sersic_unc}
\end{figure*}

To find the relationship between $\alpha$, $r_{\rm{min}}$, and $n_0$, we used \texttt{SymbolicRegression.jl}. We computed $\alpha$ for all of our test cases and used only $n_0$ and $r_{\mathrm{min}}$ to find an equation for $\alpha$ -- the fitter did not have access to $R_1$, $R_2$, or $\Delta n$. The best-scoring functional form is 

\begin{equation*}
    \alpha = 1.223 / n_0^{\sqrt{r_{\rm{min}}}}.
\end{equation*}

Using this fit, we compute predicted $\alpha$ and then the corresponding $R_2 / R_1$. The mean squared error on the ratio of effective radii is only 0.0005, indicating an excellent fit to the data, especially considering that the fitter was not given radius values themselves. Therefore, we conclude that the degeneracy between the best-fitting radius $R_2$ given $n_2$ is:

\begin{equation*}
    \frac{R_2}{R_1} \approx \exp \left[ 1.223 \Delta n / n_0^{\sqrt{r_{\rm{min}}}} \right] 
\end{equation*}
This is a purely analytic degeneracy between $R$ and $n$ that arises from the definition of the S\'ersic model itself. Values of $R_2$ computed in this way will minimize the squared error between an observed S\'ersic galaxy $(n_1, R_1, I_{01})$ and a fitted model with a fixed $n_2$. However, obviously, \textit{the best fit} will be achieved when $n_2 = n_1$. Therefore, it makes sense to then ask a question: how far can $n_2$ change and still produce a good fit? 


To investigate this, we opted to compute a more empirically motivated goodness-of-fit using the $\chi^2$ of the simulated images. We then found the \textit{maximum} $\Delta n$ range which resulted in a $p(\chi^2) > 0.05$, which would be an acceptable fit. This corresponds to a reduced $\chi^2$ between 1.05 and 1.3, depending on the number of degrees of freedom, i.e.  the size of the image. 
\begin{figure}[t]
    \centering
    \includegraphics[width=\linewidth]{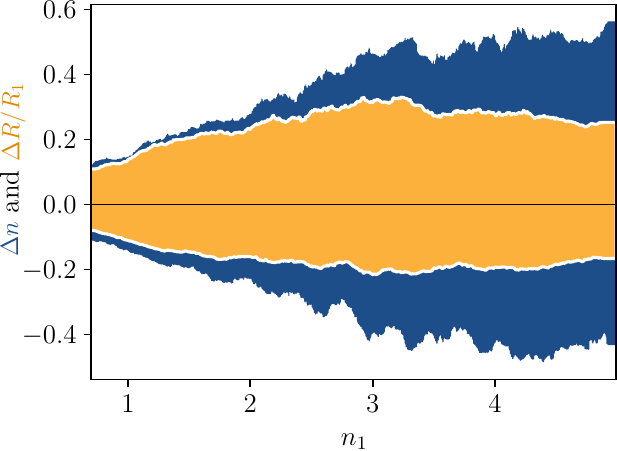}
    \vspace{-5pt}
    \caption{The uncertainty $\Delta n$ and $\Delta R/R_1$, defined the same way as in Fig. \ref{fig:sersic_unc}, for a mock galaxy with $\avgsnrm = 3$ and $\reseffm = 10$, as a function of $n_1$. While the uncertainty in $n$ grows from $0.1$ to $0.6$ at high S\'ersic indices, the uncertainty in $R_e$ stays at 20\% for an image of this quality, independent of $n_1$.}
    \label{fig:sersic_unc_n}
    \vspace{-22pt}
\end{figure}

Fig. \ref{fig:sersic_unc} shows the maximum values of $\Delta R/R$ (left) and $\Delta n$ (right) that pass these criteria as a function of \avgsnr{} and \reseff{} for a true profile with $n=4$. This is essentially a measure of uncertainty in $R$ and $n$ that is inherent to the S\'ersic profile, and so it should be treated as a \textit{lower bound} on the uncertainty in these quantities. In realistic use cases, since galaxies are complex structures with features besides a simple one-component S\'ersic, real $\chi^2$ will be higher than in our idealized case, and poorer fits might have to be accepted, which increases the intrinsic uncertainty on $R$ and $n$.

Even considering these as lower bounds, the uncertainty in $R$ for $\avgsnrm \sim 5$ is 15\% and can reach 50\% at lower \avgsnr{}, which is in excellent agreement with the scatter in S\'ersic radius at low signal-to-noise in our tests (Fig. \ref{fig:r_grid}). The analytic uncertainty is a somewhat weaker function of resolution than what we saw in Fig. \ref{fig:r_grid}, likely due to the complex structure of real galaxies. Nevertheless, our analytic approach perfectly predicts the observed scatter in  the S\'ersic radius measurements as a function of \avgsnr{}, which is quite remarkable.

The intrinsic uncertainty in $n$ predicted from our model is somewhat lower than the observational scatter seen in Fig. \ref{fig:sersic_grid}, where the scatter was $\Delta n \sim 1$ at \reseff{}$=10$ for all \avgsnr{}. In contrast, the model allows stricter constraints on $\Delta n$ when \avgsnr{} is high, reaching a $\pm 0.15$ uncertainty. This discrepancy could also be due to the underlying structure of realistic galaxies: e.g., a concentrated nuclear starburst could significantly increase the $n$ estimate when sufficiently resolved. Even our simple model, however, still provides a good indication of the \textit{lower bound} on the uncertainty in $n$ one should expect.

Finally, the range of $\Delta R/R$ and $\Delta n$ allowed by our $\chi^2$ estimates varies as a function of the intrinsic profiles, i.e. the original $n_1$. More early-type profiles are more degenerate with one another, while lower-$n$ profiles are better constrained. To demonstrate this, we plotted the allowed range of $\Delta R/R_1$ (orange shading) and $\Delta n$ (blue shading) for images with $\avgsnrm{}=3$ and $\reseffm = 10$ as a function of $n_1$. Interestingly, while $\Delta n$ increases with $n_1$ as expected, the $\Delta R/R_1$ estimate is fairly constant with $n_1$, staying at approximately 20\%.

\section{Petrosian magnitude deficit for S\'ersic profiles}
\label{sec:appendix_g}

As discussed in Section~3.2.1, Petrosian radii ($R_p$) are robust to changes in signal-to-noise but capture a decreasing fraction of a galaxy's total flux as the S\'ersic index $n$ increases \citep{Graham2005}. Because \texttt{statmorph} uses $R_p$ to define the outer edge of the galaxy for total flux measurements---specifically integrating out to a radius defined by the \texttt{petro\_extent\_cas} parameter---this profile dependence has direct implications for the measured magnitudes and downstream morphological parameters.

\begin{figure}[ht!]
    \centering
    \includegraphics[width=\columnwidth]{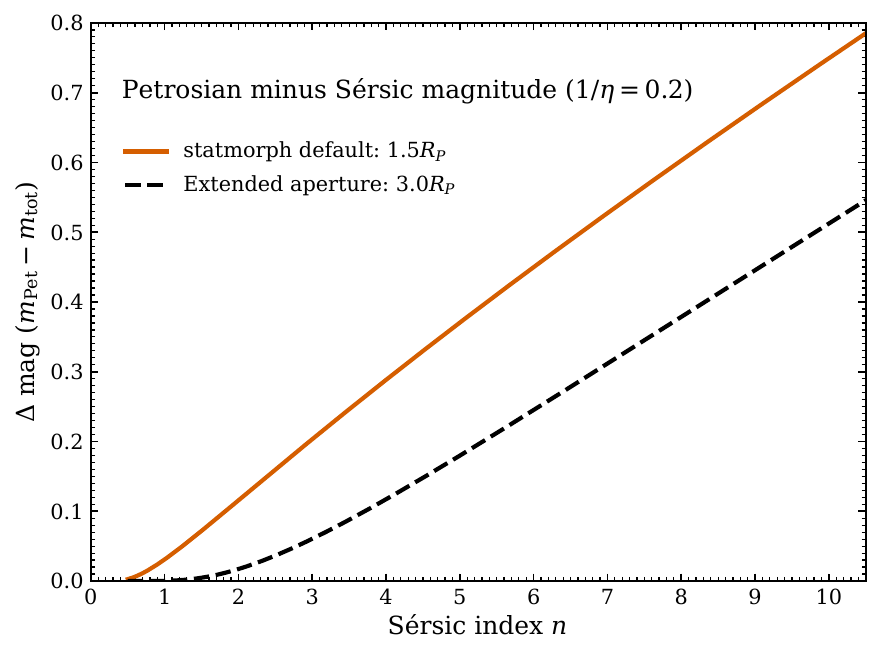}
    \caption{The magnitude deficit ($\Delta$ mag) between the measured Petrosian magnitude and the true total S\'ersic magnitude as a function of the S\'ersic index $n$. The orange line shows the deficit using the \texttt{statmorph} default aperture of $1.5 R_p$ (\texttt{petro\_extent\_cas} = 1.5), while the dashed black line shows an extended aperture of $3.0 R_p$. High-$n$ galaxies require larger apertures to avoid systematically underestimating their total flux and, consequently, their downstream size and concentration metrics. Equations used to derive these curves can be found in \citet{Graham2005}.}
    \label{fig:petro_sersic}
\end{figure}

Figure~\ref{fig:petro_sersic} quantifies the missed flux (measured as $\Delta {\rm mag} = m_{\rm Pet} - m_{\rm tot}$) for a pure S\'ersic profile using the standard Petrosian definition ($1/\eta = 0.2$). For a typical disk galaxy ($n \approx 1$), the default \texttt{statmorph} aperture of $1.5 R_p$ captures the vast majority of the light. However, for an early-type galaxy ($n=4$), the default $1.5 R_p$ aperture misses a significant fraction of the extended light envelope, resulting in a deficit of $\sim 0.28$ magnitudes. While increasing \texttt{petro\_extent\_cas} to $3.0 R_p$ mitigates this deficit, users must balance this against the increased inclusion of sky noise and potential nearby contaminants.

\end{document}